\documentclass[aps,pra,twocolumn,superscriptaddress]{revtex4}
\usepackage[utf8]{inputenc}
\usepackage{amsmath}
\usepackage{amsfonts}
\usepackage{amssymb}
\usepackage{mathtools}
\usepackage{nicefrac}
\usepackage{graphicx}
\usepackage{hyperref}
\usepackage[caption=false]{subfig}
\usepackage{xcolor}
\usepackage{booktabs}
\usepackage{comment}

\newcommand{\R}{{\mathbb{R}}}

\newcommand{\Four}{{\cal F}}

\newcommand{\rv}{\mathbf{r}}

\newcommand{\kv}{\mathbf{k}}
\newcommand{\td}{\text{d}}
\newcommand{\im}{\text{i}}

\newcommand{\dens}{\rho}
\graphicspath{{Figures/}}

\begin{document}

\author{Juri Grossi} 
\affiliation{Department of Chemistry \& Pharmaceutical Sciences and Amsterdam Institute of Molecular and Life Sciences (AIMMS), Faculty of Science, Vrije Universiteit, De Boelelaan 1083, 1081HV Amsterdam, The Netherlands}
\author{Ziad H. Musslimani}
\affiliation{Department of Chemistry \& Pharmaceutical Sciences and Amsterdam Institute of Molecular and Life Sciences (AIMMS), Faculty of Science, Vrije Universiteit, De Boelelaan 1083, 1081HV Amsterdam, The Netherlands}
\affiliation{Department of Mathematics, Florida State University, Tallahassee, FL 32306-4510, 
USA.}
\author{Michael Seidl}
\affiliation{Department of Chemistry \& Pharmaceutical Sciences and Amsterdam Institute of Molecular and Life Sciences (AIMMS), Faculty of Science, Vrije Universiteit, De Boelelaan 1083, 1081HV Amsterdam, The Netherlands}
\author{Paola Gori-Giorgi}
\affiliation{Department of Chemistry \& Pharmaceutical Sciences and Amsterdam Institute of Molecular and Life Sciences (AIMMS), Faculty of Science, Vrije Universiteit, De Boelelaan 1083, 1081HV Amsterdam, The Netherlands}

 \title{Kohn-Sham equations with functionals from the strictly-correlated regime: Investigation with a spectral renormalization method}

\date{\today}
\begin{abstract}
We re-adapt a spectral renormalization method, introduced in nonlinear optics, to solve the Kohn-Sham (KS) equations of density functional theory (DFT), with a focus on functionals based on the strictly-correlated electrons (SCE) regime, which are particularly challenging to converge. Important aspects of the  method are: (i) the eigenvalues and the density are computed simultaneously; (ii) it converges using randomized initial guesses; (iii) easy to implement. Using this method we could converge for the first time the Kohn-Sham equations with functionals that include the next leading term in the strong-interaction limit of density functional theory, the so-called zero-point energy (ZPE) functional as well as with an interaction-strength-interpolation (ISI) functional that includes both the exact SCE and ZPE terms. This work is the first building block for future studies on quantum systems confined in low dimensions with different statistics and long-range repulsions, such as localization properties of fermions and bosons with strong long-range repulsive interactions in the presence of a random external potential. 
\end{abstract}
\maketitle
\section{Introduction}\label{intro}
Capturing the effects of the interactions between the particles of a quantum system in a computationally efficient way is of crucial importance in many areas of Physics and Chemistry. Particle-particle interactions not only determine many of the fundamental physical properties of the system under study, but also play a crucial role regarding practical applications in fields ranging from materials science to theoretical and computational Chemistry, atomtronics, spintronics and quantum information, to name a few. 

Computationally efficient approximate methods that target interacting quantum particles in real space (i.e., without resorting to lattice Hamiltonians) are mainly based on single-particle equations for a set of orbitals $\phi_i(\rv)$, e.g., Gross-Pitaevskii for bosons, Hartree-Fock (HF) and Kohn-Sham (KS) Density Functional Theory (DFT) for fermions. They  all rely on an ansatz to transform the particle-particle interactions into an effective one-body potential that depends non-linearly on the orbitals $\phi_i(\rv)$. The problem is then reduced to a set of non-linear single-particle Schrödinger  equations,  requiring the search for a self-consistent solution. These methods, within the current approximations for the effective potential, typically fail, even at the qualitative level, when the physics of the system under study differs too much from the one of non-interacting particles.

A different class of approximations to transform the particle-particle interactions into an effective one-body potential has emerged in the recent years, based on the semiclassical limit of the many-electron Schrödinger equation taken at fixed single-particle density \cite{MalGor-PRL-12,MalMirCreReiGor-PRB-13,MenMalGor-PRB-14,MalMirMenBjeKarReiGor-PRL-15,KhoLinLin-ARXIV-19}. The formalism, called ``strictly-correlated-electrons'' (SCE) functional, corresponds to the strong-coupling limit of KS DFT for the many-electron problem \cite{Sei-PRA-99,SeiGorSav-PRA-07,GorVigSei-JCTC-09,Lew-CRM-18,CotFriKlu-ARMA-18}, and can be generalised to other particles (bosons and fermions) with repulsive long-ranged interactions \cite{MalMirMenBjeKarReiGor-PRL-15}. The use of the SCE one-body potential in the KS equations has a very distinctive attractive feature: the results for the total energy, for the single-particle density and for the chemical potential become asymptotically close to the ones of the exact many-body problem as the system approaches the limit in which particle-particle interactions dominate over the kinetic energy \cite{MalGor-PRL-12,MalMirCreReiGor-PRB-13,MenMalGor-PRB-14,MalMirMenBjeKarReiGor-PRL-15,MirSeiGor-PRL-13}, which is the regime where current approximations typically break down completely. In (quasi) one-dimensional systems, the SCE potential has a known form \cite{Sei-PRA-99,ColDepDim-CJM-15,MalGor-PRL-12,MalMirCreReiGor-PRB-13} in terms of integrals of the single-particle density, with computational cost, in principle, similar to the one of the local-density approximation (LDA). In 2 and 3 dimensions, an accurate (although not always exact) form is known for spherically symmetric systems \cite{SeiGorSav-PRA-07,MenMalGor-PRB-14,SeiDiMGerNenGieGor-arxiv-17}, while for general geometry one could resort to approximations inspired to the SCE mathematical structure \cite{VucGor-JPCL-17,Vuc-JCTC-19,GouVuc-JCP-19} or to algorithms from the optimal transport (OT) community \cite{MenLin-PRB-13,BenCarCutNenPey-SIAM-15,FriVog-SIMAJMA-18,KhoYin-SIAMJSC-19,LinHoCutJor-ARXIV-19,KhoLinLin-ARXIV-19,AlfCoyEhrLom-arxiv-19}, as SCE maps into a multimarginal OT problem \cite{ButDepGor-PRA-12,CotFriKlu-CPAM-13}. 
 While the systems studied in Chemistry are usually far from the limit in which KS SCE becomes accurate, this is not the case for many interesting physical systems, such as electrons confined at the interface of semiconductor heterostructures or dipolar and charged cold atoms \cite{GhoGucUmrUllBar-NP-06,RonCavBelGol-JCP-06,GhoGucUmrUllBar-PRB-07,MalMirCreReiGor-PRB-13,MenMalGor-PRB-14}. Their physics can be captured with Hamiltonians in the continuum, with long-ranged repulsive interactions, often in 1 or 2 dimensions, with external potentials that drive the system close to the regime where KS SCE becomes accurate \cite{MalMirCreReiGor-PRB-13,MenMalGor-PRB-14,MalMirMenBjeKarReiGor-PRL-15}. This opens a realm of very interesting problems that can be studied, such as ground-state and dynamical (via the time-dependent extension of DFT, TD DFT) properties of fermions and bosons with strong long-range correlations in the presence of disorder.

The non-linearity introduced by the SCE potential in the corresponding single-particle equations, however, is very different from the one of standard approximations, and, especially close to the (most interesting) strongly-correlated regime, our experience is that convergence is difficult to reach, with a crucial role played by the starting guess for the self-consistent iteration. A similar observation is also reported in Ref.~\onlinecite{KhoLinLin-ARXIV-19}. This is probably an inherent feature of KS DFT, as it was observed also with the {\em exact} potential built by reverse engineering accurate solutions of the many-body system \cite{WagBakStoBurWhi-PRB-14}. Particularly the dependence on the initial guess is a very limiting factor for the study of systems in the presence of randomness. 

Moreover, the next leading order in the strong-coupling expansion of DFT, the so called zero-point energy (ZPE) functional \cite{GorVigSei-JCTC-09}, also provides a very interesting approximation that includes kinetic-correlation effects and has an involved non-local density dependence. Only very recently we have computed its functional derivative  \cite{GroSeiGorGie-PRA-19}, showing that the resulting effective one-body potential has divergences that make the convergence of the self-consistent KS equations very challenging. 

In Ref.~\cite{AblMus-OL-05} Ablowitz and Musslimani proposed a spectral renormalization (SR) scheme (in the field of non-linear optics) to compute localized solutions in non-linear waveguides, which is quite general and converges very easily. The method has been used to solve the Gross-Pitaevskii equation for bosons in several interesting cases \cite{MusYan-JOSA-04,AblHor-EPJST-09,AblFokMus-JFM-06,AblAntBakIla-PRA-12,AkkGhoMus-JPB-08}. The core idea is to re-cast the single-particle equations in  Fourier space, which are then solved using a renormalized fixed-point iteration (see Appendix.~\ref{app:spectralrenorm} for more details). Its main strengths are:  easy to implement; the ground state density and eigenvalues are computed simultaneously; shows great robustness with respect to the initial guess (including random initial guess). 

The aim of this work is to adapt the SR method to solve the KS equations, in order to build a solid basis for studying in future works the challenging physics of systems with long-range repulsions in the presence of randomness, using the strictly-correlated functionals. We focus on (quasi) one-dimensional systems (quantum wires) interacting via the effective Coulomb repulsion renormalized at the origin, to take into account the thickness of the wire \cite{BedSzaChwAda-PRB-03,GiuVig-BOOK-05,CasSorSen-PRB-06,AbePolXiaTos-EJPB-07}, for which the SCE potential can be always constructed exactly \cite{Sei-PRA-99,ColDepDim-CJM-15,MalGor-PRL-12,MalMirCreReiGor-PRB-13}.  
Using this method, we were able to also obtain for the first time converged self-consistent KS results with the SCE+ZPE functional, and with an interactiong-strenght interpolation (ISI) functional that includes exact exchange, and the exact SCE and ZPE terms. We also analyze the local-density approximation, for which we find, in one case, a different self-consistent solution than the one which was found independently by two different groups, providing evidence that our solution is the correct one. 

The paper is organised as follows: in Sec.~\ref{sec:theorback} we provide an introductory theoretical background to KS-DFT, including the approximations we are using for the exchange-correlation potential. Next, in  Sec.~\ref{sec:numericimplementation} the SR method is formally outlined for the KS scheme, including details of the numerical implementation for one-dimensional systems. Results are presented and discussed in Sec.~\ref{sec:results}, with conclusions and perspectives in the last Sec.~\ref{sec:conclusions}.

\section{Theoretical background}\label{sec:theorback}
We consider quantum mechanical systems of $N$ identical interacting particles in a given external single-particle potential $v_{\rm ext}(\rv)$,
described by hamiltonians of the kind
\begin{equation}\label{eq:genham}
\hat{H}=\hat{T}+\hat{V}_{\mathrm{int}}+\hat{V}_{\rm ext},
\end{equation}
where, $\hat{T}=-\frac12\sum_{i=1}^N\nabla_i^2$ is the kinetic energy operator of the $N$ particles, while
\begin{equation}\label{Vint}
\hat{V}_{\mathrm{int}}=\frac12\sum_{\substack{i,j=1 \\ i\neq j}}^{N}v_{\mathrm{int}}\big(|\rv_i-\rv_j|\big)
\end{equation}
is the operator of a two-body interaction between these particles, which we consider here isotropic, $v_{\mathrm{int}}(r)$.
The external potential is a local one-body operator,
\begin{equation}
\hat{V}_{\rm ext}=\sum_{i=1}^Nv_{\rm ext}(\rv_i).
\end{equation}

\subsection{Density Functional Theory}
Given a $N$-body wave function $\Psi(u_1,\dots,u_N)$, where the symbol $u_i=\rv_is_i$ comprises both position ($\rv_i$) and (if applicable) spin 
variable ($s_i$) of the $i$th particle, the corresponding density $\rho(\rv)$ is defined as
\begin{align}\label{eq:rhodef}
\rho(\rv)=N\sum_{s_1,\dots ,s_N}\int\td\rv_2\dots\mathrm{d}\rv_N\,\Big|\Psi(\rv s_1,u_2,\dots,u_N)\Big|^2.
\end{align}
Clearly, $\rho$ integrates to the particle number,
\begin{align}
\int\td\rv\,\rho(\rv)=N.
\end{align}
The curse of dimensionality (as the number of particles increases) in the search for the ground state energy $E_0$ of Eq.~\eqref{eq:genham} is addressed in DFT by  rewriting the problem as a nested minimisation, namely
\begin{equation}\label{EminF}
E_0=\min_\rho\left\{F[\rho]+\int\td\rv\,v_{\rm ext}(\rv)\,\rho(\rv)\right\},
\end{equation}
where the Hohenberg-Kohn \cite{HohKoh-PR-64} functional $F[\rho]$, in the Levy \cite{Lev-PNAS-79} constrained-search formulation, is
\begin{equation}\label{eq:HFfunctional}
F[\rho]=\min_{\Psi\to\rho}\big\langle\Psi\big|\hat{T}+\hat{V}_{\rm int}\big|\Psi\big\rangle,
\end{equation}
with ``$\Psi\to\rho$'' meaning that the search is performed over all possible wavefunctions (with the same statistics of the particles of the many-body system under study) that yield, via Eq.~\eqref{eq:rhodef}, the density $\rho$.
The functional $F[\rho]$ is ``universal" in the sense that, once the two-body interaction $v_{\rm int}(r)$ and the particle statistics is specified, $F[\rho]$ is a pure functional of $\rho$, valid for all possible external potentials $v_{\rm ext}(\rv)$.
\subsection{Kohn-Sham Equations}
The challenge is of course to find good approximations for $F[\rho]$, able to take into account the particle statistics and the particle-particle interactions $\hat{V}_{\rm int}$. 
In KS-DFT, $F[\rho]$ is divided up into three pieces,
\begin{equation}\label{Fdiv}
F[\rho]=T_{\rm s}[\rho]+U[\rho]+E_{\rm xc}[\rho],
\end{equation}
where $T_{\rm s}[\rho]$ is defined as 
\begin{align}\label{eq:Ts}
T_{\rm s}[\rho]=\min_{\Psi\to\rho}\big\langle\Psi\big|\hat{T}\big|\Psi\big\rangle.
\end{align}
Again, the constrained search is restricted over wavefunctions having the same statistics as the one of the many-body system under study. The Hartree functional $U[\rho]$ is the usual mean-field (direct) term
\begin{align}\label{eq:hartreeenergy}
U[\rho]=\frac12\int\td\rv_1\td\rv_2\,v_{\rm int}\big(|\rv_1-\rv_2|\big)\,\rho(\rv_1)\,\rho(\rv_2),
\end{align}
and the unknown exchange-correlation (xc) energy functional $E_{\rm xc}[\rho]$ is defined by Eq.~\eqref{Fdiv}, and must be approximated (see Sec.~\ref{sec:KSApprox} below).

Since $\hat{T}$ is a one-body operator, the minimising wavefunction in Eq.~\eqref{eq:Ts} for a given $\rho$ is usually a non-interacting state $\Psi=\Phi$ formed by single-particle orbitals $\phi_i(\rv)$, whose occupation numbers are dictated by the particle statistics. The minimisation with respect to the density $\rho$ of the energy of the system under study, given by Eqs~\eqref{EminF} and \eqref{Fdiv}, is then rewritten as
\begin{equation}\label{eq:KSDFTAndreas}
	E_0=\min_{\Phi}\left\{\langle\Phi|\hat{T}+\hat{V}_{\rm ext}|\Phi\rangle+U[\rho_\Phi]+E_{\rm xc}[\rho_\Phi]\right\},
\end{equation}
where the notation $U[\rho_\Phi]$ and $E_{\rm xc}[\rho_\Phi]$ means that these functionals depend on $\Phi$ only through its density $\rho=\rho_\Phi$, computed by inserting $\Phi$ in Eq.~\eqref{eq:rhodef}. The Euler-Lagrange equations for the minimisation \eqref{eq:KSDFTAndreas} are the KS single particle equations,
\begin{equation}
\underbrace{\left(-\frac12\nabla^2 + v_{\rm ext}(\rv) + v_{\rm Hxc}\big([\rho],\rv\big) \right) }_{\equiv\hat{h}}\phi_i(\rv) = \varepsilon_i \phi_i(\rv) \;.
\label{KSeq}
\end{equation}
Here, $\rho=\rho_\Phi$ is the density of the occupied orbitals $\phi_i$ with the lowest eigenvalues $\varepsilon_i$,
\begin{equation}\label{eq:rhoF}
\rho(\rv)=\sum_{i=1}^{i_{\rm max}} n_i |\phi_i(\rv)|^2,
\end{equation}
where, for example, $i_{\rm max}=1$ and  $n_1=N$ for bosons, $i_{\rm max}=N/2$ and all the $n_i=2$ for an even number of spin-$1/2$ fermions, etc. 

In Eq.~\eqref{KSeq}, the Hartree plus exchange-correlation potential $v_{\rm Hxc}([\rho],\rv)$,
\begin{equation}
v_{\rm Hxc}\big([\rho],\rv\big)=v_{\rm H}\big([\rho],\rv\big)+v_{\rm xc}\big([\rho],\rv\big),
\label{vsKS}
\end{equation}
is the single-particle potential that should embody the effects of particle-particle interactions via a non-linear dependence on the particle density, with $v_{\rm H}\big([\rho],\rv\big)$ given by the functional derivative of Eq.~\eqref{eq:hartreeenergy} with respect to the density  
\begin{equation}\label{eq:vH}
v_{\rm H}\big([\rho],\rv\big)=\int\td\rv'\,v_{\rm int}\big(|\rv-\rv'|\big)\,\rho(\rv') \;,
\end{equation}
and, similarly, $v_{\rm xc}\big([\rho],\rv\big)$ the exchange-correlation potential defined as
\begin{equation}\label{eq:generalvxc}
v_{\rm xc}\big([\rho],\rv\big)=\frac{\delta E_{\rm xc}[\rho]}{\delta\rho(\rv)}.
\end{equation}
\subsection{Approximations for the xc functional}\label{sec:KSApprox}
The KS construction gives a way to include the main effects of particle statistics in the energy density functional, by invoking a non-interacting system with the same density and particle statistics of the physical, interacting, one. The whole problem is then reduced to finding suitable approximations for $E_{\rm xc}[\rho]$ and its functional derivative $v_{\rm xc}\big([\rho],\rv\big)$, Eq.~\eqref{eq:generalvxc}. While in Chemistry hundreds of different approximations for the case of electrons in 3D are available, here we focus on approximations that can be used to study model systems in Physics, confined in low dimensions, at low density, and with different interactions and particle statistics.

\subsubsection{Hartree approximation}
In this case, we simply set $v_{\rm xc}\big([\rho],\rv\big)\approx0$ in Eq.~\eqref{vsKS}.
The Hartree approximation corresponds to treat particles as if they were interacting only with an effective mean field generated by the charge distribution $\rho(\rv)$.
By neglecting both exchange and correlation effects, the Hartree potential of Eq.~\eqref{eq:vH} is the same regardless the statistics of the particles.

If we consider bosons interacting with a contact interaction $v_{\rm int}(\vert\mathbf{r}-\mathbf{r}'\vert)=g\,\delta(\vert\mathbf{r}-\mathbf{r}'\vert)$, via this approximation Eq.~\eqref{KSeq} reduces to the Gross-Pitaevskii equation.


\subsubsection{Local density approximation (LDA)}
In the LDA, one first computes the xc energy per particle $\epsilon_{\rm xc}(\dens)$ of a uniform quantum gas with a constant density $\dens$. The particles interact via the same $v_{\rm int}(r)$ and have the same statistics (bosons, fermions with different spins) we aim to treat. The xc energy is then obtained by replacing locally the uniform density of the quantum gas with $\dens(\mathbf{r})$ and averaging over all space:
\begin{equation}
E^{\rm LDA}_{\rm xc}[\rho]=\int\td\rv\,\rho(\rv)\,\epsilon_{\rm xc}\big(\rho(\rv)\big).
\end{equation}
Typically, the xc energy of the uniform quantum gas is computed via Quantum Monte Carlo (QMC) or other many-body methods and then parametrized as a function of $\dens$, taking into account known asymptotic properties at high- ($\rho\to \infty$) and low-density ($\rho\to 0$). 

In 3D, parametrizations based on QMC data are available for spin-$1/2$ fermions (in different spin-polarisation states)  with the Coulomb interaction \cite{CepAld-PRL-80,PerZun-PRB-81,VosWilNus-CJP-80,PerWan-PRB-92}, contact interaction~\cite{MaPilTroDai-NAT-12}, as well as Coulomb interactions screened at long-\cite{ZecGorMorBac-PRB-04} and short-range \cite{TouSavFla-IJQC-04,PazMorGorBac-PRB-06}. 

In 2D a parametrization of QMC data for the interaction $1/r$ (as one is usually interested in systems interacting with the Coulomb interaction in 3D, with a strong confinement in one direction)  is available for spin-$1/2$ fermions (again, in different spin states) \cite{AttMorGorBac-PRL-02} and for bosons \cite{DepConMor-PRB-04}. A formula for fermions having spin higher than $1/2$ (or other degrees of freedom)  is also available \cite{KarKosReiMan-PRB-03}, based on an interpolation between electrons and bosons. 

For the 1D case, if we are interested in modelling systems repelling via the Coulomb interaction and strongly confined in 2 directions, we need to regularise the $1/r$ divergence at the origin as, otherwise, i) the wavefunction is forced to have nodes at coalescence of two particles, while in a quasi-1D system this does not happen for unlike spin particles or for bosons, and ii) the Hartree potential diverges.
A possible choice, which will be adopted in this work, is the interaction \cite{GiuVig-BOOK-05,CasSorSen-PRB-06}
\begin{equation}\label{eq:vintQ1D}
v^{\rm Q1D}_{\rm int}(x)=\frac{\sqrt{\pi}}{2b}\exp\left(\frac{x^2}{4b^2}\right)\text{erfc}\left(\frac{|x|}{2b}\right),
\end{equation} 
obtained by integrating the 3D Coulomb interaction $1/r$ over oscillator wave functions of thickness $b$ in two directions. While $v^{\rm Q1D}_{\rm int}(0)=\frac{\sqrt{\pi}}{2b}$ is finite, mimicking the effect of finite thickness, the Coulomb potential is recovered from its long range asymptotics,
\begin{equation}
v^{\rm Q1D}_{\rm int}(x)\to\frac1{|x|}\qquad\big(|x|\gg b\big).
\end{equation}
A LDA parametrisation based on QMC results for this interaction is available for different values of $b$  \cite{CasSorSen-PRB-06} and is reported in Appendix~\ref{app:NumDet} for convenience, for the case $b=0.1$ considered here.
A 1D LDA is also available for soft Coulomb interaction, $v_{\rm soft}(x)=(a^2+x^2)^{-1/2}$ \cite{HelFukCasVerMarTokRub-PRA-11}.

From Eq.~\eqref{eq:generalvxc} we have then
\begin{eqnarray}
v^{\rm LDA}_{\rm xc}\big([\rho],\rv\big)&=&\frac{\delta E^{\rm LDA}_{\rm xc}[\rho]}{\delta\rho(\rv)}\nonumber\\
&=&\epsilon_{\rm xc}\big(\rho(\rv)\big)+\rho(\rv)\,\epsilon'_{\rm xc}\big(\rho(\rv)\big).
\end{eqnarray}

\subsubsection{Strictly-correlated-electrons (SCE)}
In the KS SCE scheme \cite{MalGor-PRL-12,MalMirCreReiGor-PRB-13,MenMalGor-PRB-14,KhoLinLin-ARXIV-19}, the Hartree plus exchange-correlation functional $E_{\rm Hxc}[\rho]=U[\rho]+E_{\rm xc}[\rho]$ is approximated with the strictly-correlated functional~\cite{Sei-PRA-99,SeiGorSav-PRA-07}
\begin{equation}\label{eq:defSCE}
V^{\rm SCE}_{\rm int}[\rho]=\inf_{\Psi\to\rho}\langle\Psi\vert\hat{V}_{\rm int}\vert\Psi\rangle.
\end{equation}
Accordingly, the Hartree plus exchange correlation potential is replaced by the SCE potential:
\begin{equation}
v_{\rm Hxc}\big([\rho],x\big)\;\approx\;\tilde{v}_{\rm SCE}\big([\rho],x\big)=\frac{\delta V^{\rm SCE}_{\rm int}[\rho]}{\delta\rho(x)}.
\end{equation}
The functional $V^{\rm SCE}_{\rm int}[\rho]$ has been introduced for the case of electrons (Coulomb interactions)\cite{Sei-PRA-99,SeiGorSav-PRA-07}, and describes a semi-classical problem with prescribed single-particle density. As such, the infimum in Eq.~\eqref{eq:defSCE} is reached on a distribution which does not depend on the spin variables. In other words, the functional $V^{\rm SCE}_{\rm int}[\rho]$ is the same for all particle statistics (for a rigorous proof, see Refs.~\cite{Lew-CRM-18,CotFriKlu-ARMA-18}), and can be thus combined with the $T_s[\rho]$ having the statistics we want to describe \cite{MalMirMenBjeKarReiGor-PRL-15}. Also, another advantage of the SCE functional is that it can be constructed for different long-range repulsive interactions \cite{MalMirMenBjeKarReiGor-PRL-15} without requiring a parametrisation of the corresponding uniform quantum gas. Moreover, in the limit in which the interactions among the particles become dominant, the KS equations with the SCE functional approach asymptotically the exact many-body ground-state energy, density, and chemical potential \cite{Lew-CRM-18,CotFriKlu-ARMA-18}. 

We refer the reader to Refs.~\cite{SeiGorSav-PRA-07,MalMirCreReiGor-PRB-13,MenMalGor-PRB-14} and~\cite{MalMirMenBjeKarReiGor-PRL-15} for the SCE theory, while here we outline how to construct $\tilde{v}_{\rm SCE}\big([\rho],x\big)$ for the case $D=1$ \cite{Sei-PRA-99}, which has been proven \cite{ColDepDim-CJM-15} to yield the exact solution to the problem posed by Eq.~\eqref{eq:defSCE} when the function $v_{\rm int}(x)$ is convex, as it is the case for Eq.~\eqref{eq:vintQ1D}. For a given density $\rho(x)$ with $N$ electrons in one dimension, we introduce a sequence of $N$ co-motion functions $f_1([\rho],x),\dots,f_N([\rho],x)$ defined as follows: $f_1([\rho],x)=x$; for $n\ge2$, $f_n([\rho],x)$ has a pole at location $x=a_n$ fixed by the condition
\begin{equation}
\int_{a_n}^{\infty}\td t\,\rho(t)=n-1.
\end{equation}
In terms of $a_n$, the functions $f_n([\rho],x)$ are fixed by
\begin{eqnarray}\label{eq:comodef}
\text{for }x<a_n:~~&&\int_{x}^{f_n([\rho],x)}\hspace*{-1.5mm}\td t\,\rho(t)=n-1,\nonumber\\\nonumber\\
\text{for }x>a_n:~~&&\int_{f_n([\rho],x)}^{x}\td t\,\rho(t)=N-(n-1).
\end{eqnarray}
Note that the co-motion functions form a group (with respect to composition) with $N$ elements satisfying
\begin{equation}
f_m\big(f_n(x)\big)=f_{\mathrm{mod}_N[m+n-1]}(x).
\end{equation}
The SCE potential is given in terms of the $f_n(x)$
\begin{equation}
\tilde{v}_{\rm SCE}\big([\rho],x\big)=\sum_{i=2}^N\int_{-\infty}^x\td y\,v'_{\rm int}\big(y-f_i([\rho],y)\big).
\label{vSCE}
\end{equation}

\subsubsection{SCE plus Zero Point Energy (ZPE)}
The SCE approximation treats electrons semiclassically, by neglecting any kinetic correlation contribution. The next leading term in the semiclassical expansion can be written as Zero Point Energy (ZPE) oscillations in a metric dictated by the density \cite{GorVigSei-JCTC-09}. Although there is no rigorous proof, there is numerical evidence in simple cases that for a fixed density $\rho$ the exact Hohenberg-Kohn functional approaches the SCE plus ZPE functional in the limit of strong coupling \cite{GroKooGieSeiCohMorGor-JCTC-17}.

The ZPE functional is written as $D(N-1)$ oscillator energies $\frac{1}{2}\hbar \omega_\mu$ (here we are setting $\hbar=1$) given by functionals $\omega_\mu([\rho],\rv)$ of the density $\dens$, and averaged over space \cite{GorVigSei-JCTC-09},
\begin{equation}
F^{\rm ZPE}[\rho]=\frac{1}{2}\sum_{\mu=D+1}^{ND}\int\mathrm{d}\mathbf{r}\frac{\dens(\mathbf{r})}{N}\omega_\mu([\dens],\mathbf{r}).
\end{equation}
The SCE+ZPE approximation in the KS scheme reads
\begin{equation}
E_{\rm Hxc}[\dens]\approx V^{\rm SCE}_{\rm int}[\rho]+F^{\rm ZPE}[\rho].
\end{equation}
Consequently, in Eq.~\eqref{eq:generalvxc} we approximate
\begin{equation}\label{eq:VHXCZPE}
v_{\mathrm{Hxc}}([\dens],\mathbf{r})\approx \tilde{v}_{\mathrm{SCE}}(\mathbf{r})+\tilde{v}_{\mathrm{ZPE}}([\dens],\mathbf{r})
\end{equation}
with 
\begin{equation}
\tilde{v}_{\mathrm{ZPE}}([\dens],\mathbf{r})=\frac{\delta F^{\rm ZPE}[\dens]}{\delta\rho(\mathbf{r})}.
\end{equation}
This functional derivative is quite involved to compute, and has been obtained analytically only for the simple case of $N=2$ electrons in 1D, see Ref.~\cite{GroSeiGorGie-PRA-19}. In this case, there is only one frequency $\omega_\mu=\omega$ and \cite{GroSeiGorGie-PRA-19}
\begin{equation}\label{eq:funcder}
\tilde{v}_{\mathrm{ZPE}}([\dens],x)=\frac{\omega([\dens],x)}{4}+\frac{1}{4}\int_x^{f(x)}\Lambda([\dens],y)\mathrm{d}y,
\end{equation}
where $\omega([\dens],x)$  reads \cite{Sei-PRA-99,GroSeiGorGie-PRA-19}
\begin{equation}
\omega([\dens],x)=\sqrt{v''_{int}(\vert x-f(x)\vert)\left(\frac{\dens(x)}{\dens(f(x))}+\frac{\dens(f(x))}{\dens(x)}\right)}.
\end{equation}
The functional $\Lambda([\dens],y)$ is defined in terms of the co-motion function $f(x)$ and the density $\dens(x)$,
\begin{align}\label{eq:explilambda}
&\Lambda([\dens],y)\nonumber\\&=\frac{v'''_{\mathrm{int}}(f(y)-y)}{\omega(y)}+\frac{v''_{\mathrm{int}}(f(y)-y)}{\omega(y)}\frac{\dens'(f(y))}{\dens(f(y))}\frac{3f'(y)^2+1}{f'(y)^2+1}
\end{align} Although not immediate from Eq.~\eqref{eq:explilambda}, it can be shown~\cite{GroSeiGorGie-PRA-19} that $\Lambda([\dens],y)$ is a bounded function. 
Therefore, the second term in Eq.~\eqref{eq:funcder} is subleading with respect to $\omega([\dens],x)$ both at $x\sim a_1$ and $x\sim\pm\infty$, since typically $\omega([\dens],x)$ diverges at those points.  As discussed at length in Sec.~\ref{sec:results}, this can have quite relevant consequences on the converged result of a KS scheme.

\subsubsection{ZPE with interaction strength interpolation (ZPEisi)}
The SCE functional is the limit of the Hohenberg-Kohn functional when $\hbar\to 0$. Physically speaking, the SCE approximation provides more and more accurate pieces of information the more the particle-paticle interactions are predominant with respect to the kinetic energy (effective Bohr radius much smaller than the average particle-particle distance).
Since chemical systems are usually not in this regime, in quantum chemistry the SCE limit finds a useful application when combined with an interpolation  along the so-called adiabatic connection \cite{LanPer-SSC-75}, modeling the exact xc energy by connecting the SCE system to the non interacting one \cite{SeiPerLev-PRA-99,SeiPerKur-PRL-00}.
This interaction strength interpolation (ISI) idea, using different forms for the interpolation function \cite{SeiPerLev-PRA-99,SeiPerKur-PRL-00,GorVigSei-JCTC-09,LiuBur-JCP-09} has been extensively tested on chemical systems \cite{SeiPerKur-PRL-00,FabGorSeiDel-JCTC-16,GiaGorDelFab-JCP-18,VucGorDelFab-JPCL-18,FabSmiGiaDaaDelGraGor-JCTC-19,Con-PRB-19,VucFabGorBur-arxiv-20}. It has also been applied successfully to the two-valley electron gas \cite{ZarNeiParPee-PRB-17}. Here we will test a simplified form of the ISI scheme that uses both the SCE and ZPE functionals to interpolate between weak (exact exchange) and strong interaction, proposed in Ref.~\onlinecite{MalMirGieWagGor-PCCP-14}. Within this approximation, that we call here ZPEisi, the Hartree and xc correlation functional reads\cite{MalMirGieWagGor-PCCP-14,GroSeiGorGie-PRA-19} 
\begin{equation}\label{exisi}
E^{\mathrm{ZPEisi}}_{\rm Hxc}[\rho]\approx V_{\mathrm{int}}^{\mathrm{SCE}}[\rho]+\underbrace{F^{\mathrm{ZPE}}[\rho]\left(\sqrt{1+a[\dens]}-\sqrt{a[\rho]}\right)}_{F^{\mathrm{ZPE}}_{\mathrm{isi}}[\rho]},
\end{equation}
with
\begin{align}
a[\dens] &= \left(\frac{F^{\mathrm{ZPE}}[\dens]}{2(E_{\mathrm{x}}[\dens]-(V_{\mathrm{int}}^{\mathrm{SCE}}[\dens]-U_H[\dens]))}\right)^2,\nonumber\\E_\mathrm{x}[\rho]&=\langle\Phi\vert\hat{V}_{\mathrm{int}}\vert\Phi\rangle-U_\mathrm{H}[\rho],
\end{align}
and $\Phi$ the non-interacting wavefunction built with the orbitals solutions of the self-consistent KS equations.
While in the Chemistry literature the ISI functionals have been always used with semilocal approximations for the SCE and the ZPE functionals \cite{SeiPerKur-PRL-00,FabGorSeiDel-JCTC-16,GiaGorDelFab-JCP-18,VucGorDelFab-JPCL-18,FabSmiGiaDaaDelGraGor-JCTC-19,Con-PRB-19,VucFabGorBur-arxiv-20}, here we can test them for the first time, at least in a very simple case, with the full non-local exact functionals from the strictly-correlated regime, using their functional derivatives in the self-consistent KS equations.


\section{Numerical Implementation and the Spectral Renormalization algorithm}\label{sec:numericimplementation}

The SR algorithm we use is a readaptation of the method of Ref.~\cite{AblMus-OL-05}.
Given the $D$-dimensional forward Fourier transform
\begin{equation}\label{eq:FT}
\hat{\phi}(\kv)\equiv\Four[\phi(\rv)]=\int\td\rv\,\phi(\rv)\,e^{-\im\kv\cdot\rv} \;, 
\end{equation}
and its inverse
\begin{equation}
\label{IFT}
\phi(\rv) \equiv\Four^{-1}[\phi(\kv)]=\frac{1}{(2\pi)^{D/2}}\int\td\kv\, \hat{\phi}(\kv)\,e^{+\im\kv\cdot\rv} \;,
\end{equation}
the KS equation \eqref{KSeq} for each orbital in Fourier space reads
\begin{equation}
\frac{|\kv|^2}2\,\hat{\phi}(\kv)\,+\,\Four\Big[v_\mathrm{KS}\big([\rho],\rv\big)\,\phi(\rv)\Big]=\varepsilon\,\hat{\phi}(\kv),
\label{KSeqFT}\end{equation}
with the full (external plus Hxc) potential given by
\begin{equation}
	v_\mathrm{KS}\big([\rho],\rv\big)=v_{\rm ext}(\rv)+v_{\rm Hxc}\big([\rho],\rv\big).
\end{equation}
Multiplying Eq.~\eqref{KSeqFT} by $(\hat{\phi})^*(\kv)$ and integrating over all space results in
\begin{equation}
\varepsilon =\int\mathrm{d}\mathbf{k}\bigg\lbrace \frac{1}{2} \vert\kv\vert^2 \vert\hat{\phi}(\kv)\vert^2 + (\hat{\phi})^*(\kv)\Four\Big[v_\mathrm{KS}\big([\rho],\rv\big)\,\phi(\rv)\Big]\bigg\rbrace.
\label{epsFT}
\end{equation}
Depending on the type of external potential, the second term on the right hand side of Eq.~\eqref{epsFT} can be either positive (for example for harmonic confinement) or negative (for example for Coulomb attractive external potential).
We thus distinguish between two scenarios: (i) When $\varepsilon<0$, we have $\vert\kv\vert^2 -2\varepsilon\ne0$ for all $\kv\in\R^D$, and Eq.~\eqref{KSeqFT} can be rewritten as
\begin{equation}
\hat{\phi}(\kv)=-\frac{\;\Four\Big[v_\mathrm{KS}\big([\rho],\rv\big)\,\phi(\rv)\Big]\;}{\displaystyle\frac{|\kv|^2}2-\varepsilon}.
\label{KSeqFTresA}
\end{equation}
When the condition $\varepsilon<0$ is not guaranteed, we choose arbitrarily a number $c>0$ and add $c\,\hat{\phi}(\kv)$ on both sides of Eq.~\eqref{KSeqFT}.
Then, instead of Eq.~\eqref{KSeqFTresA}, we can write
\begin{equation}
\hat{\phi}(\kv)=-\frac{\;\Four\Big[v_\mathrm{KS}\big([\rho],\rv\big)\,\phi(\rv)\Big]\;-\;(\varepsilon+c)\,\hat{\phi}(\kv)}{\displaystyle\frac{\vert\kv\vert^2}2+c}.
\label{KSeqFTresb}
\end{equation}
Equations~\eqref{KSeqFTresA} or \eqref{KSeqFTresb}, together with eq.~\eqref{epsFT}, are used for a fixed-point iteration, as schematically shown in Fig.\ref{fig:SRalgorithm}. A more detailed explanation of all the steps is given in Appendix~\ref{app:spectralrenorm}.
\begin{figure}[ht]
{\includegraphics[width=0.45\textwidth]{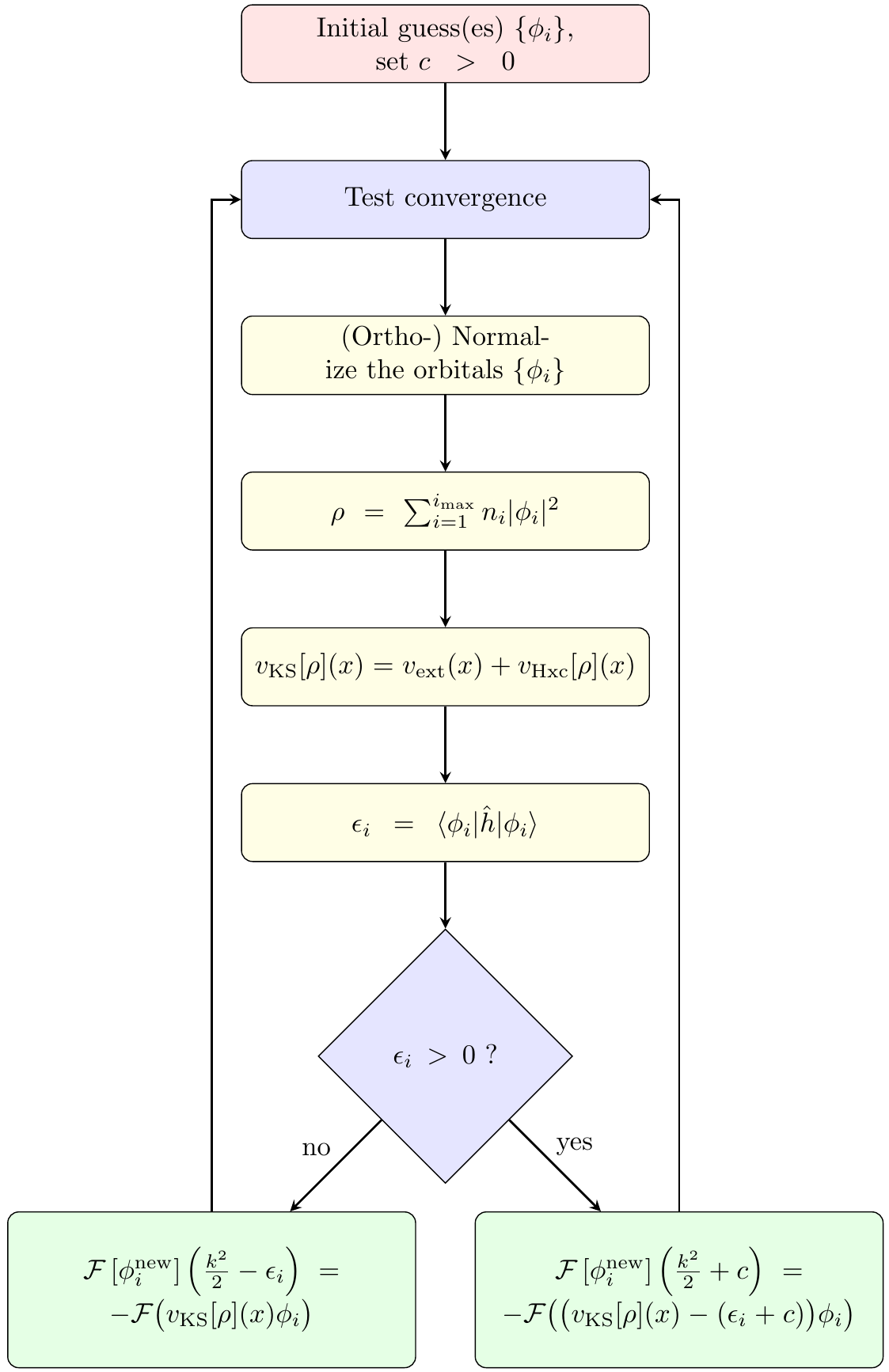}}
\caption{SR algorithm in a nutshell. $\mathcal{F}$ denotes the Fourier transform.}
\label{fig:SRalgorithm}
\end{figure}

The approximations for the xc potential discussed in the previous section, in particular the SCE and the SCE+ZPE ones, introduce a complex non-linearity in the KS equations, making their convergence rather challenging, with a delicate dependence on the initial guess. These aspects of the problem are tackled by
two key features of the SR algorithm, namely (i) the initial guesses for the orbitals (red block in Fig.~\ref{fig:SRalgorithm}) can be taken to be random noise in the interval $[0,1]$ over the whole grid, without affecting the convergence of the algorithm and (ii) at each step, the Schrödinger Equation is \textit{not} solved: instead, by inversion of the Kohn-Sham Hamiltonian in Fourier space, it is used to generate the set of orbitals for the next iteration (green blocks in Fig.~\ref{fig:SRalgorithm}).

As the algorithm converges, the norm of the non-normalised orbitals $\lbrace\phi_i^{\mathrm{new}}\rbrace$ converges to $1$. This can be used as test of convergence. Another option is to compute the Hausdorff distance between two subsequent iterations $\phi^{(j)}$ and $\phi^{(j+1)}$, exiting the loop upon reaching a certain threshold. A final possibility is to check whether the ratio $\frac{\hat{h}\phi}{\epsilon\phi}\approx 1$ everywhere in the domain of interest, or a suitable combination of the three options.
\subsection{Details of the implementation for 1D systems}

We consider 1D systems interacting via Eq.~\eqref{eq:vintQ1D}.
Such interaction is numerically unstable for large arguments, but there are at least two ways to circumvent this issue. The first is to truncate the function $v^{\rm Q1D}(x)$ at some arbitrary \textit{small} value $x_0$ and glue it with its large-$x$ expansion, including the Coulomb tail $1/\vert x\vert$.
An alternative approach (proposed by Weideman and Reddy in Ref.~\cite{WeiRed-ACM-00}) is to derive a first order differential equation for $v_{\rm int}^{\rm Q1D}(x)$  given by 
\begin{equation}
\frac{\mathrm{d}v_{\rm int}^{\rm Q1D}}{\mathrm{d}x} - \frac{x}{2b^2} v_{\rm int}^{\rm Q1D} = -\frac{1}{2b^2} \;,
\end{equation}
which upon a change of variables $x=\frac{s(1+t)}{1-t}$ takes the form \cite{WeiRed-ACM-00}
\begin{equation}
\label{eq:veediffeq}
(1-t)^3 \frac{\mathrm{d}v_{\rm int}^{\rm Q1D}}{\mathrm{d}t} - \frac{s^2}{b^2} v_{\rm int}^{\rm Q1D}  = ( v_{\rm int}^{\rm Q1D} - 1 ) \frac{s}{b^2} \;.
\end{equation}
Equation \eqref{eq:veediffeq} is solved on the domain $t\in [0, 1]$ with the derivative computed spectrally using Chebychev differentiation matrices. We chose the latter in all of our simulations.

\begin{figure}[ht]
\includegraphics[width=.45\textwidth]{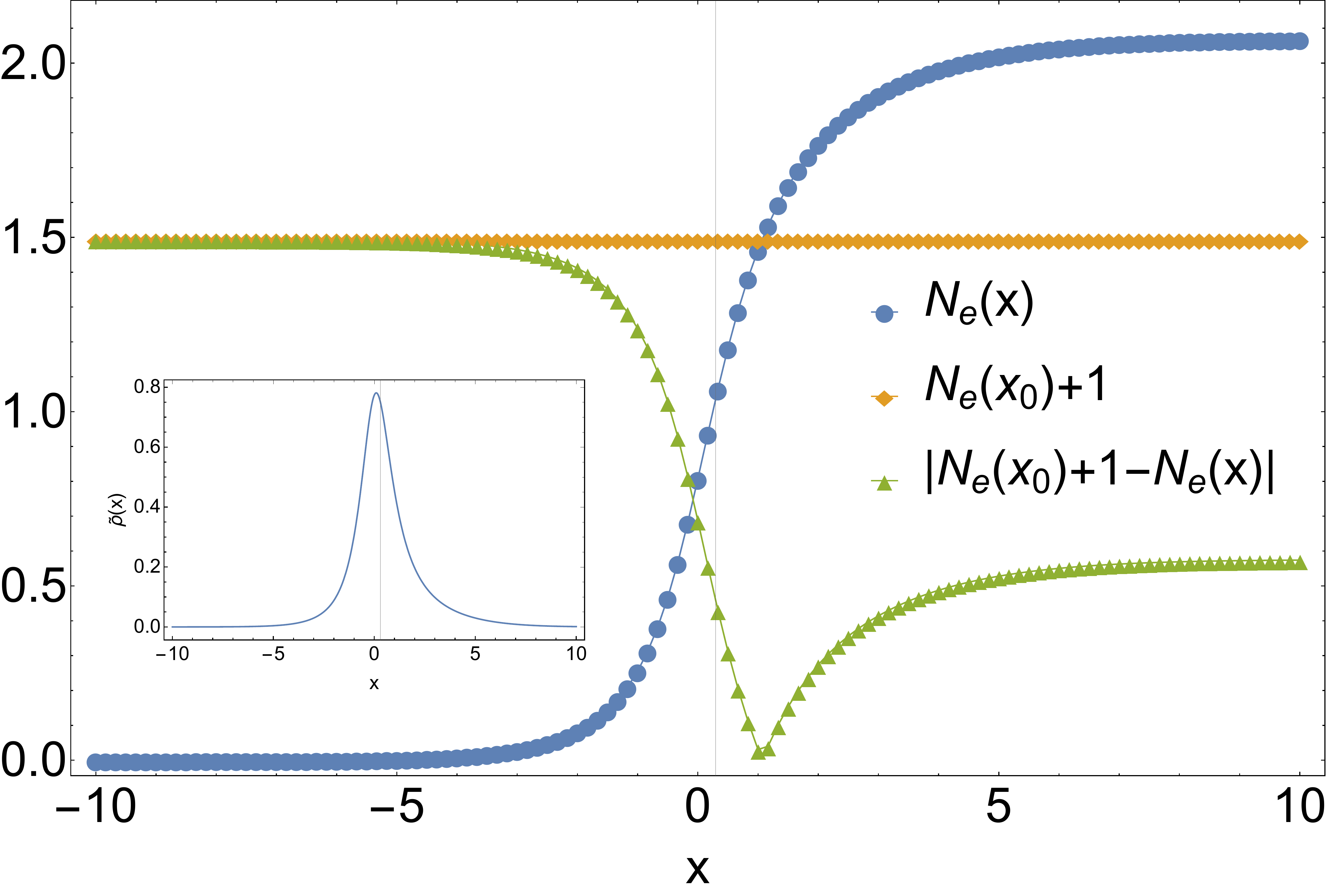}
\caption{Construction of the co-motion function for ${\tilde{\rho}}$ in the point $x_0= -9/20$, for which we can read $f(x_0)\approx 1.15$. Inset: plot of the  density $\tilde{\rho}$ (see main text).
Notice that, being the density non symmetric, $a_1\neq 0$ (vertical line both in the main plot and in the inset).}
\label{fig:comoconstrucplot}
\end{figure}

To explain the numerical implementation of the SCE functional we consider here the special case of $N=2$ (the general case follows straightforwardly).
The co-motion function $f(x)$ must satisfy (see Eq. \eqref{eq:comodef}):
\begin{equation}\label{eq:comodefN2}
\int_x^{f(x)}\rho(t)\mathrm{d} t=\mathrm{sign}(a_1-x)
\end{equation}
In terms of the cumulant function
\begin{equation}
N_e(x)=\int_{-\infty}^x\rho(t)\mathrm{d}t,
\end{equation} 
Eq. \eqref{eq:comodefN2} reads
\begin{equation}
N_e(f(x))=N_e(x)+\mathrm{sign}(a_1-x),
\end{equation}
and $a_1$ is defined by $N_e(a_1)=1$.

For each point $x_0$ we can then find the corresponding $f(x_0)$ by minimisation:
\begin{equation}
f(x_0)=\mathrm{arg}\min_t\bigg\{\vert N_e(x_0)+\mathrm{sign}(a_1-x_0)-N_e(t)\vert\bigg\},
\end{equation}
avoiding the computation of the inverse function $N_e^{-1}(x)$, which is problematic in regions where the density is very small, with the cumulant  $N_e(x)$ approximately constant.
As an example, in Fig.~\ref{fig:comoconstrucplot} we illustrate this procedure for a two electron density $\tilde{\rho}(x)\sim 0.96~\frac{e^{-(0.2x-0.5)^2}}{1+x^2}$. For completness, we report more details on our 1D implementation in Appendix~\ref{app:morenum1D}.

\section{Results}\label{sec:results}
In this section we report numerical test results that are obtained with the external parabolic potential
\begin{equation}\label{eq:extpotexplicit}
v^L_{\mathrm{ext}}(x)=\frac{8}{L^4}x^2,
\end{equation} 
which has been used to model quantum wires \cite{AbePolXiaTos-EJPB-07,BedSzaChwAda-PRB-03}. Furthermore, it was also used, for the first time, in Refs.~\cite{MalMirCreReiGor-PRB-13,MalMirMenBjeKarReiGor-PRL-15} to test the SCE functional as an approximation to the true Hartree and exchange correlation potential in a self-consistent calculation.

The parameter $L$ allows us to adjust the scale of the parabolic external potential which in turn  drives the system continuously   from the weakly correlated regime ($L\ll 1$) to the highly correlated one ($L\gg 1$). Typical values for the constant $c$ range between $c=25$ (for the case $L=1$) and $c=1$ (for $L=70$).

In particular, upon the scaling $x_i\mapsto\tilde{x}_iL$, $b\mapsto\tilde{b}L$, the Hamiltonian takes the form
\begin{equation}\label{eq:scaledhamL}
L^2\hat{H}_L=\sum_i\left(-\frac{1}{2}\frac{\mathrm{d}^2}{\mathrm{d}\tilde{x}_i^2}+v^1_{\mathrm{ext}}(\tilde{x}_i)+L\sum_{j>i}v^{\mathrm{Q1D}}_{\mathrm{int}}(\vert \tilde{x}_i-\tilde{x}_j\vert)\right).
\end{equation}
From Eq.~\eqref{eq:scaledhamL}, we see clearly that as $L$ increases, the interaction term becomes dominant and therefore the system becomes more correlated.
\subsection{LDA and SCE approximations}
To probe the robustness of the algorithm with respect to the initial guess, we solve the KS LDA and KS SCE  equations by starting in both cases the iteration  with random initial orbitals sampled from a uniformly distributed density function.

\begin{figure*}
\centering
\subfloat[][$L=1$]
   {\includegraphics[width=0.4\textwidth]{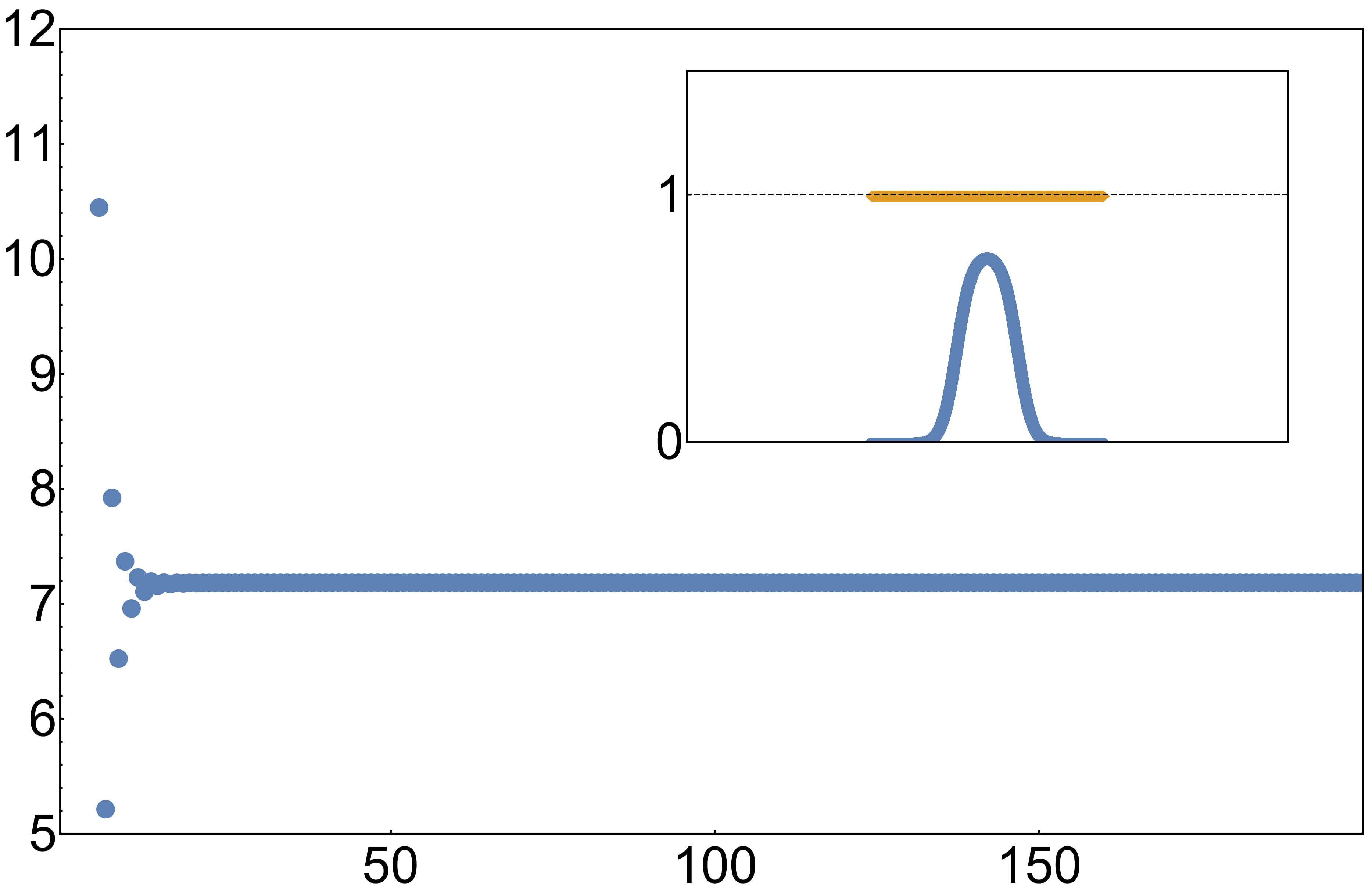}}~
\subfloat[][$L=1$]
   {\includegraphics[width=0.4\textwidth]{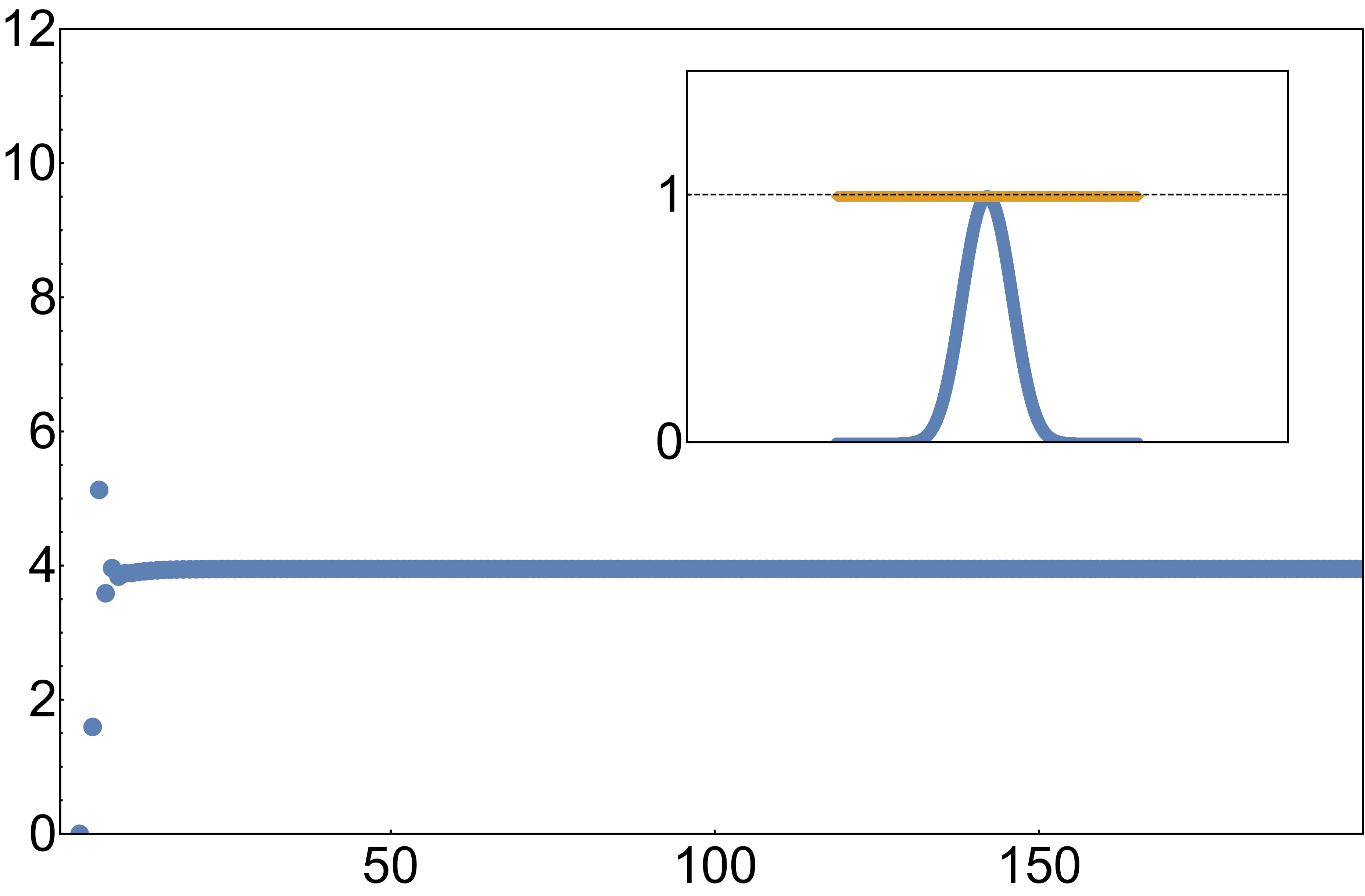}}\\
   \subfloat[][$L=12$]
   {\includegraphics[width=0.4\textwidth]{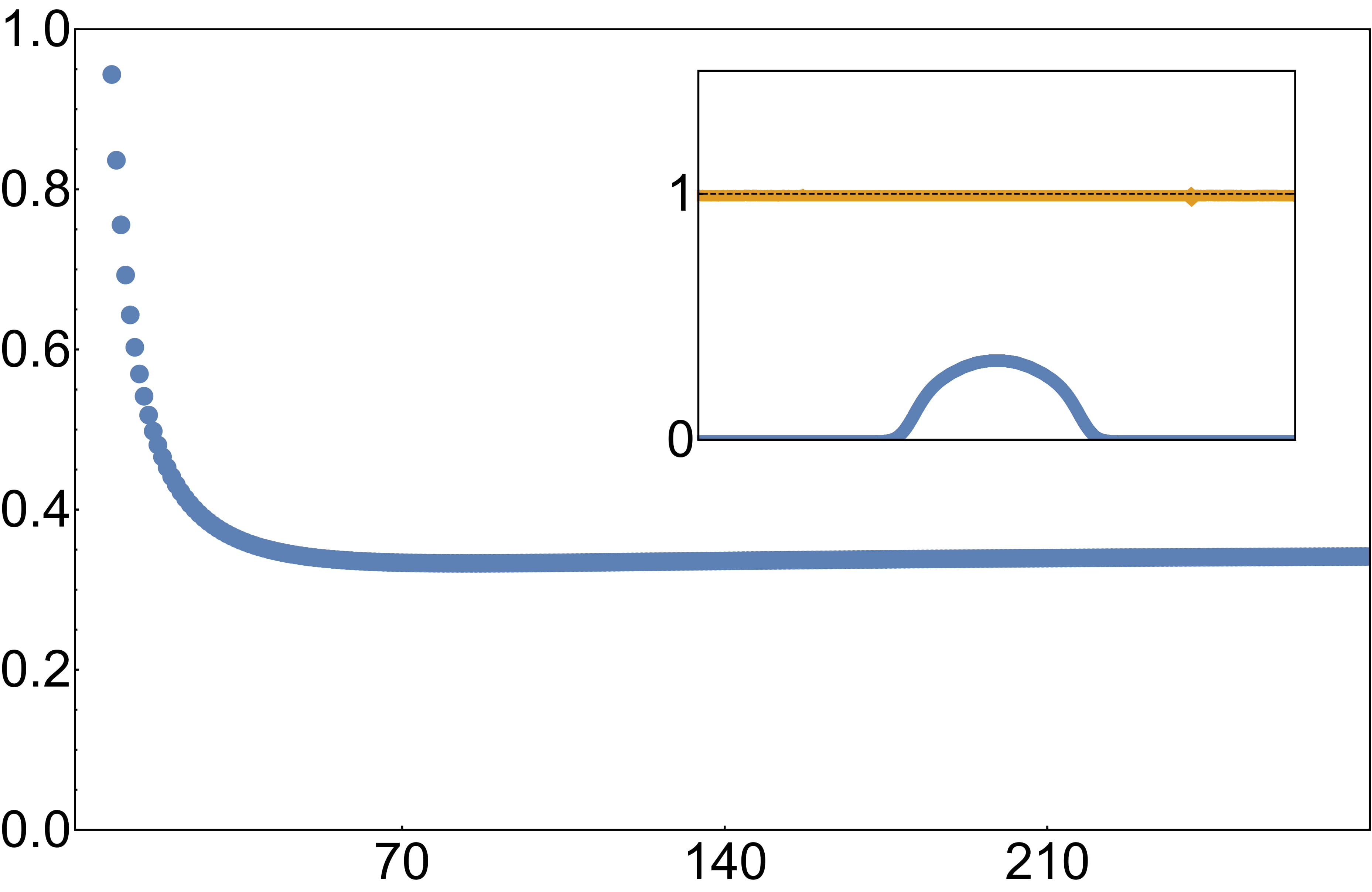}}~
\subfloat[][$L=12$]
   {\includegraphics[width=0.4\textwidth]{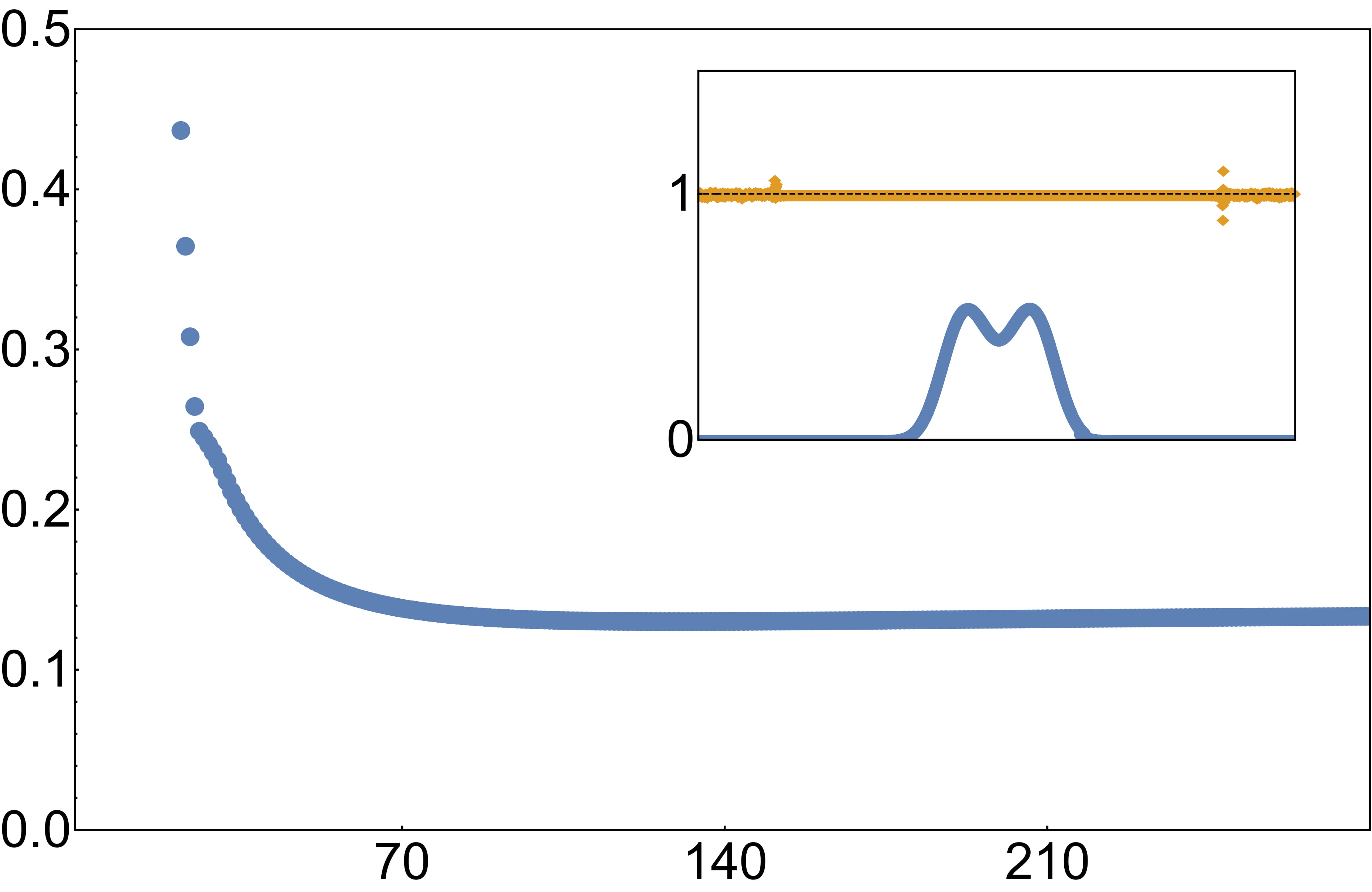}}\\
   \subfloat[][$L=29$]
   {\includegraphics[width=0.4\textwidth]{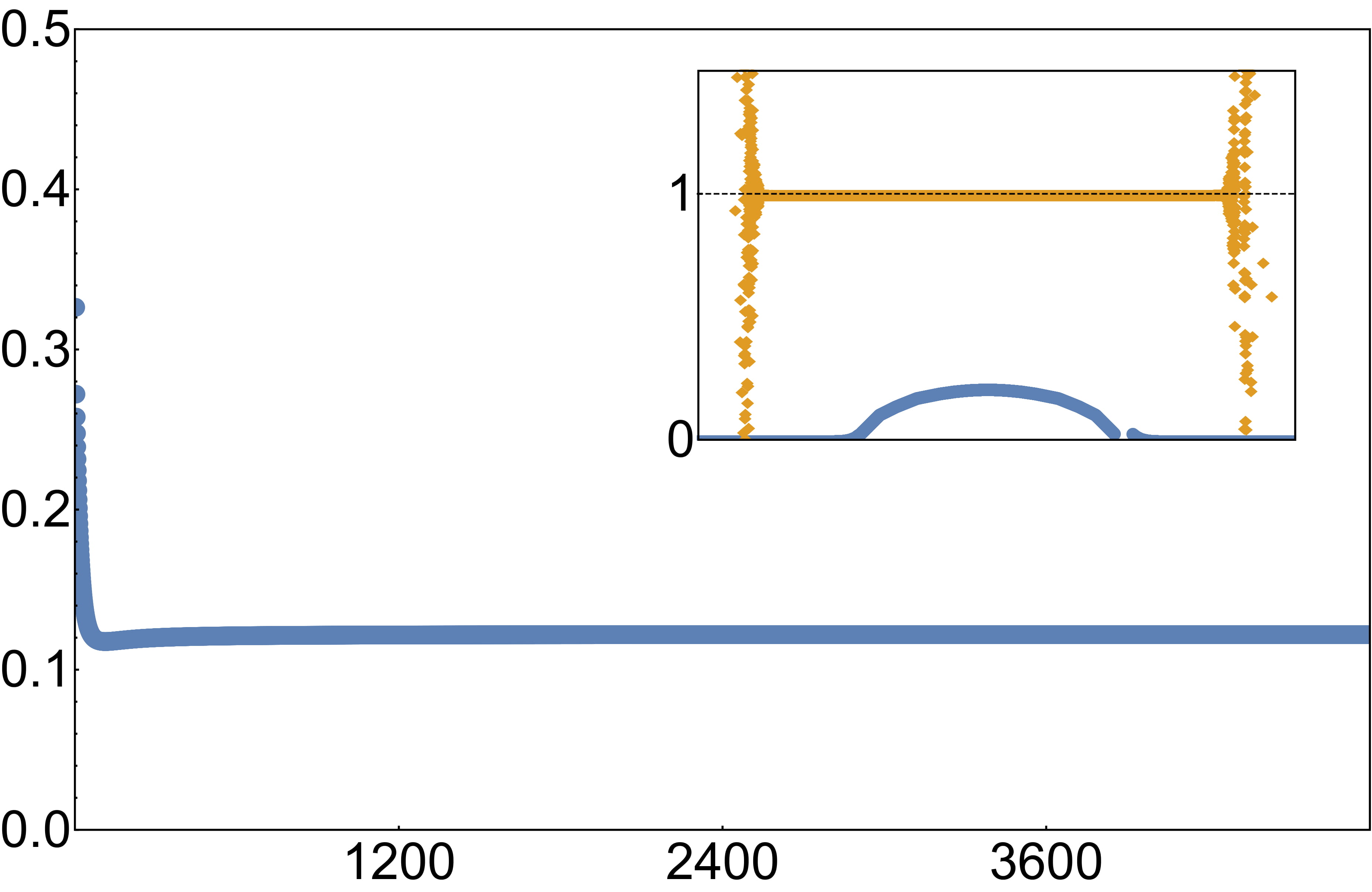}}~
\subfloat[][$L=29$]
   {\includegraphics[width=0.4\textwidth]{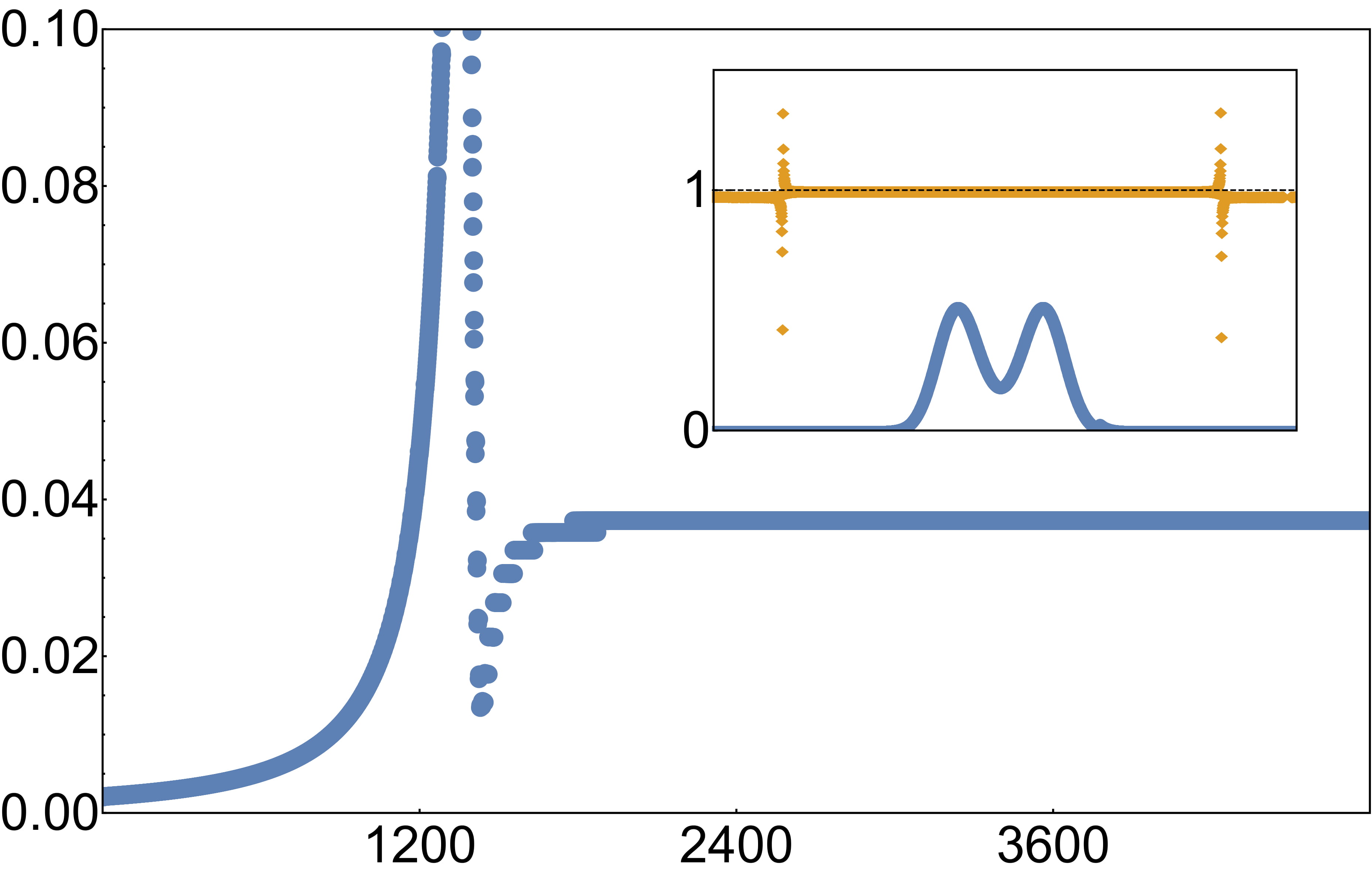}}\\
   \subfloat[][$L=70$]
   {\includegraphics[width=0.4\textwidth]{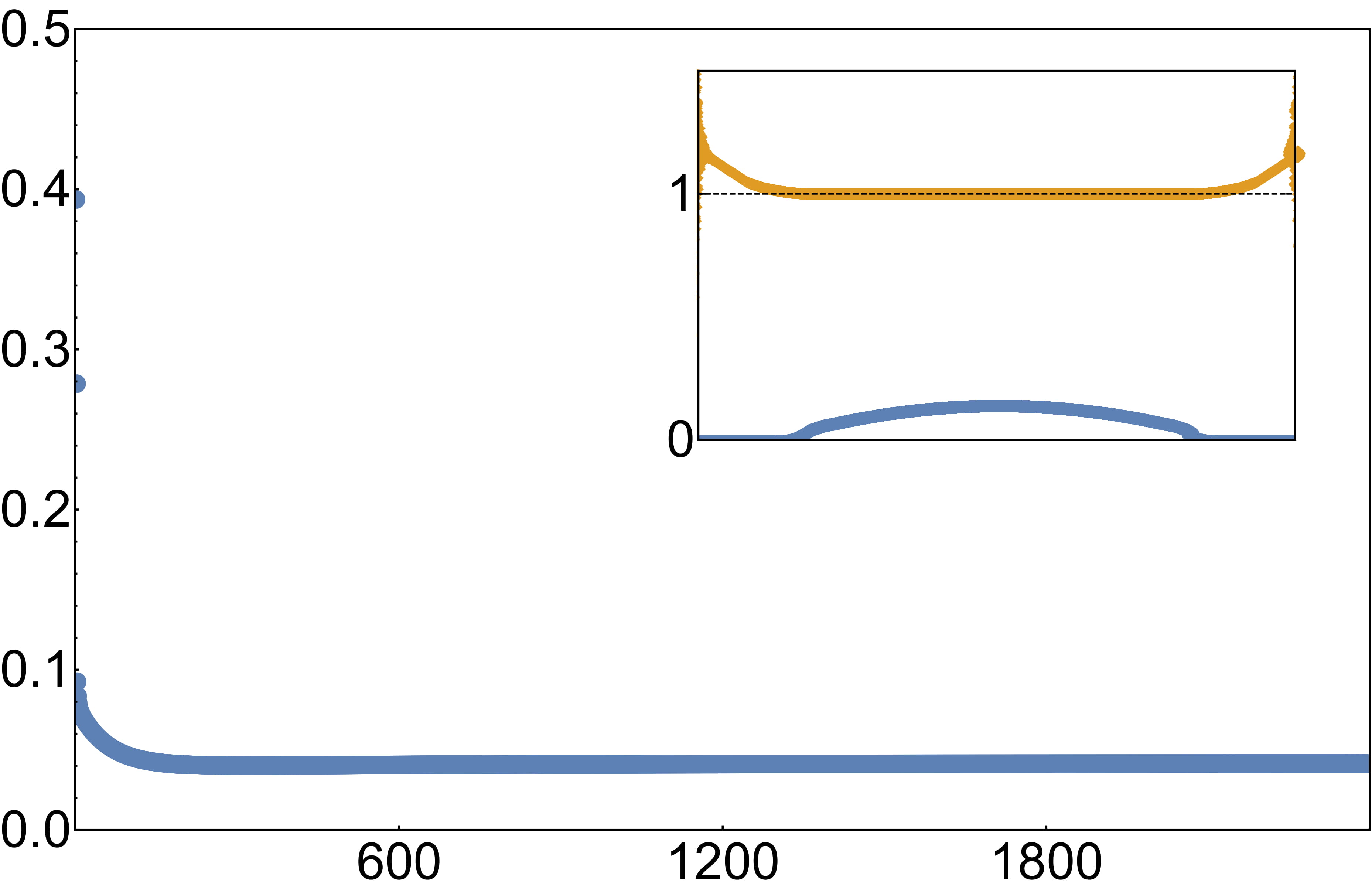}}~
\subfloat[][$L=70$]
   {\includegraphics[width=0.4\textwidth]{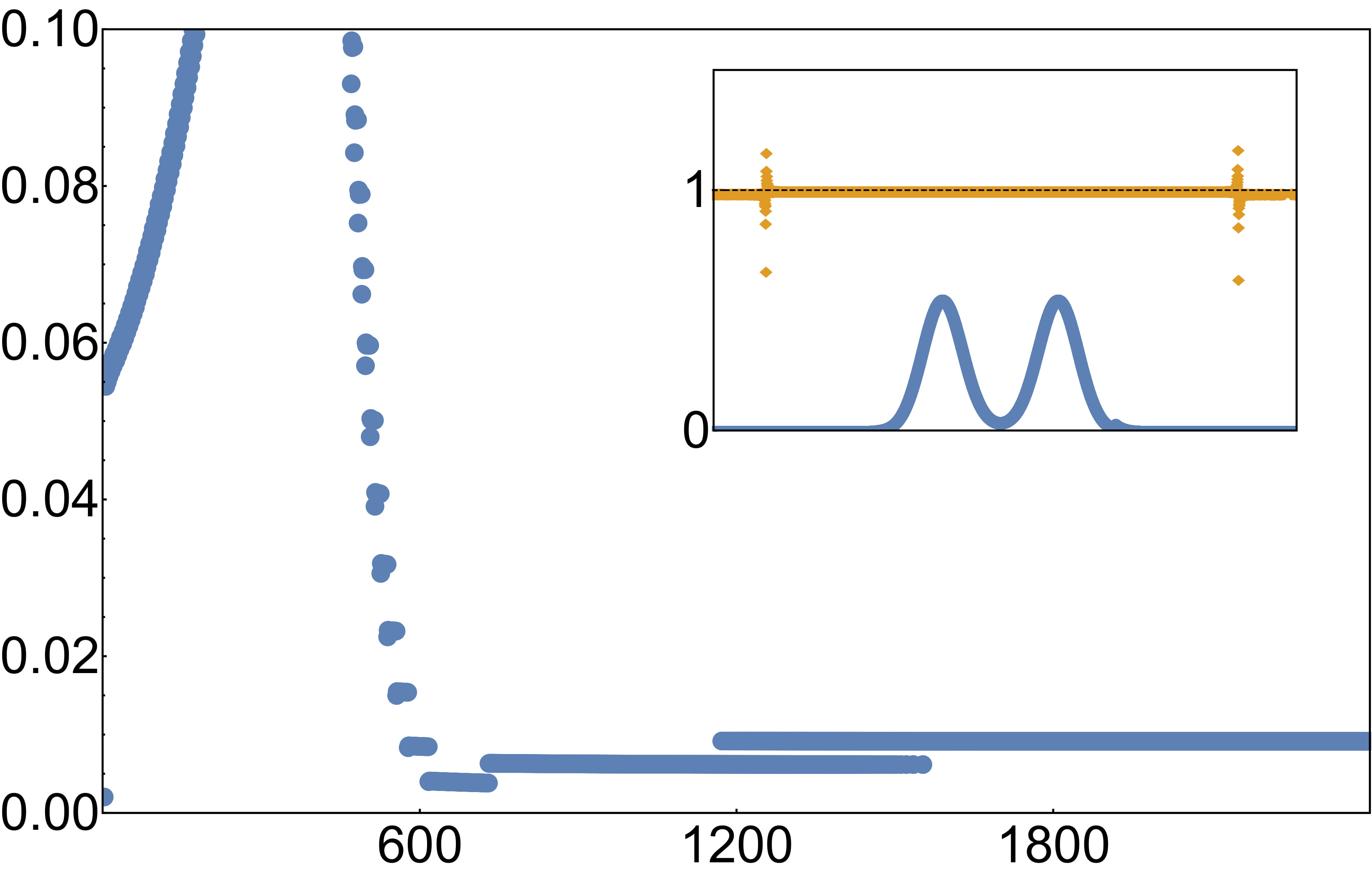}}
\caption{KS eigenvalue in the LDA approximation (left) and in the SCE approximation (right), at different correlation regimes (different values of $L$) as a function of the number of iterations. Insets: plot of the ratio $\frac{\hat{h}\phi}{\varepsilon\phi}$ (orange) and the corresponding density in uniformly scaled coordinates (blue).}
\label{fig:SCELDAcomputation}
\end{figure*}

In Fig.~\ref{fig:SCELDAcomputation}, we show, for the case $N=2$, the KS eigenvalue $\varepsilon$ as a function of the number of iterations within the LDA and the SCE approximation. To check that we have solved the KS equations, we plot in the inset the ratio $\frac{\hat{h}[\rho]\phi_(x)}{\varepsilon\phi(x)}$ at convergence, with $\hat{h}$ the KS single-particle hamiltonian, Eq.\eqref{KSeq}.
We clearly see from this figure that convergence becomes slower (more iterations are needed) as the system becomes more and more correlated (larger $L$). This is true for both the LDA and the SCE cases. 

\begin{figure*}
\centering
\subfloat[][$N=4$]
   {\includegraphics[width=0.4\textwidth]{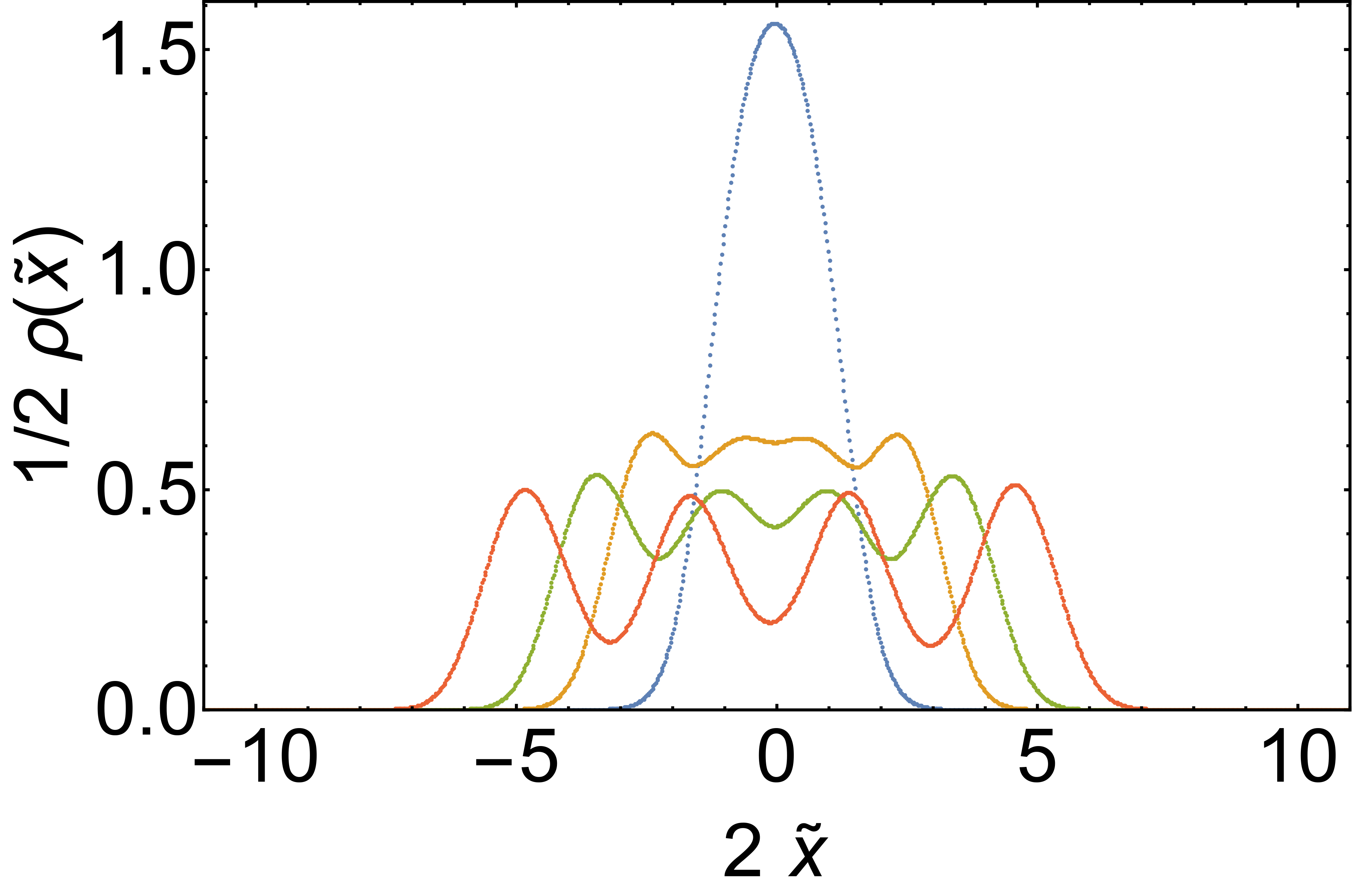}}~
\subfloat[][$N=4$]
   {\includegraphics[width=0.4\textwidth]{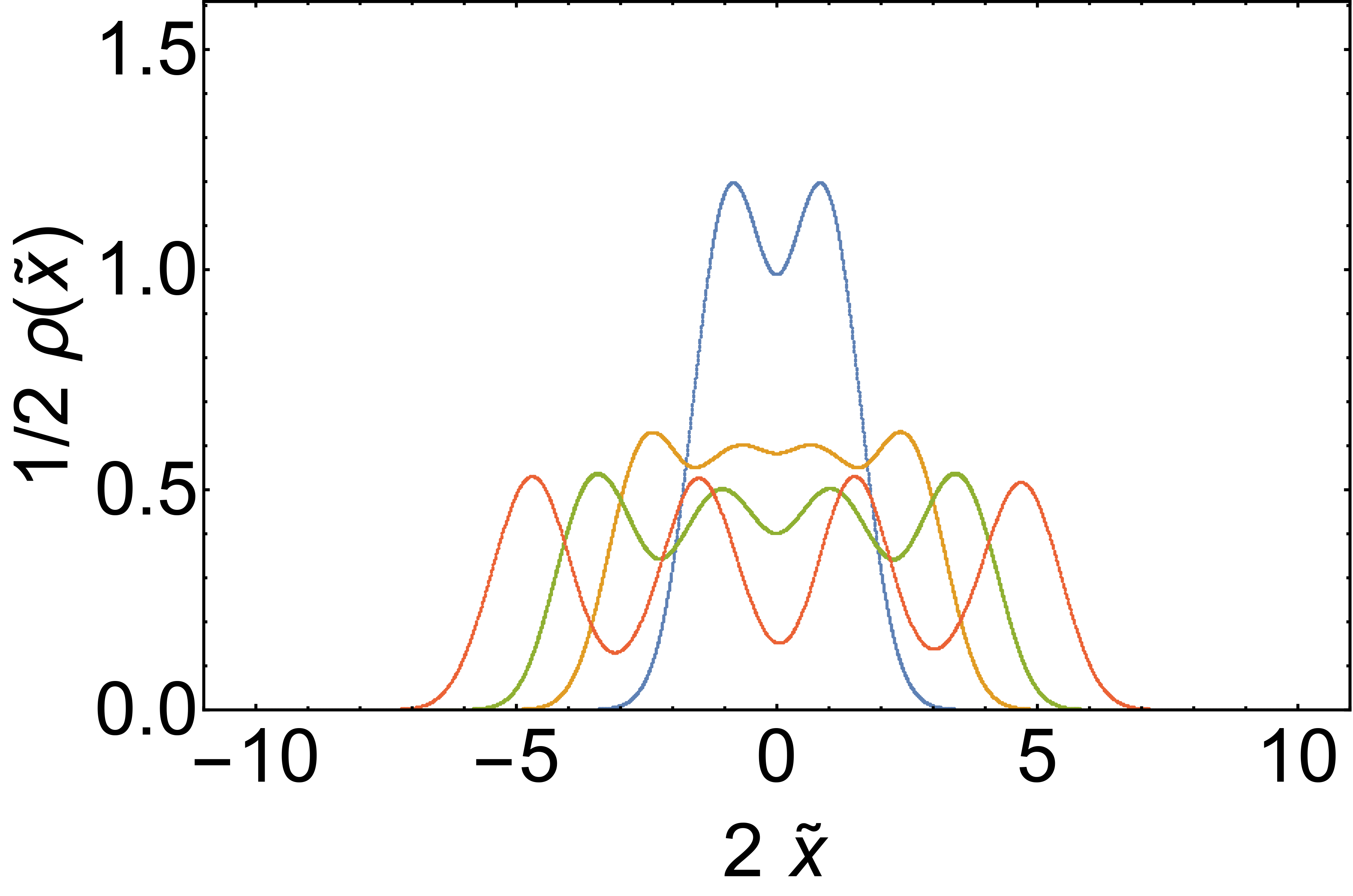}}\\
   \subfloat[][$N=8$]
   {\includegraphics[width=0.4\textwidth]{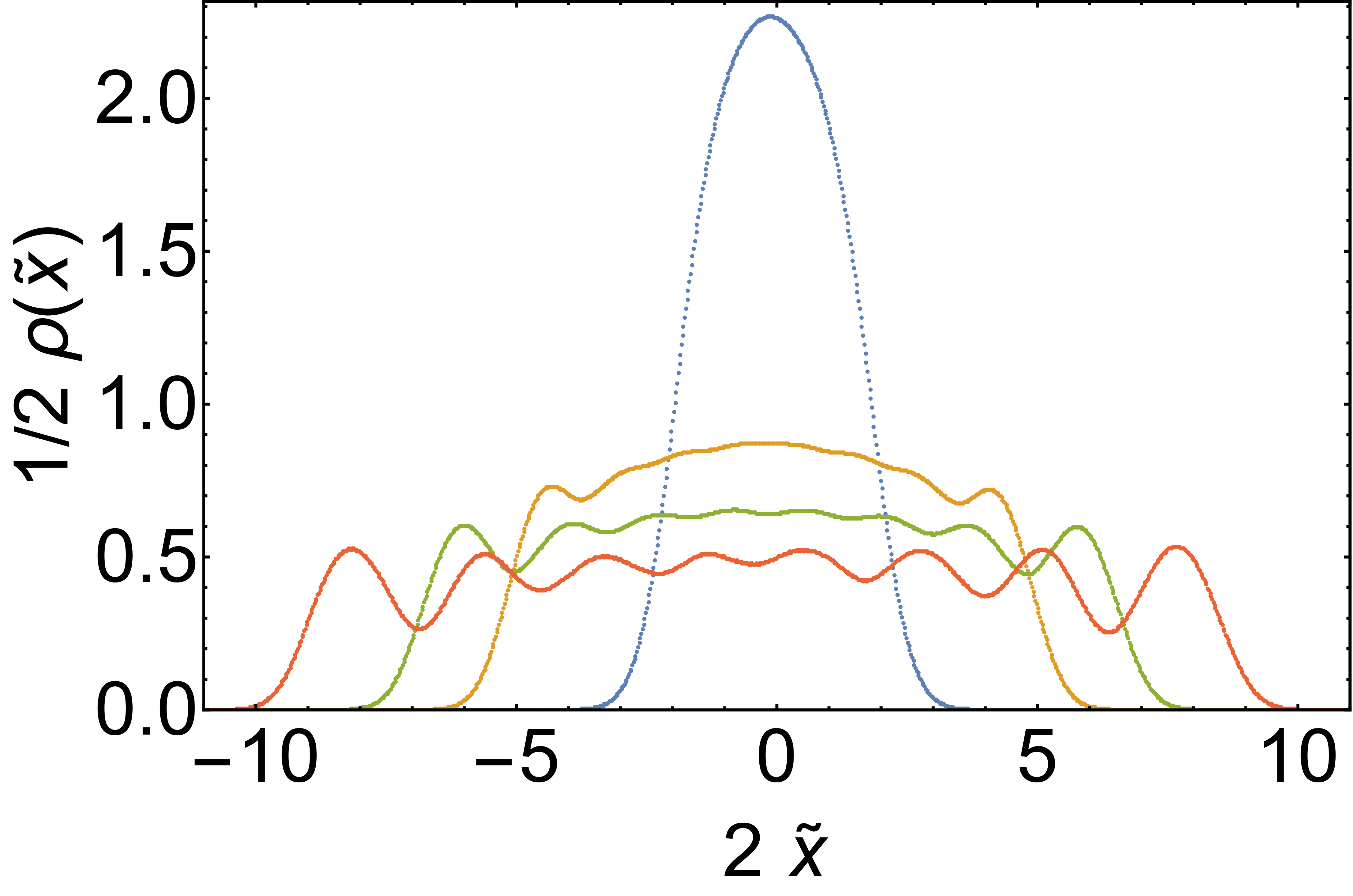}}~
\subfloat[][$N=8$]
   {\includegraphics[width=0.4\textwidth]{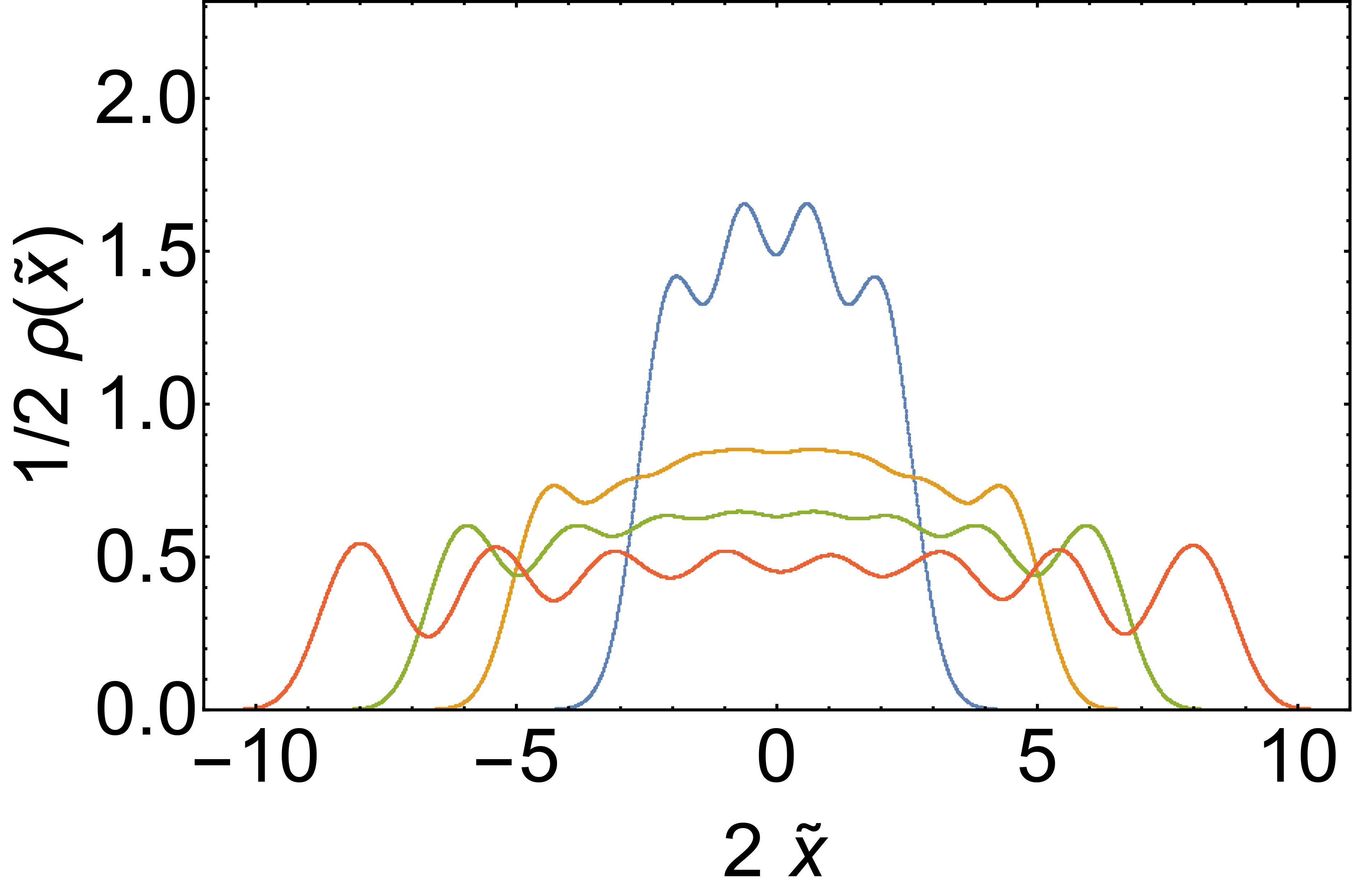}}\\
   \subfloat[][$N=16$]
   {\includegraphics[width=0.4\textwidth]{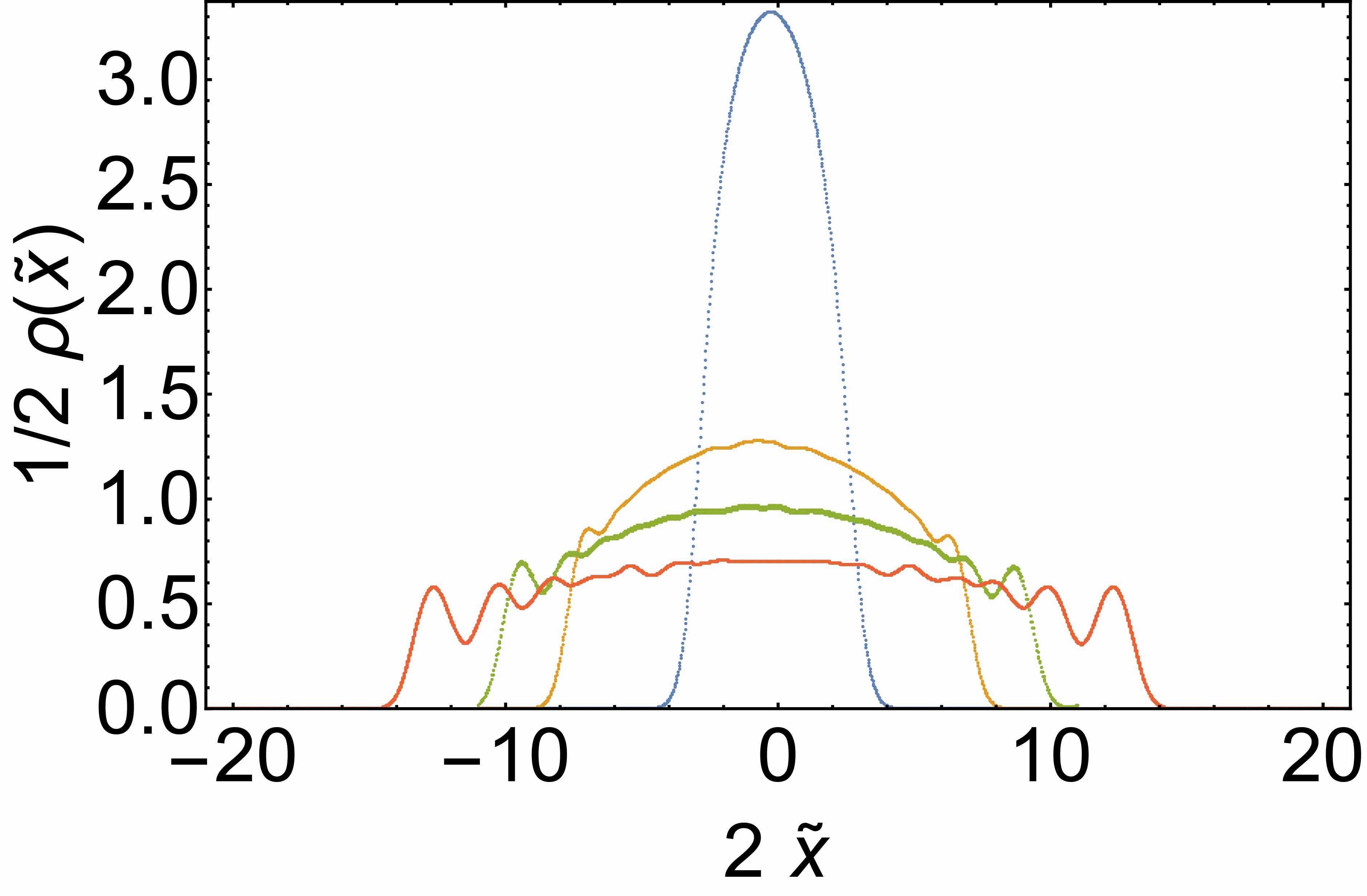}}~
\subfloat[][$N=16$]
   {\includegraphics[width=0.4\textwidth]{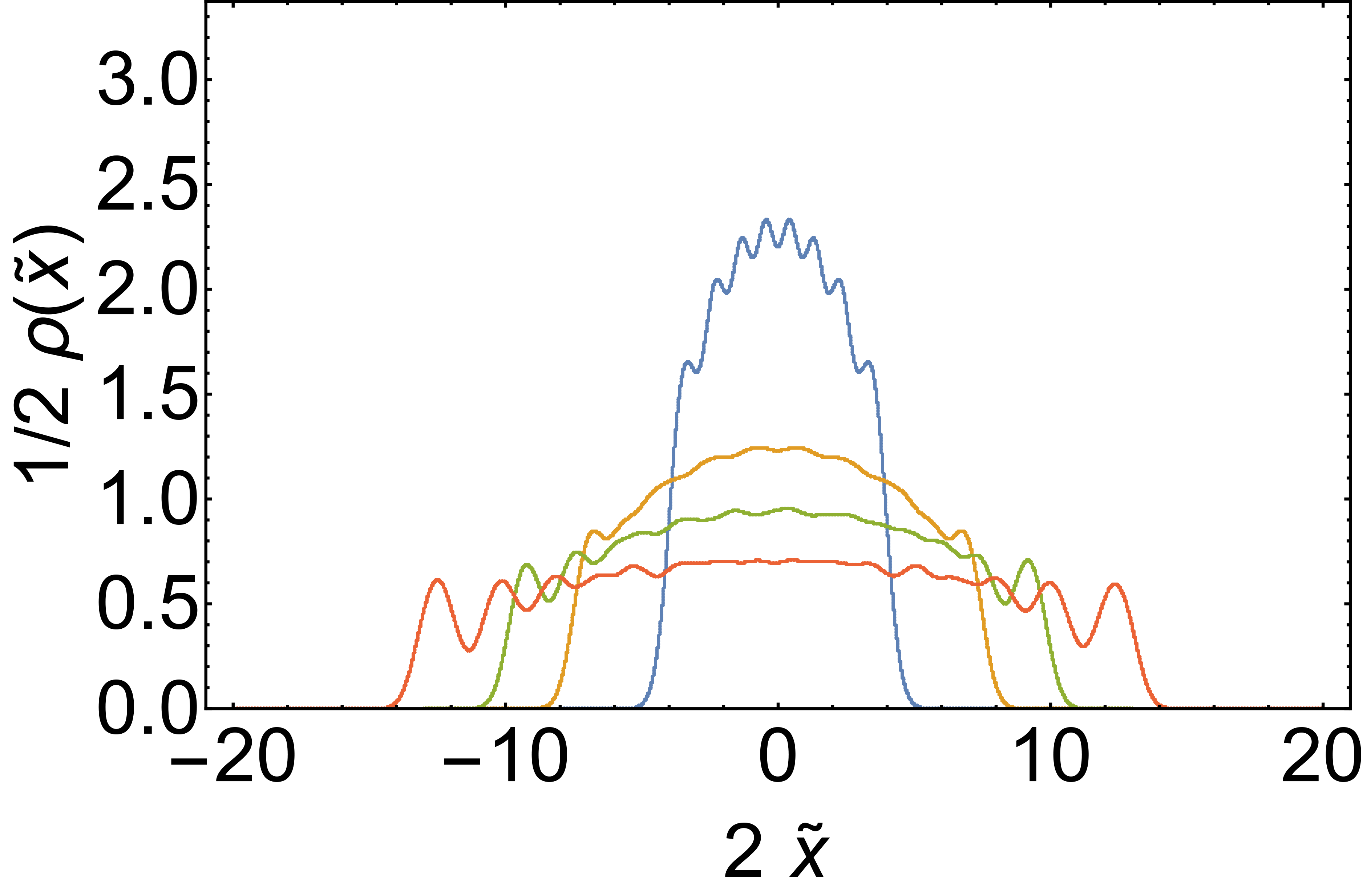}}\\
\caption{KS SCE computation, in scaled units (see main text, Eq.\eqref{eq:scaledhamL}), for different $N$ at different correlation regimes ($L=1,~12,~29,~70$). Left: bosons. Right: fermions.}
\label{fig:manyN}
\end{figure*}

In Fig.~\ref{fig:manyN} we also plot the KS SCE computation for $4$, $8$, and $16$ particles, for both bosons (left column) and fermions (right column), at different correlation regimes. 

\begin{figure*}
\centering
\subfloat[][$L=1$]
   {\includegraphics[width=0.5\textwidth]{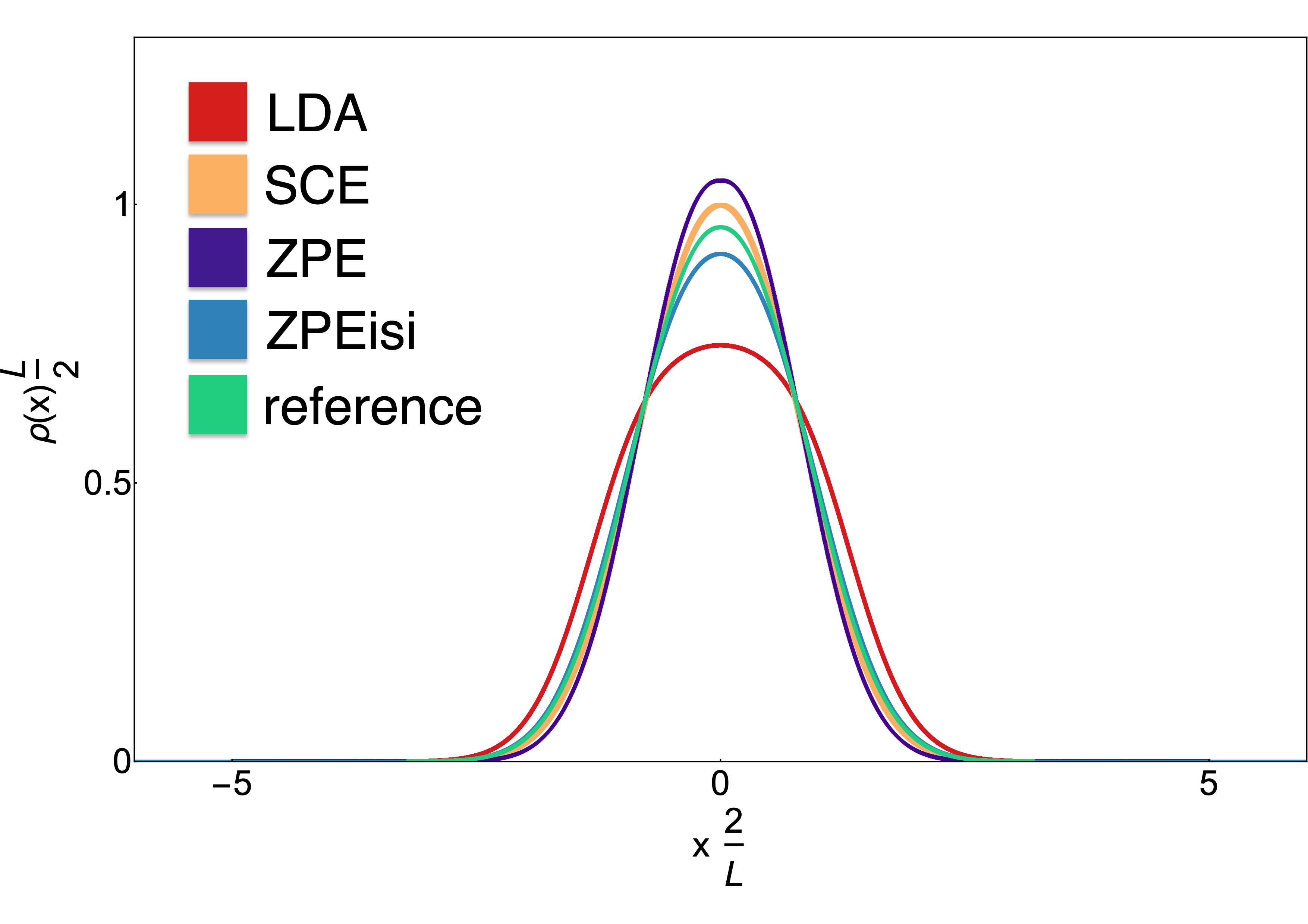}}~
\subfloat[][$L=12$]
   {\includegraphics[width=0.5\textwidth]{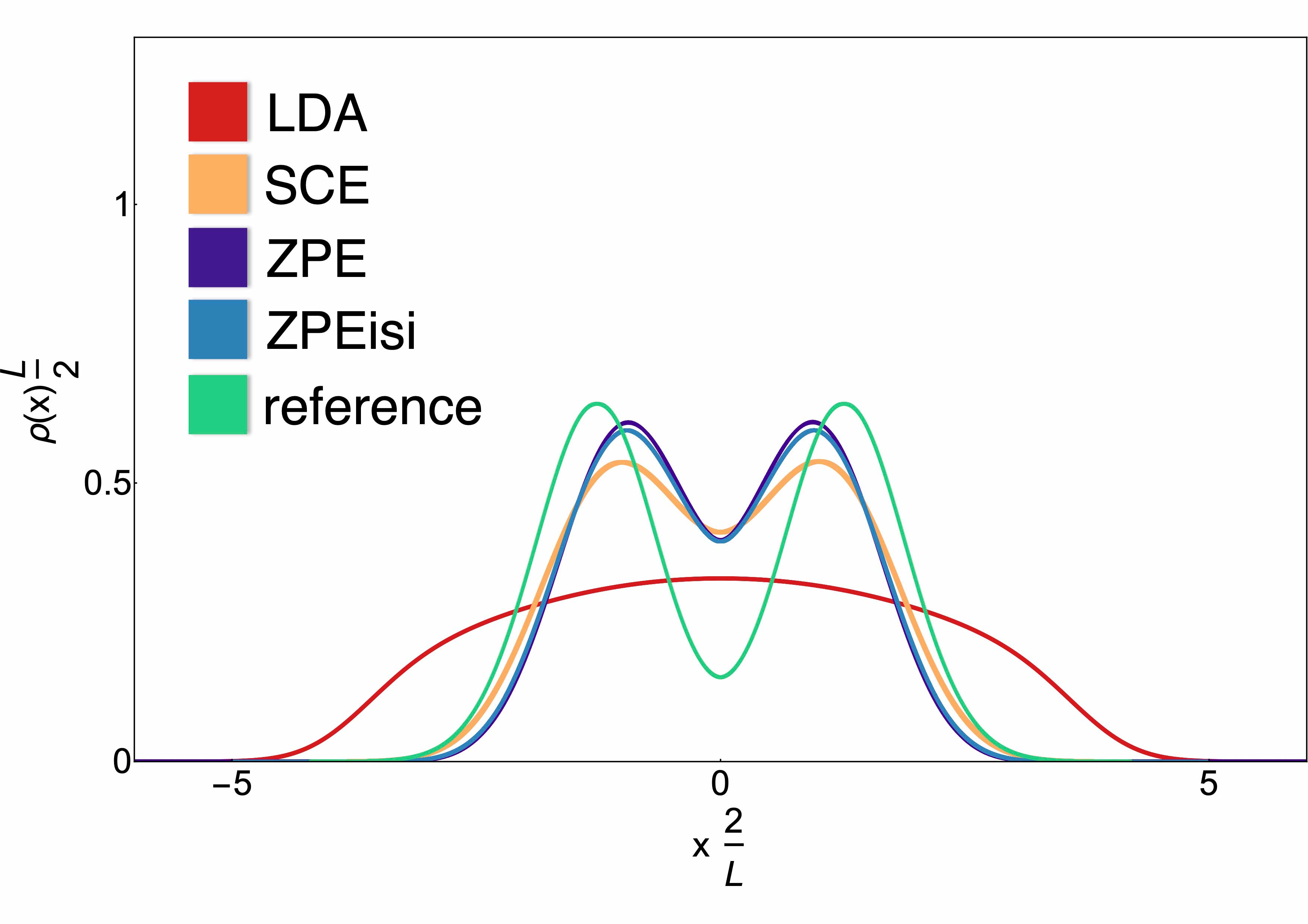}}\\
\subfloat[][$L=29$]
   {\includegraphics[width=0.5\textwidth]{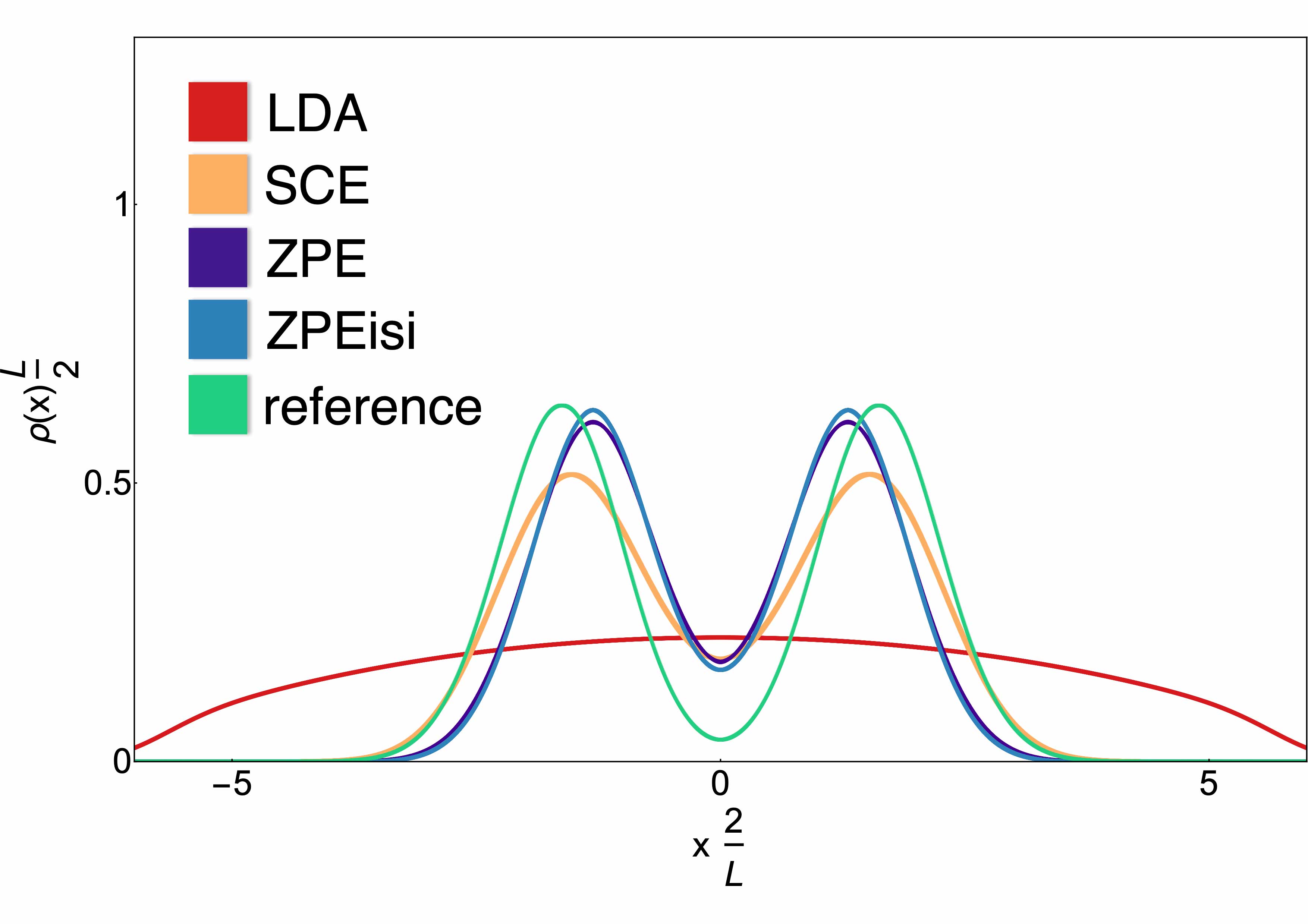}}~
\subfloat[][$L=70$]
   {\includegraphics[width=0.5\textwidth]{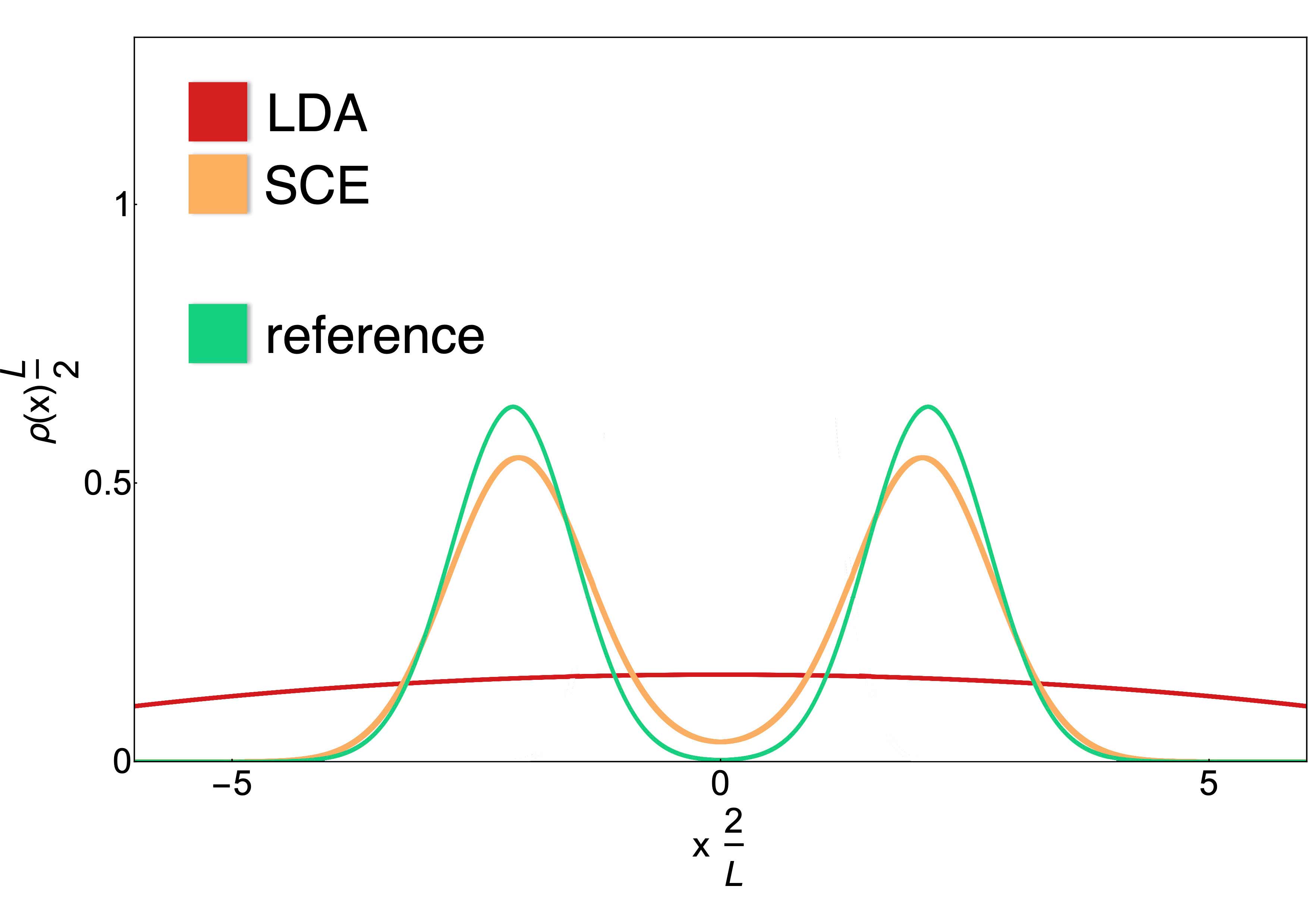}}
\caption{Comparison between  the self-consistent  ground state density in the KS LDA approximation, the KS SCE, the KS SCE+ZPE and KS SCE+ZPEisi approximations with the exact many-body result (labeled ``reference'').}
\label{fig:densitiesSF}
\end{figure*}

In Fig.~\ref{fig:densitiesSF} we show the densities obtained self-consistently for different values of $L$, comparing them with those obtained by an exact diagonalisation of the many-body hamiltonian. 
Both for the LDA and the SCE case we could confirm the results of Refs.~\cite{MalGor-PRL-12,MalMirCreReiGor-PRB-13} (obtained with the Numerov algorithm, using a shooting method and linear mixing for the self-consistency), with a single exception. 
In fact, we note that for the case $L=1$ (panel a) our LDA computation gives a density which is sensibly different from the ones found, independently, in Fig.~1 of Ref.~\cite{MalGor-PRL-12} and in Fig.~7 of Ref.~\cite{AbePolXiaTos-EJPB-07}, which were both very similar to each other and much closer to the exact many-body density. However, the corresponding inset in our Fig.~\ref{fig:SCELDAcomputation} clearly shows that our result does solve the KS-LDA equation, while we tested the density of Ref~\cite{MalGor-PRL-12} and we found that $\frac{\hat{h}[\rho]\phi(x)}{\varepsilon\phi(x)}$ is not as close to 1 as our new result. 

As already explained in the introduction, we see that the LDA breaks down as the system becomes more and more correlated (large $L$), while KS SCE gives densities that are closer and closer to the exact many-body ones.

\begin{figure*}
\centering
\subfloat[][$L=1$]
   {\includegraphics[width=.32\textwidth]{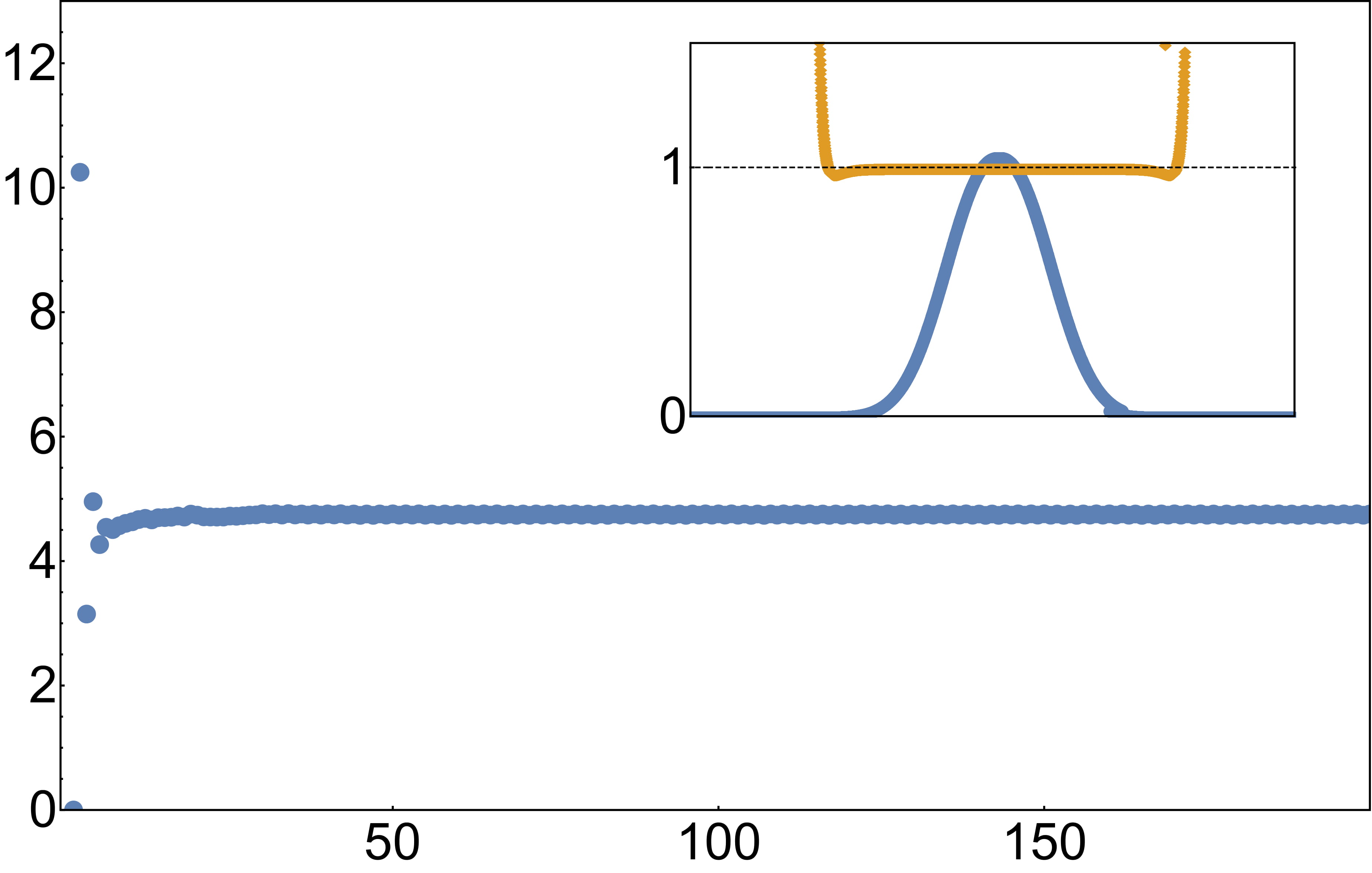}}~
   \subfloat[][$L=12$]
   {\includegraphics[width=.32\textwidth]{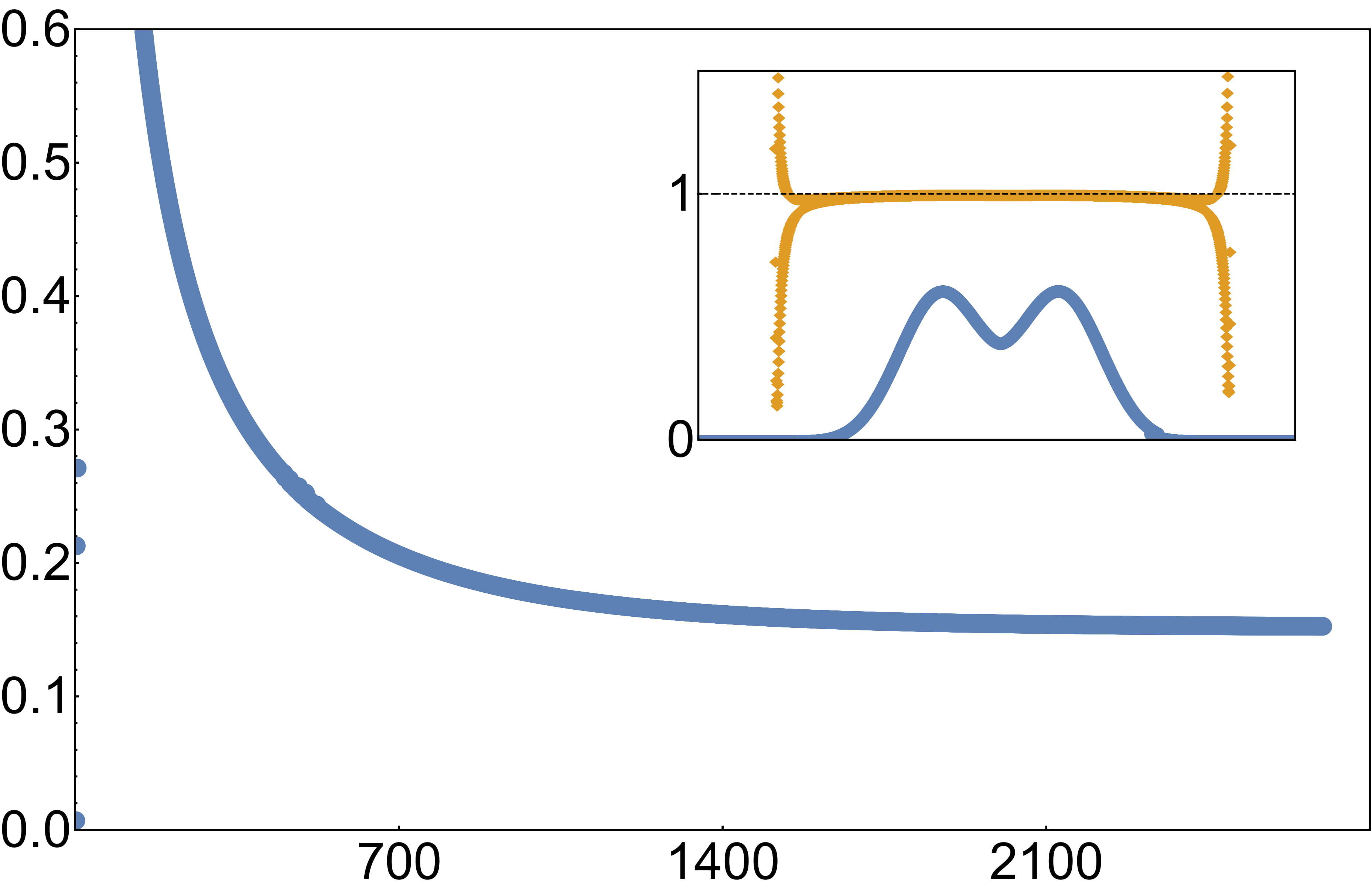}}~
   \subfloat[][$L=29$]
   {\includegraphics[width=.32\textwidth]{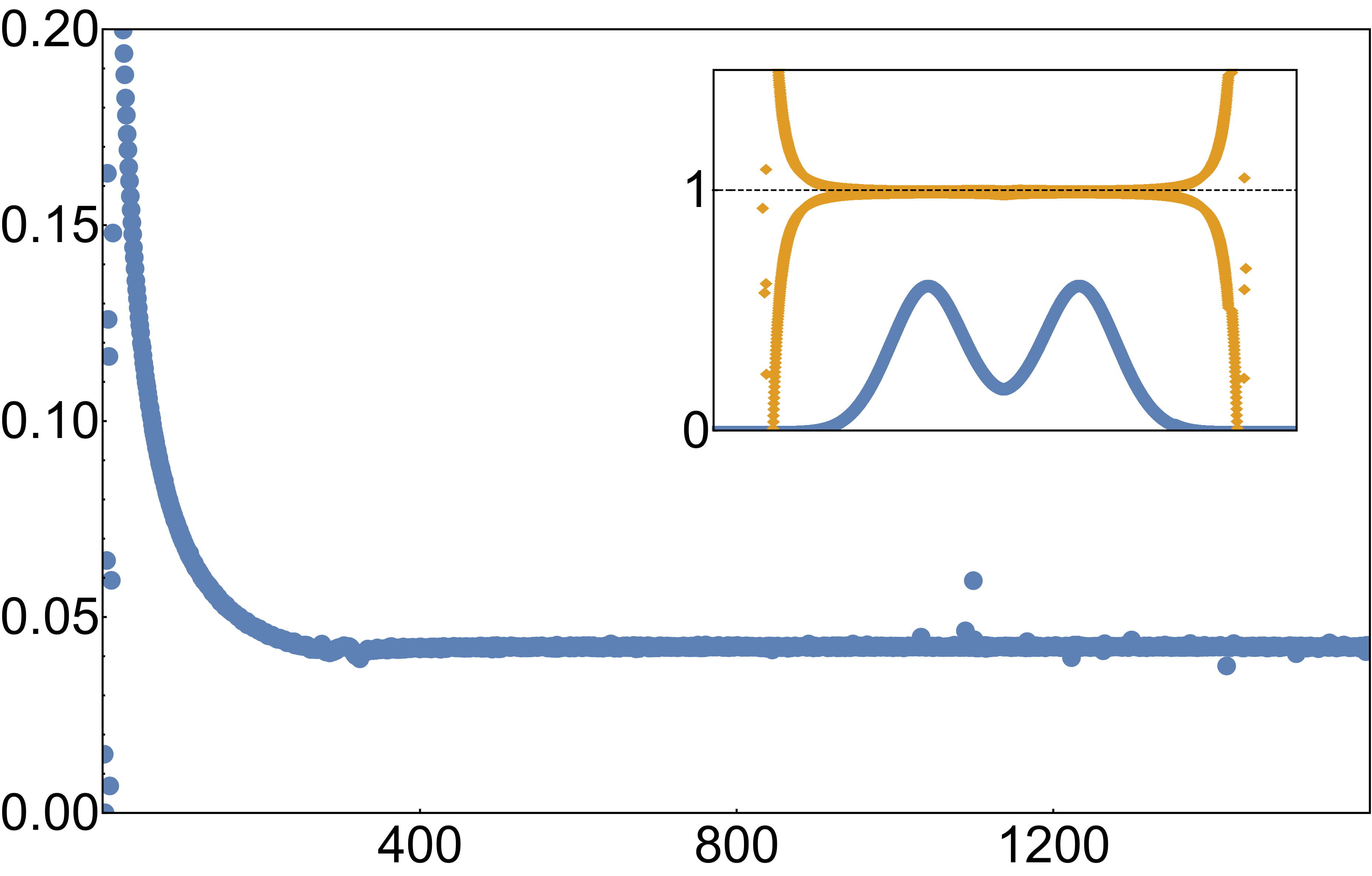}}\\
   \subfloat[][$L=1$]
   {\includegraphics[width=.32\textwidth]{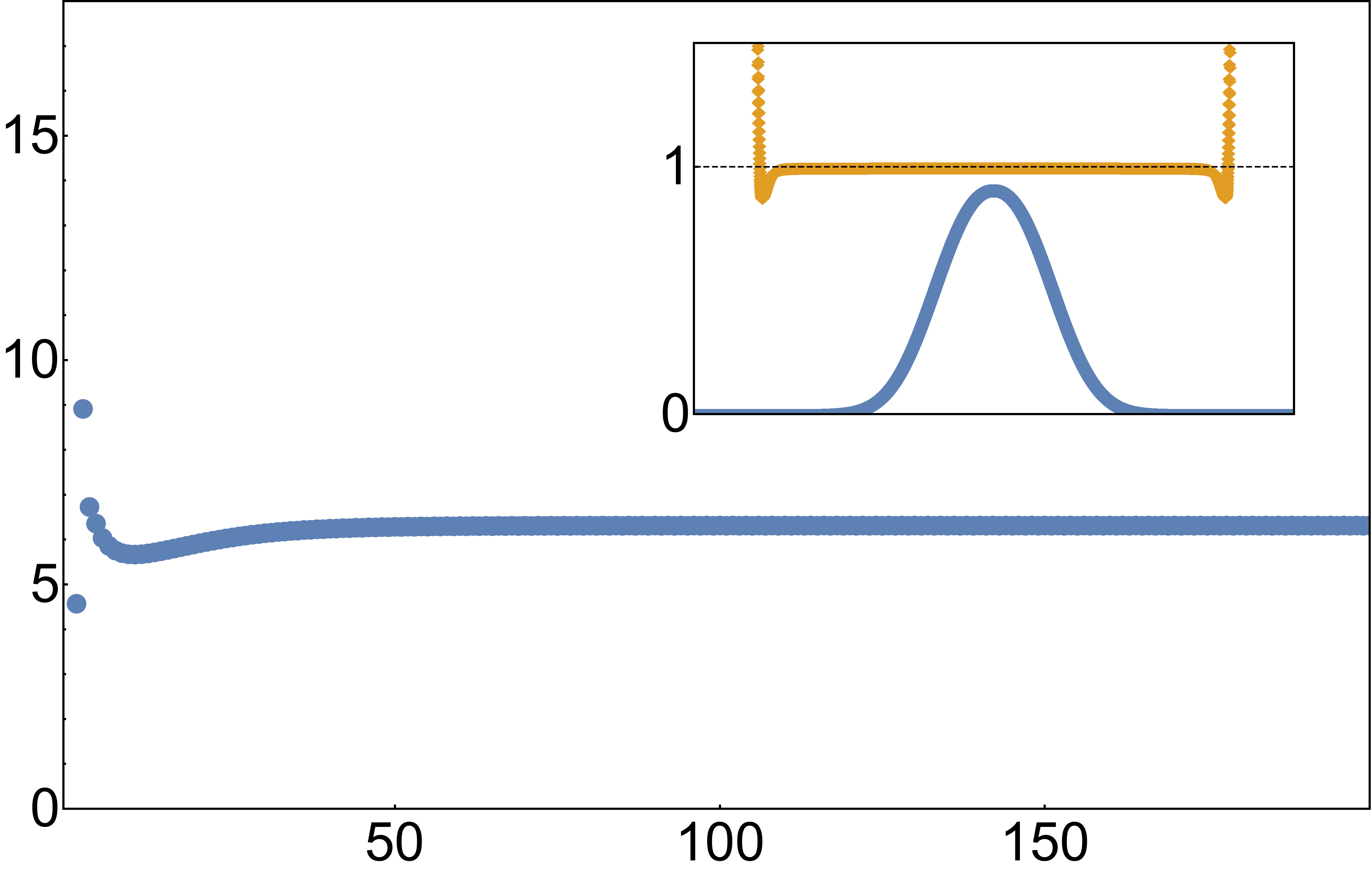}}~
   \subfloat[][$L=12$]
   {\includegraphics[width=.32\textwidth]{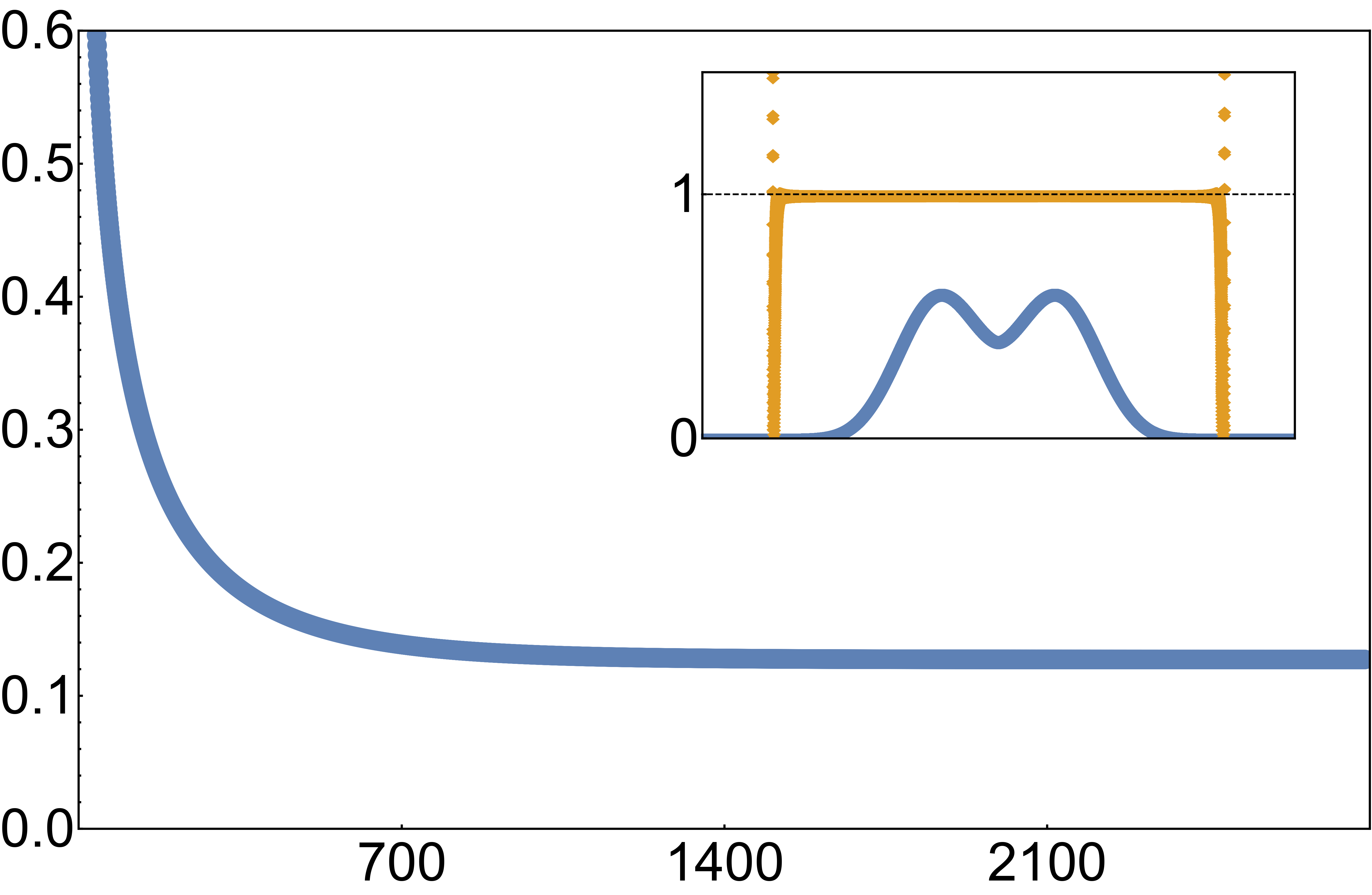}}~
   \subfloat[][$L=29$]
   {\includegraphics[width=.32\textwidth]{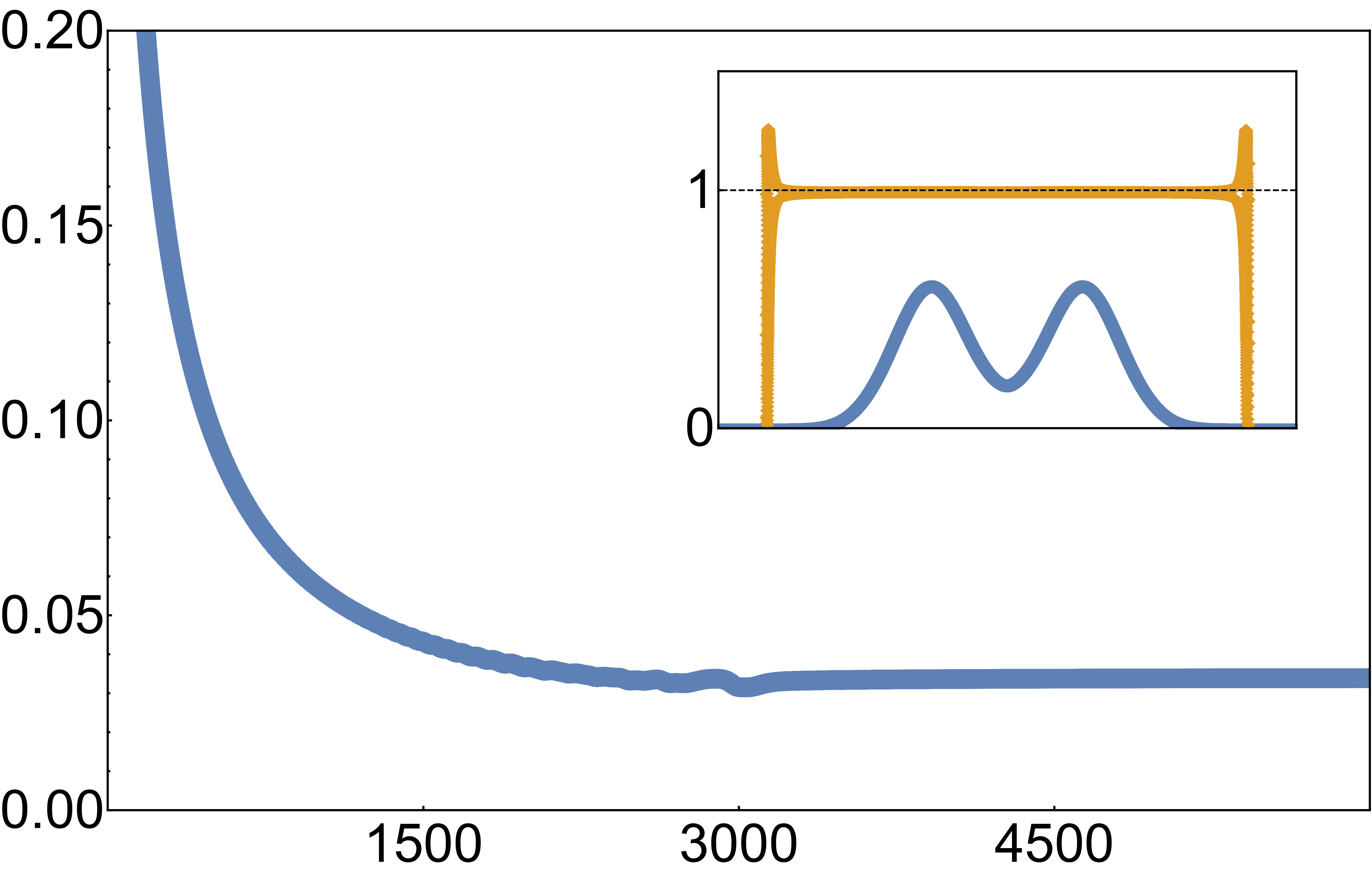}}
\caption{KS eigenvalue in the KS SCE+ZPE (top) and KS SCE+ZPEisi (bottom) scheme at different correlation regimes as a function of the number of iterations. Insets: plot of the ratio $\frac{\hat{h}[\rho]\phi(x)}{\varepsilon\phi(x)}$ and the corresponding density in uniformly scaled coordinates.}
\label{fig:computation2}
\end{figure*}

\begin{figure}
{\includegraphics[width=0.45\textwidth]{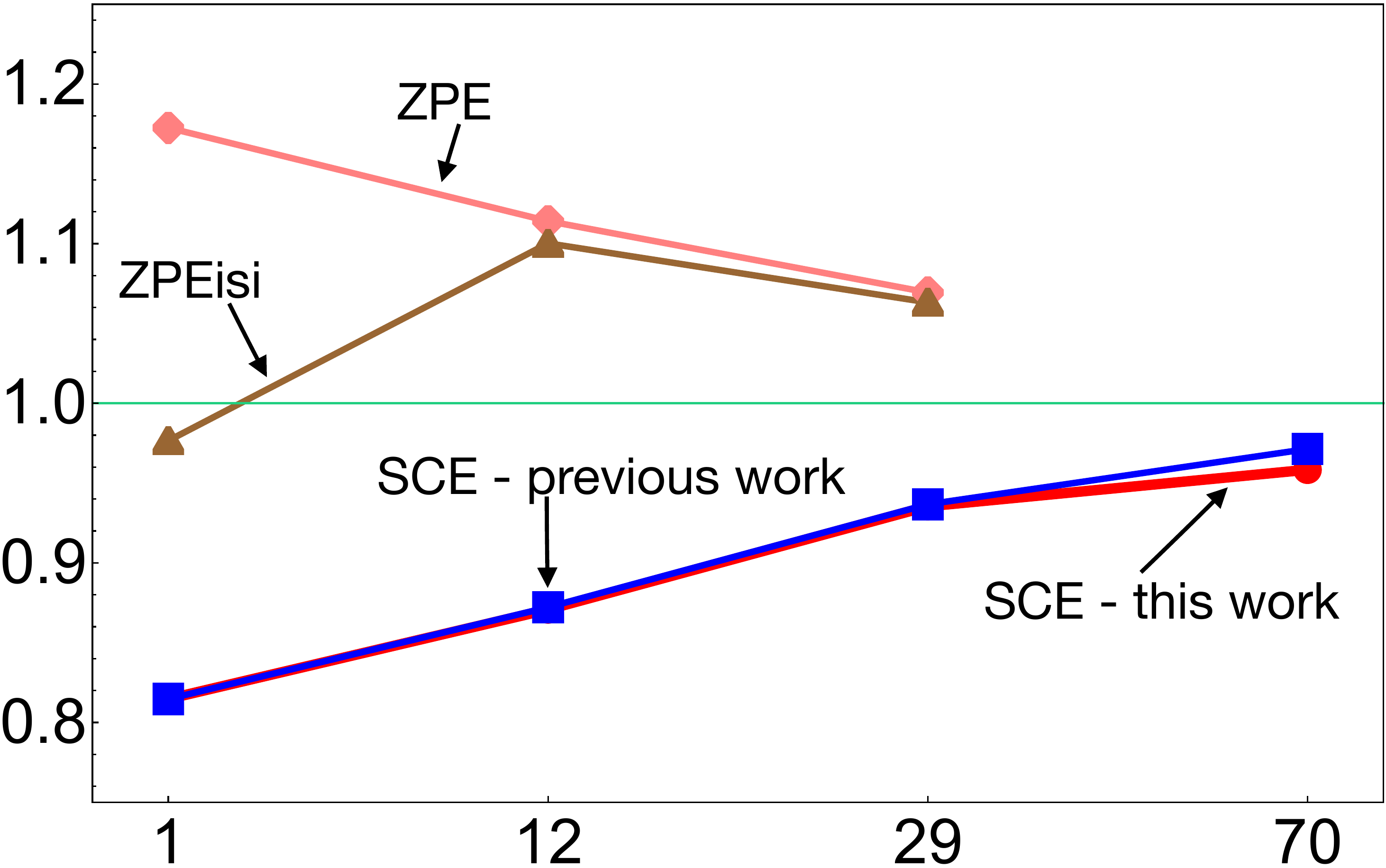}}
\caption{Ratio between the total energies from the approximations discussed in the main text and the numerically accurate one, for increasing correlation regimes $(L=1,~12,~29,~70)$ for $N=2$.}
\label{fig:energies}
\end{figure}

\begin{table}
	\caption{\label{tab:totalenergies}Total energies for different approximations for $N=2$ compared to the exact many-body case.}
\begin{ruledtabular}
\begin{tabular}{ccccc}
$L$  & 1 & 12 & 29 & 70\\
{\textbf{exact}} & 6.92367 & 0.13943 & 0.04028 & 0.01152\\
{SCE - Ref.\cite{MalMirCreReiGor-PRB-13}} & 5.64063 & 0.12161 & 0.03773 & 0.01119\\
{SCE - this work} & 5.64119 & 0.12142 & 0.03768 & 0.01104\\
{ZPE} & 8.11878 & 0.15535 & 0.04308 &\\
{ZPEisi} & 6.76139 &  0.15340 & 0.04283 &\\
\end{tabular}
\end{ruledtabular}
\end{table}

\begin{table}
	\caption{\label{tab:KSenergies}KS highest occupied orbital energy for the LDA and SCE approximations compared with the exact many-body chemical potential for $N=2$.}
\begin{ruledtabular}
\begin{tabular}{ccccc}
$L$  & 1 & 12 & 29 & 70\\
{\textbf{exact}} & 4.92567 & 0.13443 & 0.03790 & 0.01112 \\
{SCE} & 3.94900 & 0.13664 & 0.03730 & 0.01126\\
{LDA} & 7.18306 & 0.34539 & 0.12192 &0.04148 \\
\end{tabular}
\end{ruledtabular}
\end{table}

\subsection{The SCE+ZPE and ZPEisi approximations}
The potential $v_{\mathrm{ZPE}}([\dens],x)$ has divergences that make the convergence of the KS SCE+ZPE equations extremely challenging. In this work we were able for the first time to reach convergence in some cases, but we had to use a more regular initial guess, namely $\phi_0(x)\sim \cosh^{-1}(x)$.

We report in Fig.~\ref{fig:computation2} the KS eigenvalue as a function of the number of iterations, and, again, in the inset we show the ratio $\frac{\hat{h}[\rho]\phi(x)}{\varepsilon\phi(x)}$ at convergence. We clearly see that it is much harder to converge the KS equations with this functional. 

\begin{figure}
\centering
   {\includegraphics[width=0.45\textwidth]{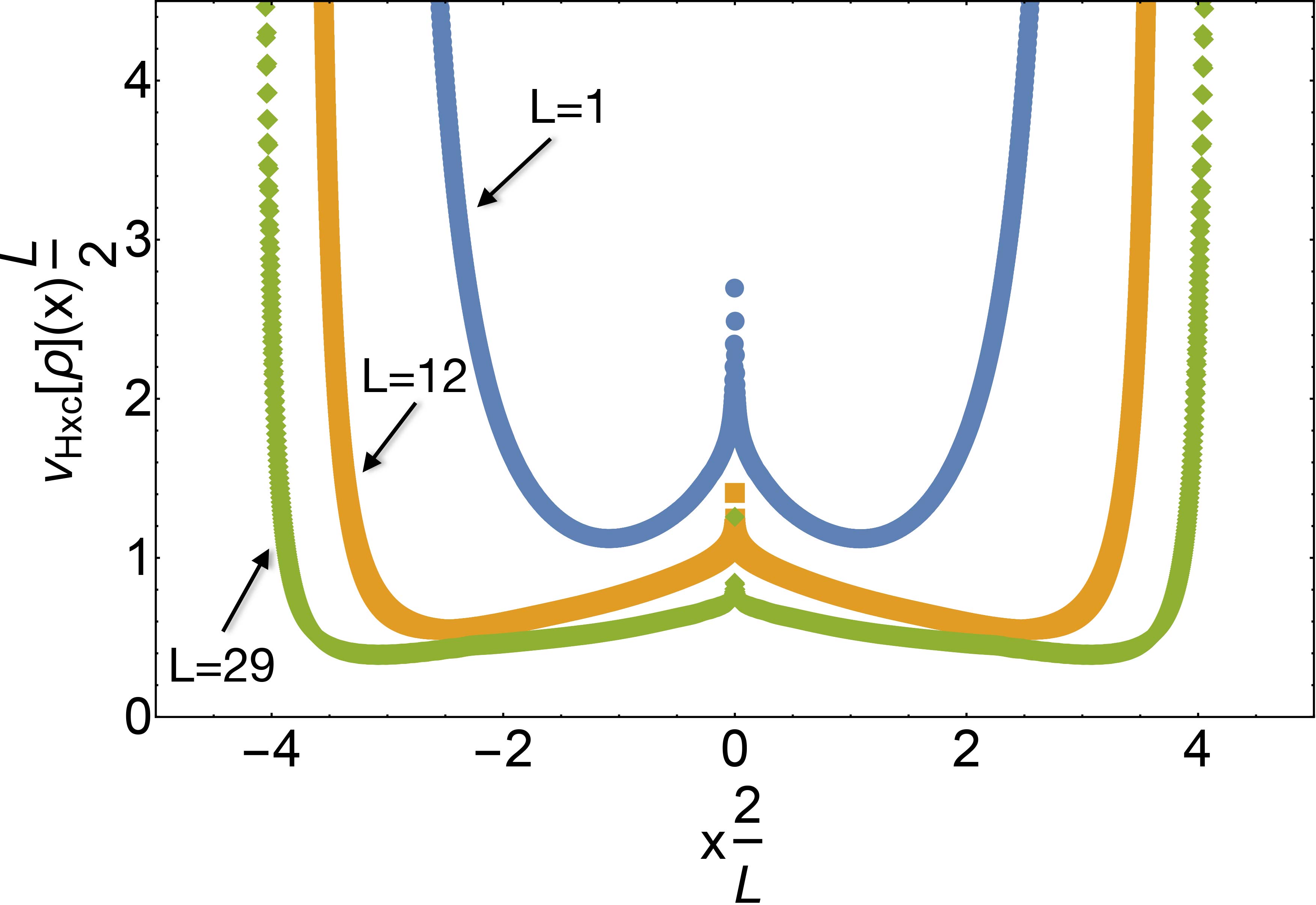}}
\caption{$v_{Hxc}([\dens],x)$ in the SCE+ZPE approximation for different values of $L$. Inset: we plot the $v_{Hxc}([\dens],x)$ in the whole computational box: as it can be seen, it diverges very rapidly at all characteristic length $L$, and becomes numerically unstable at large $x$, in a region where the density is $\approx 0$.}
\label{fig:vHxcSF}
\end{figure}

Looking at Fig.~\ref{fig:densitiesSF}, it is clear that the KS SCE+ZPE functional localizes the density more, compared to SCE alone, in all the cases studied.
 In the following, we will argue that this is due to the predominance of $v_{\rm Hxc}([\dens],x)$, in the SCE+ZPE approximation, with respect to the external potential in Eq.~\eqref{eq:extpotexplicit} at large~$x$. 
To illustrate this, we first show in Fig.~\ref{fig:vHxcSF} the self-consistent $v_{\rm Hxc}(x)$ of Eq.~\eqref{eq:VHXCZPE} for various $L$.  Notice that for the case $L=70$, for the reasons just outlined, we were not been able to reach convergence. Therefore, we omitted such case from our discussion of the results, both for ZPE and ZPEisi approaches. 

We can clearly see that for $x\approx 0$ and $x\approx\pm\infty$, $v_{\rm Hxc}([\dens],x)$ diverges: this is a consequence of the predominance of the frequency $\omega([\dens],x)$  which diverges where either $\rho(x)\to 0$ or $\rho(f(x))\to 0$,
\begin{equation}\label{eq:expansionomega}
\omega([\dens],x)\sim\frac{1}{x^{3/2}}\sqrt{\frac{\dens(0)}{\dens(x)}}\sim \frac{e^{a\frac{x^2}{2}}}{x^{3/2}}\quad x\to \infty.
\end{equation} 

Defining the scaling \begin{equation}
x\to \frac{t}{\alpha^{2/3}},\qquad\alpha=\frac{\sqrt{8}}{L^2}
\end{equation}
and defining the scaled density $\dens_\gamma(x)=\gamma\dens(\gamma x)$, the single particle KS-SCE+ZPE Hamiltonian can be expanded at large $x$ according to
\begin{equation}\label{eq:scaledham}
\frac{\hat{h}_\alpha}{\alpha^{2/3}}=-\frac{\alpha^{2/3}}{2}\frac{\mathrm{d}^2}{\mathrm{d}t^2}+t^2+\tilde{v}_{\rm SCE}[\dens_{\alpha^{2/3}}](t)+\alpha^{1/3}\omega([\dens_{\alpha^{2/3}}],t),
\end{equation}
where we have used the fact that \cite{GorSei-PCCP-10,GroSeiGorGie-PRA-19}
\begin{subequations}
\begin{eqnarray}
\tilde{v}_{\rm SCE}([\dens_\gamma], x/\gamma)&=& \gamma\tilde{v}_{\rm SCE}([\dens], x),\\ 
v_{\rm ZPE}([\dens],x)&\to &\omega([\dens],x)\quad x\to \infty,\\f_i([\dens_\gamma],x)&=&\frac{1}{\gamma}f_i([\dens],\gamma x).
\end{eqnarray}
\end{subequations}
Looking at Eq.~\eqref{eq:scaledham}, we can see that, although as $L$ increases ($\alpha\to 0$) the kinetic energy and the ZPE term become negligible with respect to the external and the SCE potentials, for any \textit{fixed} $\alpha$. However, due to equation \eqref{eq:expansionomega}, it is always possible to find $t$ large enough such that the ZPE term becomes dominant with respect to the external potential, resulting in a stronger confinement of the density. 

It is also very interesting to note that, as Fig.~\ref{fig:densitiesSF} shows, the self-consistent KS SCE+ZPE densities do not improve systematically towards the exact many-body ones with respect to the bare KS SCE case. While for the energy evaluated at a given density the SCE+ZPE has been shown to approximate the exact many body energy \cite{GroKooGieSeiCohMorGor-JCTC-17} very closely at strong coupling (and better than SCE alone), the exact KS potential clearly is not well approximated at strong coupling by the SCE+ZPE. Most probably, the expansion of the potential at strong coupling is not uniform, having different relevant scaled variables in different regions of space (e.g., classically allowed and classically forbidden regions). This point needs further investigation, which will be the object of future work.

As for the ZPEisi approximation, it seems that it weakly acts as a correction to the SCE+ZPE approximation although towards the exact result, with a greater effect at lower correlation regimes. In table~\ref{tab:totalenergies}, we compare the total energies within the different approximations (see also Fig.~\ref{fig:energies}).
Finally, in table~\ref{tab:KSenergies} we report the highest occupied KS eigenvalue comparing it with the exact many-body chemical potential for SCE and LDA (for functionals with the ZPE case this comparison would not be meanigful as the Hxc potential does not go to zero when $|x|\to \infty$).

\section{Conclusions and Outlook}\label{sec:conclusions}
Building upon a successful use in other fields (mainly nonlinear photonics), in this paper we suggest and implement the spectral renormalization method as a mean to obtain numerical solutions for Kohn-Sham-type equations, focussing on the challenging case of xc functionals based on the strictly-correlated regime. 

We have implemented this scheme on benchmark problems using the KS equation as a test bed, obtaining for the first time self-consistent results with the SCE+ZPE functional, at least for systems not too close to the strong-coupling regime. These results showed that the ZPE functional implemented self consistently does not improve results with respect to the bare SCE case alone, suggesting the exact KS potential at strong coupling must have a different kind of  expansion. 
We also have implemented the interaction-strength interpolation showing that it corrects partially the extreme features of the ZPE regime; finally, we obtained a better converged result for an LDA computation appeared in previous works.
The key features of the algorithm used in this work can be summarized as follows. (i) {\it Ease of implementation.} Most real-space algorithms are based on either eigenvalue type solvers or shooting methods where finite-difference scheme is the method of choice to discretize space (kinetic energy). As such, low numerical accuracy is often used (second order as an example) as a trade off to numerical implementation. In our proposed scheme, the accuracy is spectral, all potentials are computed pseudo-spectrally and the implementation is straightforward making the coding process easy and simple. (ii) {\it Dependence on initial guesses.} It is a well-known fact that many fixed point iteration algorithms either fail to converge or show poor dependence on initial guesses. To test the robustness of the SR algorithm against initial guesses, we ran many simulations where (a) Gaussian-type (narrow or wide with either low or high amplitude) and (b) random function with uniform distribution were used to initialize the iteration. These tests were successful by using both the LDA and the SCE functional as approximations for the xc energy.
The work reported in this paper gives us a solid basis to explore in future work new physics related to strongly correlated many-body Anderson localization by using the KS SCE approach. 
\begin{acknowledgments}
	Financial support was provided by the H2020/MSCA-IF ``SCP-Disorder'' [grant 797247] and the European Research Council under H2020/ERC Consolidator Grant ``corr-DFT'' [grant 648932]. 
\end{acknowledgments}


\appendix
\section{Spectral Renormalization Algorithm}\label{app:spectralrenorm}
For clarity, we shall now write up the fixed-point scheme in detail for both bosons and fermions.
\subsection{Bosons}
In this case only one orbital $\phi(\rv)$ is needed. Let $\phi^{(n)}({\bf r})$ be the approximation for the orbital at iteration $n$. Then we define
\begin{align}
\phi^{(n+1/2)}(\rv)  & = \frac{\phi^{(n)}({\bf r})}{{||\phi^{(n)}||}}, \\
\rho^{(n+1/2)}(\rv)  & = N |\phi^{(n+1/2)}(\rv)|^2,
\end{align}
and
and the updated (new) eigenvalue $\varepsilon^{(n+1/2)}$ and orbital $\phi^{(n+1)}(\rv)$ are obtained in Fourier space from the following fixed point iteration:
\begin{widetext}
\begin{align}\label{eq:specbos}
\varepsilon^{(n+1/2)}&= \int\mathrm{d}\mathbf{r}\bigg\lbrace\frac{1}{2}\vert \nabla\phi^{(n+1/2)}(\mathbf{r})\vert^2+v_\mathrm{KS}\left([\rho^{(n+1/2)}],\mathbf{r}\right)\vert \phi^{(n+1/2)}(\mathbf{r})\vert^2\bigg\rbrace,\\\nonumber\\
\hat{\phi}^{(n+1)}(\kv)  &=- \,\frac{\;\Four\Big[\;v_\mathrm{KS}\Big([\rho^{(n+1/2)}],\rv\Big)\;\phi^{(n+1/2)}(\rv)\;\Big]\;-\theta\left(\varepsilon^{(n+1/2)}\right)\left(\varepsilon^{(n+1/2)}+c\right)\hat{\phi}^{(n+1/2)}(\kv) }
{\displaystyle\frac{\vert\kv\vert ^2}2-\varepsilon^{(n+1/2)}+\theta\left(\varepsilon^{(n+1/2)}\right)\left(\varepsilon^{(n+1/2)}+c\right)},
\end{align}
\end{widetext}
where $||\phi^{(j)}||^2$ is the norm of the orbital defined by $||\phi^{(j)}||^2 = \int|\phi^{(j)}(\mathbf{r})|^2d(\mathbf{r})$ and  $\theta(x)$ is the Heaviside step function. After each iteration step the orbital in real space is obtained from
\begin{equation}
\phi^{(n+1)}(\rv) = \Four^{-1} \left[\hat{\phi}^{(n+1)}(\kv)\right] \;.
\end{equation}
\subsection{Fermions}
For simplicity we consider an even number of fermions with spin $1/2$. In contrast to the bosonic case, now $N/2$ orbitals $\phi_i(\rv)$ are needed (see Eq.~\eqref{eq:rhoF}).
At each iteration step $n$ we normalize each orbital 
\begin{equation}
\phi_i^{(n+1/2)}(\rv) = \frac{\phi_i^{(n)}({\bf r})}{{||\phi_i^{(n)}||}} \;, \qquad(i=1,\dots,N/2)\;.
\end{equation}
followed by proper orthogonalization procedure such as as the Gram-Schmidt or the Löwdin scheme.
Then,
\begin{equation}
 \rho^{(n+1/2)}=2\sum\limits_{i=1}^{N/2}\big|\phi_i^{(n+1/2)}(\rv)\big|^2.
 \end{equation}
 Now, the fixed-point iteration reads
\begin{widetext}
\begin{align}
\varepsilon_i^{(n+1/2)}
&=\int\td\rv\left\{\frac{\,\big|\nabla\phi_i^{(n+1/2)}(\rv)\big|^2\,}2\,+\,v_\mathrm{KS}\Big([\rho^{(n+1/2)}],\rv\Big)\,\big|\phi_i^{(n+1/2)}(\rv)\big|^2\right\},\\\nonumber\\
\hat{\phi_i}^{(n+1)}(\kv)  &=- \,\frac{\;\Four\Big[\;v_\mathrm{KS}\Big([\rho^{(n+1/2)}],\rv\Big)\;\phi_i^{(n+1/2)}(\rv)\;\Big]\;-\theta\left(\varepsilon_i^{(n+1/2)}\right)\left(\varepsilon_i^{(n+1/2)}+c\right)\hat{\phi}^{(n+1/2)}(\kv) }
{\displaystyle\frac{\vert\kv\vert ^2}2-\varepsilon_i^{(n+1/2)}+\theta\left(\varepsilon_i^{(n+1/2)}\right)\left(\varepsilon_i^{(n+1/2)}+c\right)}.
\end{align}
\end{widetext}
As in the previous case, the iteration is concluded by reverting to real space
\begin{equation}
\phi_i^{(n+1)}(\rv) = \Four^{-1} \left[\hat{\phi}_i^{(n+1)}(\kv)\right].
\end{equation}

\section{Details of the 1D numerical implementation}\label{app:morenum1D}
\subsection{Computation of the Hartree potential}
\begin{figure}
{\includegraphics[width=0.45\textwidth]{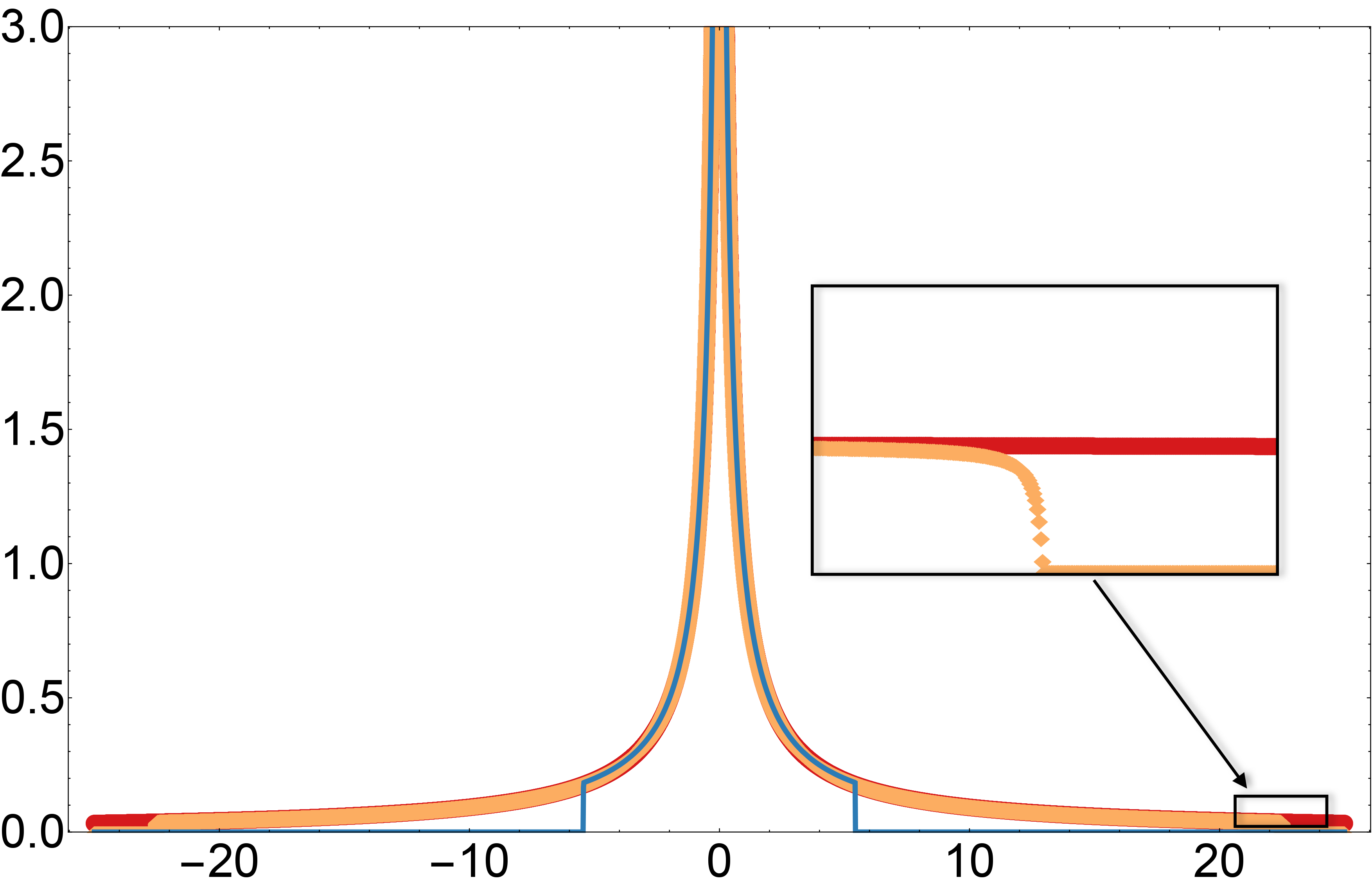}}
\caption{The interaction $v_\mathrm{int}^{\mathrm{Q1D}}(x)$ is numerically unstable already at $x\approx 6$ (solid line, in blue). On the other hand, solution of Eq.~\eqref{eq:veediffeq} (in red) remains numerically stable at greater distances. Finally, we plot in orange the  ``mollified" effective Coulomb interaction $v^p_{\rm int}$ which overlaps with $v_\mathrm{int}^{\mathrm{Q1D}}(x)$ in the inner part of the box, while smoothly going to 0 at the boundaries.}
\label{fig:veecomparisons}
\end{figure}

Another aspect, with respect to previous implementations of the SR algorithm for the Gross-Pitaevskii equation \cite{MusYan-JOSA-04,AblHor-EPJST-09,AblFokMus-JFM-06,AblAntBakIla-PRA-12,AkkGhoMus-JPB-08} is the numerical evaluation of the Hartree term 
\begin{align}
v_{\rm H}\big([\rho],\rv\big)=\int\td\rv'\,v_{\rm int}\big(|\rv-\rv'|\big)\,\rho(\rv') \;,
\end{align}
which, due to the slow decay of the effective Coulomb interaction ($v_{\rm int}$), makes the implementation of Fourier based convolutional schemes problematic. However, by properly adjusting the effective Coulomb interaction (while keeping its true far field Coulomb characteristic) one can still take advantage of the fast discrete Fourier transform algorithm. 
One way to achieve this goal is to multiply the Hartree interaction by the so-called mollifier $p(x)$ defined by
\begin{equation}
p(x) =\left\{ \begin{array}{rlr}
  \exp {\left[\frac{\xi}{x^2-L_c^2}\right]} & \mbox{for} & |x| < L_c\\ \\ 0 \;\;\;\;\;\;\;\;\;\;\;\;\;\;\;\;\;\; & \mbox{for} & |x| > L_c
  \end{array}\right.
\end{equation}
with $L_c$ denoting an arbitrary cutoff which, for a computational domain of size $L_d$, can be chosen as $0.9L_d$ and $\xi$ is a positive small parameter of the order of $10^{-4}$.
The ``mollified" effective Coulomb interaction $v^p_{\rm int}$ is now given by $v^p_{\rm int} \equiv p(x)v_{\rm int}$ with the Hartree potential
\begin{align}
v_{\rm H}\big([\rho],\rv\big)=\int\td\rv'\,v^p_{\rm int}\big(|\rv-\rv'|\big)\,\rho(\rv') \;.
\end{align}
With this at hand, the numerical evaluation of the Hartree potential follows from the convolutional FFT algorithm, i.e.,
\begin{align}
v_{\rm H}\big([\rho],\rv\big) = 2\pi \Four^{-1}\left[ \Four \left( v^p_{\rm int} \right)  \Four \left( \rho \right) \right]\;.
\end{align}
In Fig.~\ref{fig:veecomparisons}, we show a typical behavior of the effective Coulomb potential with (orange) and without (red) a mollifier. For comparison, we also add the direct evaluation of $v_{\rm int}^{\rm Q1D}(x)$ (in blue).

\subsection{LDA for electrons in 1D}\label{app:NumDet}
\begin{figure}[h!]
\includegraphics[width=.5\textwidth]{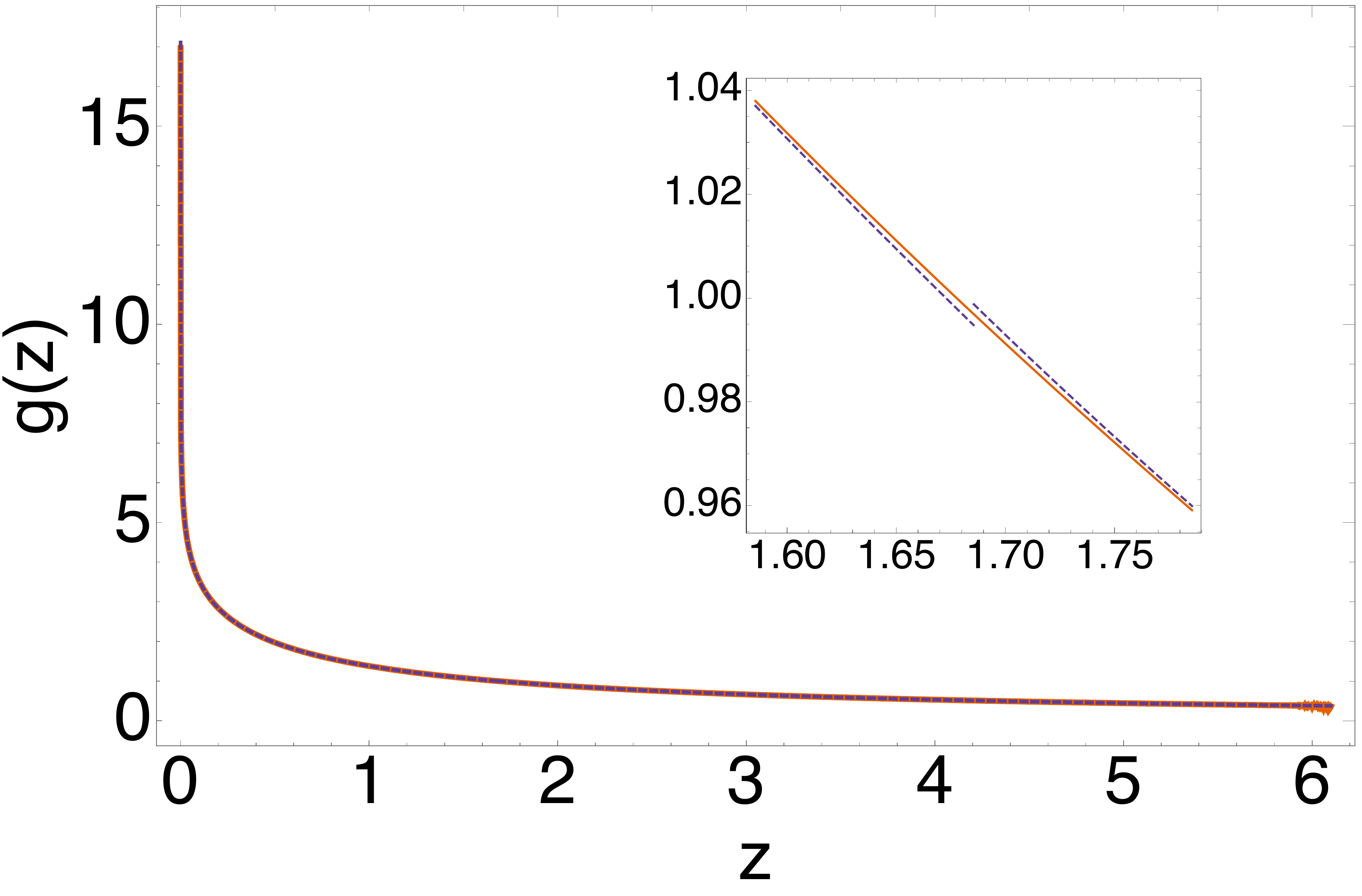}
\caption{Comparison between $g(z)$ (red, solid) and $\tilde{g}(z)$ (violet, dashed). The two curves are almost on top. Moreover, 
$\tilde{g}(z)$ is still evaluated exactly at large arguments, while $g(z)$ shows numerical instability already at $z\sim 6$.}
\label{fig:gapprox}
\end{figure}

The energy density $\epsilon_{\rm xc}(\dens)$ can be decomposed in its exchange and correlation part respectively
\begin{equation}
\epsilon_{\rm xc}(\dens)=\epsilon_{\rm x}(\dens)+\epsilon_{\rm c}(\dens).
\end{equation}
For electrons in 1D with the interaction given by Eq.~\eqref{eq:vintQ1D}, the exchange energy is known analytically,
\begin{align}
\epsilon_{\mathrm{x}}(\dens)=-\frac12\,\dens\,g\big(b\pi\,\dens\big),
\label{exLDAQ1D}\end{align}
with the function
\begin{eqnarray}
g(z)&=&\frac1{2z^2}\bigg\{-\gamma+\exp(z^2)\text{Ei}(-z^2)-2\ln z\nonumber\\
&&\hspace*{20mm}+\,G^{2,2}_{2,3}\Big(z^2\Big|\begin{array}{c}1,\frac32\\1,1,2\end{array}\Big)\bigg\}.
\end{eqnarray}
Here, $\gamma=-0.577216$ is Euler's constant, $\mathrm{Ei}(u)=-{\mathcal{P}}\int_{-u}^\infty\frac{e^{-z}}{z}\mathrm{d} t$ is the exponential integral function,
and $G$ denotes the Meijer $G$ function.
As the analytical $g(z)$ is numerically unstable, we expand $g(z)=g_<(z)+\mathcal{O}(z^{14})$ for small $z$ and
$g(z)=g_>(z)+\mathcal{O}(z^{-16})$ for large $t$,
\begin{subequations}\label{eq:gsmgla}
\begin{eqnarray}
g_<(z)&=&\sum_{m=0}^{7}\Big[a_m-b_m\log(z)\Big]z^{2m},\\
g_>(z)&=&\frac{\pi^{3/2}}{2z}-\frac{\log(z)}{z^2}+\sum_{m=0}^{7}c_mz^{-2m}.
\end{eqnarray}
\end{subequations}
As the minimum difference $|g_<(z)-g_>(z)|$, occuring at $z=z_0\approx1.68$, is extremely small, we simply truncate the two expansions, to obtain the 
approximation
\begin{equation}
g(z)\approx\tilde{g}(z)\equiv\begin{cases}&g_<(z)\quad z\leq z_0,\\&g_>(z)\quad z>z_0. \end{cases}
\end{equation}
Notice the small discontinuity of $\tilde{g}(z)$ at $z_0$ in Fig.~\ref{fig:gapprox}.
The coefficients $a_m,b_m,c_m$ of Eq.~\eqref{eq:gsmgla} are listed in Table \ref{tab:coefftab}.

\begin{table*}
\caption{List of coefficients used in Eq.~\eqref{eq:gsmgla}}
\label{tab:coefftab}
\begin{ruledtabular}
\begin{tabular}{ccccccccc}
   & $m=0$ & $m=1$ & $m=2$ & $m=3$  & $m=4$ & $m=5$ & $m=6$ & $m=7$\\
$a_m$ & $1.21139$ & $0.132454$ & $0.0276020$ & $0.00533283$  & $0.000892750$ & $0.0001297100$ & $0.0000165559$ & $0$\\
$b_m$ & $1.00000$ & $0.166667$ & $0.0333333$ & $0.00595238$ & $0.000126263$ & $0.0000152625$ & $0$ & $0$ \\
$c_m$ & $0$ & $-1.2886100$ & $-0.16666667$ & $0.1$ & $-0.142857$ & $0.3333333333$ & $-1.09090909091$ & $4.61538$
\end{tabular}
\end{ruledtabular}
\end{table*}

For the correlation energy, we use the parametrization from Ref.~\cite{CasSorSen-PRB-06},
\begin{equation}\label{eq:ecLDAQ1D}
\epsilon_{\mathrm{c}}(r_s)=-\frac{1}{2}\frac{r_s}{A+Br_s^n+Cr_s^2}\,\ln(1+\alpha r_s+\beta r_s^m),
\end{equation}
with the 1D density parameter $r_s=\frac1{2\dens}$.  
The values of the 7 parameters $A,B,C,n,\alpha,\beta,m$ are different for different values of $b$, 
see Table IV of Ref.~\cite{CasSorSen-PRB-06}. For convenience of the reader, we report them in Table \ref{tab:corrtab}.
\begin{table}
\centering
\caption{The set of parameters used in Eq.~\eqref{eq:ecLDAQ1D} for $b=0.1$ in the interaction \eqref{eq:vintQ1D}}
\label{tab:corrtab}
\begin{ruledtabular}
\begin{tabular}{ccccccc}
   $A$ &  $B$ &  $C$ & $\alpha$ & $\beta$ & $n$ & $m$  \\
$4.66$  & $2.092$  & $3.735$  & $23.63$ & $109.9$ & $1.379$ & $1.837$  \\
\end{tabular}
\end{ruledtabular}
\end{table}


\bibliography{bib_clean}

\begin{thebibliography}{68}
\expandafter\ifx\csname natexlab\endcsname\relax\def\natexlab#1{#1}\fi
\expandafter\ifx\csname bibnamefont\endcsname\relax
  \def\bibnamefont#1{#1}\fi
\expandafter\ifx\csname bibfnamefont\endcsname\relax
  \def\bibfnamefont#1{#1}\fi
\expandafter\ifx\csname citenamefont\endcsname\relax
  \def\citenamefont#1{#1}\fi
\expandafter\ifx\csname url\endcsname\relax
  \def\url#1{\texttt{#1}}\fi
\expandafter\ifx\csname urlprefix\endcsname\relax\def\urlprefix{URL }\fi
\providecommand{\bibinfo}[2]{#2}
\providecommand{\eprint}[2][]{\url{#2}}

\bibitem[{\citenamefont{Malet and Gori-Giorgi}(2012)}]{MalGor-PRL-12}
\bibinfo{author}{\bibfnamefont{F.}~\bibnamefont{Malet}} \bibnamefont{and}
  \bibinfo{author}{\bibfnamefont{P.}~\bibnamefont{Gori-Giorgi}},
  \bibinfo{journal}{Phys. Rev. Lett.} \textbf{\bibinfo{volume}{109}},
  \bibinfo{pages}{246402} (\bibinfo{year}{2012}).

\bibitem[{\citenamefont{Malet et~al.}(2013)\citenamefont{Malet, Mirtschink,
  Cremon, Reimann, and Gori-Giorgi}}]{MalMirCreReiGor-PRB-13}
\bibinfo{author}{\bibfnamefont{F.}~\bibnamefont{Malet}},
  \bibinfo{author}{\bibfnamefont{A.}~\bibnamefont{Mirtschink}},
  \bibinfo{author}{\bibfnamefont{J.~C.} \bibnamefont{Cremon}},
  \bibinfo{author}{\bibfnamefont{S.~M.} \bibnamefont{Reimann}},
  \bibnamefont{and}
  \bibinfo{author}{\bibfnamefont{P.}~\bibnamefont{Gori-Giorgi}},
  \bibinfo{journal}{Phys. Rev. B} \textbf{\bibinfo{volume}{87}},
  \bibinfo{pages}{115146} (\bibinfo{year}{2013}).

\bibitem[{\citenamefont{Mendl et~al.}(2014)\citenamefont{Mendl, Malet, and
  Gori-Giorgi}}]{MenMalGor-PRB-14}
\bibinfo{author}{\bibfnamefont{C.~B.} \bibnamefont{Mendl}},
  \bibinfo{author}{\bibfnamefont{F.}~\bibnamefont{Malet}}, \bibnamefont{and}
  \bibinfo{author}{\bibfnamefont{P.}~\bibnamefont{Gori-Giorgi}},
  \bibinfo{journal}{Phys. Rev. B} \textbf{\bibinfo{volume}{89}},
  \bibinfo{pages}{125106} (\bibinfo{year}{2014}).

\bibitem[{\citenamefont{Malet et~al.}(2015)\citenamefont{Malet, Mirtschink,
  Mendl, Bjerlin, Karabulut, Reimann, and
  Gori-Giorgi}}]{MalMirMenBjeKarReiGor-PRL-15}
\bibinfo{author}{\bibfnamefont{F.}~\bibnamefont{Malet}},
  \bibinfo{author}{\bibfnamefont{A.}~\bibnamefont{Mirtschink}},
  \bibinfo{author}{\bibfnamefont{C.}~\bibnamefont{Mendl}},
  \bibinfo{author}{\bibfnamefont{J.}~\bibnamefont{Bjerlin}},
  \bibinfo{author}{\bibfnamefont{E.}~\bibnamefont{Karabulut}},
  \bibinfo{author}{\bibfnamefont{S.}~\bibnamefont{Reimann}}, \bibnamefont{and}
  \bibinfo{author}{\bibfnamefont{P.}~\bibnamefont{Gori-Giorgi}},
  \bibinfo{journal}{Physical Review Letters} \textbf{\bibinfo{volume}{115}},
  \bibinfo{pages}{033006} (\bibinfo{year}{2015}).

\bibitem[{\citenamefont{Khoo et~al.}(2019)\citenamefont{Khoo, Lin, Lindsey, and
  Ying}}]{KhoLinLin-ARXIV-19}
\bibinfo{author}{\bibfnamefont{Y.}~\bibnamefont{Khoo}},
  \bibinfo{author}{\bibfnamefont{L.}~\bibnamefont{Lin}},
  \bibinfo{author}{\bibfnamefont{M.}~\bibnamefont{Lindsey}}, \bibnamefont{and}
  \bibinfo{author}{\bibfnamefont{L.}~\bibnamefont{Ying}},
  \bibinfo{journal}{arXiv preprint arXiv:1905.08322}  (\bibinfo{year}{2019}).

\bibitem[{\citenamefont{Seidl}(1999)}]{Sei-PRA-99}
\bibinfo{author}{\bibfnamefont{M.}~\bibnamefont{Seidl}},
  \bibinfo{journal}{Physical Review A} \textbf{\bibinfo{volume}{60}},
  \bibinfo{pages}{4387} (\bibinfo{year}{1999}).

\bibitem[{\citenamefont{Seidl et~al.}(2007)\citenamefont{Seidl, Gori-Giorgi,
  and Savin}}]{SeiGorSav-PRA-07}
\bibinfo{author}{\bibfnamefont{M.}~\bibnamefont{Seidl}},
  \bibinfo{author}{\bibfnamefont{P.}~\bibnamefont{Gori-Giorgi}},
  \bibnamefont{and} \bibinfo{author}{\bibfnamefont{A.}~\bibnamefont{Savin}},
  \bibinfo{journal}{Physical Review A} \textbf{\bibinfo{volume}{75}},
  \bibinfo{pages}{042511} (\bibinfo{year}{2007}).

\bibitem[{\citenamefont{Gori-Giorgi et~al.}(2009)\citenamefont{Gori-Giorgi,
  Vignale, and Seidl}}]{GorVigSei-JCTC-09}
\bibinfo{author}{\bibfnamefont{P.}~\bibnamefont{Gori-Giorgi}},
  \bibinfo{author}{\bibfnamefont{G.}~\bibnamefont{Vignale}}, \bibnamefont{and}
  \bibinfo{author}{\bibfnamefont{M.}~\bibnamefont{Seidl}},
  \bibinfo{journal}{Journal of chemical theory and computation}
  \textbf{\bibinfo{volume}{5}}, \bibinfo{pages}{743} (\bibinfo{year}{2009}).

\bibitem[{\citenamefont{Lewin}(2018)}]{Lew-CRM-18}
\bibinfo{author}{\bibfnamefont{M.}~\bibnamefont{Lewin}}, \bibinfo{journal}{C.
  R. Math.} \textbf{\bibinfo{volume}{356}}, \bibinfo{pages}{449}
  (\bibinfo{year}{2018}).

\bibitem[{\citenamefont{Cotar et~al.}(2018)\citenamefont{Cotar, Friesecke, and
  Kl{\"u}ppelberg}}]{CotFriKlu-ARMA-18}
\bibinfo{author}{\bibfnamefont{C.}~\bibnamefont{Cotar}},
  \bibinfo{author}{\bibfnamefont{G.}~\bibnamefont{Friesecke}},
  \bibnamefont{and}
  \bibinfo{author}{\bibfnamefont{C.}~\bibnamefont{Kl{\"u}ppelberg}},
  \bibinfo{journal}{Arch. Ration. Mech. An.} \textbf{\bibinfo{volume}{228}},
  \bibinfo{pages}{891} (\bibinfo{year}{2018}).

\bibitem[{\citenamefont{Mirtschink et~al.}(2013)\citenamefont{Mirtschink,
  Seidl, and Gori-Giorgi}}]{MirSeiGor-PRL-13}
\bibinfo{author}{\bibfnamefont{A.}~\bibnamefont{Mirtschink}},
  \bibinfo{author}{\bibfnamefont{M.}~\bibnamefont{Seidl}}, \bibnamefont{and}
  \bibinfo{author}{\bibfnamefont{P.}~\bibnamefont{Gori-Giorgi}},
  \bibinfo{journal}{Phys. Rev. Lett.} \textbf{\bibinfo{volume}{111}},
  \bibinfo{pages}{126402} (\bibinfo{year}{2013}).

\bibitem[{\citenamefont{Colombo et~al.}(2015)\citenamefont{Colombo, De~Pascale,
  and Di~Marino}}]{ColDepDim-CJM-15}
\bibinfo{author}{\bibfnamefont{M.}~\bibnamefont{Colombo}},
  \bibinfo{author}{\bibfnamefont{L.}~\bibnamefont{De~Pascale}},
  \bibnamefont{and}
  \bibinfo{author}{\bibfnamefont{S.}~\bibnamefont{Di~Marino}},
  \bibinfo{journal}{Canad. J. Math} \textbf{\bibinfo{volume}{67}},
  \bibinfo{pages}{350} (\bibinfo{year}{2015}).

\bibitem[{\citenamefont{Seidl et~al.}(2017)\citenamefont{Seidl, Di~Marino,
  Gerolin, Nenna, Giesbertz, and Gori-Giorgi}}]{SeiDiMGerNenGieGor-arxiv-17}
\bibinfo{author}{\bibfnamefont{M.}~\bibnamefont{Seidl}},
  \bibinfo{author}{\bibfnamefont{S.}~\bibnamefont{Di~Marino}},
  \bibinfo{author}{\bibfnamefont{A.}~\bibnamefont{Gerolin}},
  \bibinfo{author}{\bibfnamefont{L.}~\bibnamefont{Nenna}},
  \bibinfo{author}{\bibfnamefont{K.~J.} \bibnamefont{Giesbertz}},
  \bibnamefont{and}
  \bibinfo{author}{\bibfnamefont{P.}~\bibnamefont{Gori-Giorgi}},
  \bibinfo{journal}{arXiv preprint arXiv:1702.05022}  (\bibinfo{year}{2017}).

\bibitem[{\citenamefont{Vuckovic and Gori-Giorgi}(2017)}]{VucGor-JPCL-17}
\bibinfo{author}{\bibfnamefont{S.}~\bibnamefont{Vuckovic}} \bibnamefont{and}
  \bibinfo{author}{\bibfnamefont{P.}~\bibnamefont{Gori-Giorgi}},
  \bibinfo{journal}{The Journal of Physical Chemistry Letters}
  \textbf{\bibinfo{volume}{8}}, \bibinfo{pages}{2799} (\bibinfo{year}{2017}),
  \bibinfo{note}{pMID: 28581751}.

\bibitem[{\citenamefont{Vuckovic}(2019)}]{Vuc-JCTC-19}
\bibinfo{author}{\bibfnamefont{S.}~\bibnamefont{Vuckovic}},
  \bibinfo{journal}{Journal of Chemical Theory and Computation}
  \textbf{\bibinfo{volume}{15}}, \bibinfo{pages}{3580} (\bibinfo{year}{2019}).

\bibitem[{\citenamefont{Gould and Vuckovic}(2019)}]{GouVuc-JCP-19}
\bibinfo{author}{\bibfnamefont{T.}~\bibnamefont{Gould}} \bibnamefont{and}
  \bibinfo{author}{\bibfnamefont{S.}~\bibnamefont{Vuckovic}},
  \bibinfo{journal}{The Journal of Chemical Physics}
  \textbf{\bibinfo{volume}{151}}, \bibinfo{pages}{184101}
  (\bibinfo{year}{2019}).

\bibitem[{\citenamefont{Mendl and Lin}(2013)}]{MenLin-PRB-13}
\bibinfo{author}{\bibfnamefont{C.~B.} \bibnamefont{Mendl}} \bibnamefont{and}
  \bibinfo{author}{\bibfnamefont{L.}~\bibnamefont{Lin}},
  \bibinfo{journal}{Physical Review B} \textbf{\bibinfo{volume}{87}},
  \bibinfo{pages}{125106} (\bibinfo{year}{2013}).

\bibitem[{\citenamefont{Benamou et~al.}(2015)\citenamefont{Benamou, Carlier,
  Cuturi, Nenna, and Peyr{\'e}}}]{BenCarCutNenPey-SIAM-15}
\bibinfo{author}{\bibfnamefont{J.-D.} \bibnamefont{Benamou}},
  \bibinfo{author}{\bibfnamefont{G.}~\bibnamefont{Carlier}},
  \bibinfo{author}{\bibfnamefont{M.}~\bibnamefont{Cuturi}},
  \bibinfo{author}{\bibfnamefont{L.}~\bibnamefont{Nenna}}, \bibnamefont{and}
  \bibinfo{author}{\bibfnamefont{G.}~\bibnamefont{Peyr{\'e}}},
  \bibinfo{journal}{SIAM J. on Sci. Comput.} \textbf{\bibinfo{volume}{37}},
  \bibinfo{pages}{A1111} (\bibinfo{year}{2015}).

\bibitem[{\citenamefont{Friesecke and V\"ogler}(2018)}]{FriVog-SIMAJMA-18}
\bibinfo{author}{\bibfnamefont{G.}~\bibnamefont{Friesecke}} \bibnamefont{and}
  \bibinfo{author}{\bibfnamefont{D.}~\bibnamefont{V\"ogler}},
  \bibinfo{journal}{SIAM Journal on Mathematical Analysis}
  \textbf{\bibinfo{volume}{50}}, \bibinfo{pages}{3996} (\bibinfo{year}{2018}),
  \eprint{https://doi.org/10.1137/17M1150025},
  \urlprefix\url{https://doi.org/10.1137/17M1150025}.

\bibitem[{\citenamefont{Khoo and Ying}(2019)}]{KhoYin-SIAMJSC-19}
\bibinfo{author}{\bibfnamefont{Y.}~\bibnamefont{Khoo}} \bibnamefont{and}
  \bibinfo{author}{\bibfnamefont{L.}~\bibnamefont{Ying}},
  \bibinfo{journal}{SIAM Journal on Scientific Computing}
  \textbf{\bibinfo{volume}{41}}, \bibinfo{pages}{B773} (\bibinfo{year}{2019}),
  \eprint{https://doi.org/10.1137/18M1207478},
  \urlprefix\url{https://doi.org/10.1137/18M1207478}.

\bibitem[{\citenamefont{Lin et~al.}(2019)\citenamefont{Lin, Ho, Cuturi, and
  Jordan}}]{LinHoCutJor-ARXIV-19}
\bibinfo{author}{\bibfnamefont{T.}~\bibnamefont{Lin}},
  \bibinfo{author}{\bibfnamefont{N.}~\bibnamefont{Ho}},
  \bibinfo{author}{\bibfnamefont{M.}~\bibnamefont{Cuturi}}, \bibnamefont{and}
  \bibinfo{author}{\bibfnamefont{M.~I.} \bibnamefont{Jordan}},
  \bibinfo{journal}{arXiv preprint arXiv:1910.00152}  (\bibinfo{year}{2019}).

\bibitem[{\citenamefont{Alfonsi et~al.}(2019)\citenamefont{Alfonsi, Coyaud,
  Ehrlacher, and Lombardi}}]{AlfCoyEhrLom-arxiv-19}
\bibinfo{author}{\bibfnamefont{A.}~\bibnamefont{Alfonsi}},
  \bibinfo{author}{\bibfnamefont{R.}~\bibnamefont{Coyaud}},
  \bibinfo{author}{\bibfnamefont{V.}~\bibnamefont{Ehrlacher}},
  \bibnamefont{and} \bibinfo{author}{\bibfnamefont{D.}~\bibnamefont{Lombardi}},
  \bibinfo{journal}{arXiv preprint arXiv:1905.05663}  (\bibinfo{year}{2019}).

\bibitem[{\citenamefont{Buttazzo et~al.}(2012)\citenamefont{Buttazzo,
  De~Pascale, and Gori-Giorgi}}]{ButDepGor-PRA-12}
\bibinfo{author}{\bibfnamefont{G.}~\bibnamefont{Buttazzo}},
  \bibinfo{author}{\bibfnamefont{L.}~\bibnamefont{De~Pascale}},
  \bibnamefont{and}
  \bibinfo{author}{\bibfnamefont{P.}~\bibnamefont{Gori-Giorgi}},
  \bibinfo{journal}{Phys. Rev. A} \textbf{\bibinfo{volume}{85}},
  \bibinfo{pages}{062502} (\bibinfo{year}{2012}).

\bibitem[{\citenamefont{Cotar et~al.}(2013)\citenamefont{Cotar, Friesecke, and
  Kl\"uppelberg}}]{CotFriKlu-CPAM-13}
\bibinfo{author}{\bibfnamefont{C.}~\bibnamefont{Cotar}},
  \bibinfo{author}{\bibfnamefont{G.}~\bibnamefont{Friesecke}},
  \bibnamefont{and}
  \bibinfo{author}{\bibfnamefont{C.}~\bibnamefont{Kl\"uppelberg}},
  \bibinfo{journal}{Comm. Pure Appl. Math.} \textbf{\bibinfo{volume}{66}},
  \bibinfo{pages}{548} (\bibinfo{year}{2013}).

\bibitem[{\citenamefont{Ghosal et~al.}(2006)\citenamefont{Ghosal,
  G{\"u}{\c{c}}l{\"u}, Umrigar, Ullmo, and Baranger}}]{GhoGucUmrUllBar-NP-06}
\bibinfo{author}{\bibfnamefont{A.}~\bibnamefont{Ghosal}},
  \bibinfo{author}{\bibfnamefont{A.}~\bibnamefont{G{\"u}{\c{c}}l{\"u}}},
  \bibinfo{author}{\bibfnamefont{C.}~\bibnamefont{Umrigar}},
  \bibinfo{author}{\bibfnamefont{D.}~\bibnamefont{Ullmo}}, \bibnamefont{and}
  \bibinfo{author}{\bibfnamefont{H.~U.} \bibnamefont{Baranger}},
  \bibinfo{journal}{Nature Physics} \textbf{\bibinfo{volume}{2}},
  \bibinfo{pages}{336} (\bibinfo{year}{2006}).

\bibitem[{\citenamefont{Rontani et~al.}(2006)\citenamefont{Rontani, Cavazzoni,
  Bellucci, and Goldoni}}]{RonCavBelGol-JCP-06}
\bibinfo{author}{\bibfnamefont{M.}~\bibnamefont{Rontani}},
  \bibinfo{author}{\bibfnamefont{C.}~\bibnamefont{Cavazzoni}},
  \bibinfo{author}{\bibfnamefont{D.}~\bibnamefont{Bellucci}}, \bibnamefont{and}
  \bibinfo{author}{\bibfnamefont{G.}~\bibnamefont{Goldoni}},
  \bibinfo{journal}{The Journal of chemical physics}
  \textbf{\bibinfo{volume}{124}}, \bibinfo{pages}{124102}
  (\bibinfo{year}{2006}).

\bibitem[{\citenamefont{Ghosal et~al.}(2007)\citenamefont{Ghosal,
  G{\"u}{\c{c}}l{\"u}, Umrigar, Ullmo, and Baranger}}]{GhoGucUmrUllBar-PRB-07}
\bibinfo{author}{\bibfnamefont{A.}~\bibnamefont{Ghosal}},
  \bibinfo{author}{\bibfnamefont{A.}~\bibnamefont{G{\"u}{\c{c}}l{\"u}}},
  \bibinfo{author}{\bibfnamefont{C.}~\bibnamefont{Umrigar}},
  \bibinfo{author}{\bibfnamefont{D.}~\bibnamefont{Ullmo}}, \bibnamefont{and}
  \bibinfo{author}{\bibfnamefont{H.~U.} \bibnamefont{Baranger}},
  \bibinfo{journal}{Physical Review B} \textbf{\bibinfo{volume}{76}},
  \bibinfo{pages}{085341} (\bibinfo{year}{2007}).

\bibitem[{\citenamefont{Wagner et~al.}(2014)\citenamefont{Wagner, Baker,
  Stoudenmire, Burke, and White}}]{WagBakStoBurWhi-PRB-14}
\bibinfo{author}{\bibfnamefont{L.~O.} \bibnamefont{Wagner}},
  \bibinfo{author}{\bibfnamefont{T.~E.} \bibnamefont{Baker}},
  \bibinfo{author}{\bibfnamefont{E.~M.} \bibnamefont{Stoudenmire}},
  \bibinfo{author}{\bibfnamefont{K.}~\bibnamefont{Burke}}, \bibnamefont{and}
  \bibinfo{author}{\bibfnamefont{S.~R.} \bibnamefont{White}},
  \bibinfo{journal}{Phys. Rev. B} \textbf{\bibinfo{volume}{90}},
  \bibinfo{pages}{045109} (\bibinfo{year}{2014}),
  \urlprefix\url{https://link.aps.org/doi/10.1103/PhysRevB.90.045109}.

\bibitem[{\citenamefont{Grossi et~al.}(2019)\citenamefont{Grossi, Seidl,
  Gori-Giorgi, and Giesbertz}}]{GroSeiGorGie-PRA-19}
\bibinfo{author}{\bibfnamefont{J.}~\bibnamefont{Grossi}},
  \bibinfo{author}{\bibfnamefont{M.}~\bibnamefont{Seidl}},
  \bibinfo{author}{\bibfnamefont{P.}~\bibnamefont{Gori-Giorgi}},
  \bibnamefont{and} \bibinfo{author}{\bibfnamefont{K.~J.}
  \bibnamefont{Giesbertz}}, \bibinfo{journal}{Physical Review A}
  \textbf{\bibinfo{volume}{99}}, \bibinfo{pages}{052504}
  (\bibinfo{year}{2019}).

\bibitem[{\citenamefont{Ablowitz and Musslimani}(2005)}]{AblMus-OL-05}
\bibinfo{author}{\bibfnamefont{M.~J.} \bibnamefont{Ablowitz}} \bibnamefont{and}
  \bibinfo{author}{\bibfnamefont{Z.~H.} \bibnamefont{Musslimani}},
  \bibinfo{journal}{Optics letters} \textbf{\bibinfo{volume}{30}},
  \bibinfo{pages}{2140} (\bibinfo{year}{2005}).

\bibitem[{\citenamefont{Musslimani and Yang}(2004)}]{MusYan-JOSA-04}
\bibinfo{author}{\bibfnamefont{Z.~H.} \bibnamefont{Musslimani}}
  \bibnamefont{and} \bibinfo{author}{\bibfnamefont{J.}~\bibnamefont{Yang}},
  \bibinfo{journal}{JOSA B} \textbf{\bibinfo{volume}{21}}, \bibinfo{pages}{973}
  (\bibinfo{year}{2004}).

\bibitem[{\citenamefont{Ablowitz and Horikis}(2009)}]{AblHor-EPJST-09}
\bibinfo{author}{\bibfnamefont{M.}~\bibnamefont{Ablowitz}} \bibnamefont{and}
  \bibinfo{author}{\bibfnamefont{T.}~\bibnamefont{Horikis}},
  \bibinfo{journal}{The European Physical Journal Special Topics}
  \textbf{\bibinfo{volume}{173}}, \bibinfo{pages}{147} (\bibinfo{year}{2009}).

\bibitem[{\citenamefont{Ablowitz et~al.}(2006)\citenamefont{Ablowitz, Fokas,
  and Musslimani}}]{AblFokMus-JFM-06}
\bibinfo{author}{\bibfnamefont{M.}~\bibnamefont{Ablowitz}},
  \bibinfo{author}{\bibfnamefont{A.}~\bibnamefont{Fokas}}, \bibnamefont{and}
  \bibinfo{author}{\bibfnamefont{Z.}~\bibnamefont{Musslimani}},
  \bibinfo{journal}{Journal of Fluid Mechanics} \textbf{\bibinfo{volume}{562}},
  \bibinfo{pages}{313} (\bibinfo{year}{2006}).

\bibitem[{\citenamefont{Ablowitz et~al.}(2012)\citenamefont{Ablowitz, Antar,
  Bak{\i}rta{\c{s}}, and Ilan}}]{AblAntBakIla-PRA-12}
\bibinfo{author}{\bibfnamefont{M.~J.} \bibnamefont{Ablowitz}},
  \bibinfo{author}{\bibfnamefont{N.}~\bibnamefont{Antar}},
  \bibinfo{author}{\bibfnamefont{{\.I}.}~\bibnamefont{Bak{\i}rta{\c{s}}}},
  \bibnamefont{and} \bibinfo{author}{\bibfnamefont{B.}~\bibnamefont{Ilan}},
  \bibinfo{journal}{Physical Review A} \textbf{\bibinfo{volume}{86}},
  \bibinfo{pages}{033804} (\bibinfo{year}{2012}).

\bibitem[{\citenamefont{Akkermans et~al.}(2008)\citenamefont{Akkermans, Ghosh,
  and Musslimani}}]{AkkGhoMus-JPB-08}
\bibinfo{author}{\bibfnamefont{E.}~\bibnamefont{Akkermans}},
  \bibinfo{author}{\bibfnamefont{S.}~\bibnamefont{Ghosh}}, \bibnamefont{and}
  \bibinfo{author}{\bibfnamefont{Z.~H.} \bibnamefont{Musslimani}},
  \bibinfo{journal}{Journal of Physics B: Atomic, Molecular and Optical
  Physics} \textbf{\bibinfo{volume}{41}}, \bibinfo{pages}{045302}
  (\bibinfo{year}{2008}).

\bibitem[{\citenamefont{Bednarek et~al.}(2003)\citenamefont{Bednarek, Szafran,
  Chwiej, and Adamowski}}]{BedSzaChwAda-PRB-03}
\bibinfo{author}{\bibfnamefont{S.}~\bibnamefont{Bednarek}},
  \bibinfo{author}{\bibfnamefont{B.}~\bibnamefont{Szafran}},
  \bibinfo{author}{\bibfnamefont{T.}~\bibnamefont{Chwiej}}, \bibnamefont{and}
  \bibinfo{author}{\bibfnamefont{J.}~\bibnamefont{Adamowski}},
  \bibinfo{journal}{Physical Review B} \textbf{\bibinfo{volume}{68}},
  \bibinfo{pages}{045328} (\bibinfo{year}{2003}).

\bibitem[{\citenamefont{Giuliani and Vignale}(2005)}]{GiuVig-BOOK-05}
\bibinfo{author}{\bibfnamefont{G.~F.} \bibnamefont{Giuliani}} \bibnamefont{and}
  \bibinfo{author}{\bibfnamefont{G.}~\bibnamefont{Vignale}},
  \emph{\bibinfo{title}{Quantum Theory of the Electron Liquid}}
  (\bibinfo{publisher}{Cambridge University Press}, \bibinfo{address}{New
  York}, \bibinfo{year}{2005}).

\bibitem[{\citenamefont{Casula et~al.}(2006)\citenamefont{Casula, Sorella, and
  Senatore}}]{CasSorSen-PRB-06}
\bibinfo{author}{\bibfnamefont{M.}~\bibnamefont{Casula}},
  \bibinfo{author}{\bibfnamefont{S.}~\bibnamefont{Sorella}}, \bibnamefont{and}
  \bibinfo{author}{\bibfnamefont{G.}~\bibnamefont{Senatore}},
  \bibinfo{journal}{Physical Review B} \textbf{\bibinfo{volume}{74}},
  \bibinfo{pages}{245427} (\bibinfo{year}{2006}).

\bibitem[{\citenamefont{Abedinpour et~al.}(2007)\citenamefont{Abedinpour,
  Polini, Xianlong, and Tosi}}]{AbePolXiaTos-EJPB-07}
\bibinfo{author}{\bibfnamefont{S.~H.} \bibnamefont{Abedinpour}},
  \bibinfo{author}{\bibfnamefont{M.}~\bibnamefont{Polini}},
  \bibinfo{author}{\bibfnamefont{G.}~\bibnamefont{Xianlong}}, \bibnamefont{and}
  \bibinfo{author}{\bibfnamefont{M.}~\bibnamefont{Tosi}}, \bibinfo{journal}{The
  European Physical Journal B} \textbf{\bibinfo{volume}{56}},
  \bibinfo{pages}{127} (\bibinfo{year}{2007}).

\bibitem[{\citenamefont{Hohenberg and Kohn}(1964)}]{HohKoh-PR-64}
\bibinfo{author}{\bibfnamefont{P.}~\bibnamefont{Hohenberg}} \bibnamefont{and}
  \bibinfo{author}{\bibfnamefont{W.}~\bibnamefont{Kohn}},
  \bibinfo{journal}{Physical review} \textbf{\bibinfo{volume}{136}},
  \bibinfo{pages}{B864} (\bibinfo{year}{1964}).

\bibitem[{\citenamefont{Levy}(1979)}]{Lev-PNAS-79}
\bibinfo{author}{\bibfnamefont{M.}~\bibnamefont{Levy}},
  \bibinfo{journal}{Proceedings of the National Academy of Sciences}
  \textbf{\bibinfo{volume}{76}}, \bibinfo{pages}{6062} (\bibinfo{year}{1979}).

\bibitem[{\citenamefont{Ceperley and Alder}(1980)}]{CepAld-PRL-80}
\bibinfo{author}{\bibfnamefont{D.~M.} \bibnamefont{Ceperley}} \bibnamefont{and}
  \bibinfo{author}{\bibfnamefont{B.~J.} \bibnamefont{Alder}},
  \bibinfo{journal}{Phys. Rev. Lett.} \textbf{\bibinfo{volume}{45}},
  \bibinfo{pages}{566} (\bibinfo{year}{1980}).

\bibitem[{\citenamefont{Perdew and Zunger}(1981)}]{PerZun-PRB-81}
\bibinfo{author}{\bibfnamefont{J.~P.} \bibnamefont{Perdew}} \bibnamefont{and}
  \bibinfo{author}{\bibfnamefont{A.}~\bibnamefont{Zunger}},
  \bibinfo{journal}{Physical Review B} \textbf{\bibinfo{volume}{23}},
  \bibinfo{pages}{5048} (\bibinfo{year}{1981}).

\bibitem[{\citenamefont{Vosko et~al.}(1980)\citenamefont{Vosko, Wilk, and
  Nusair}}]{VosWilNus-CJP-80}
\bibinfo{author}{\bibfnamefont{S.~J.} \bibnamefont{Vosko}},
  \bibinfo{author}{\bibfnamefont{L.}~\bibnamefont{Wilk}}, \bibnamefont{and}
  \bibinfo{author}{\bibfnamefont{M.}~\bibnamefont{Nusair}},
  \bibinfo{journal}{Can. J. Phys.} \textbf{\bibinfo{volume}{{58}}},
  \bibinfo{pages}{1200} (\bibinfo{year}{1980}).

\bibitem[{\citenamefont{Perdew and Wang}(1992)}]{PerWan-PRB-92}
\bibinfo{author}{\bibfnamefont{J.~P.} \bibnamefont{Perdew}} \bibnamefont{and}
  \bibinfo{author}{\bibfnamefont{Y.}~\bibnamefont{Wang}},
  \bibinfo{journal}{Physical Review B} \textbf{\bibinfo{volume}{45}},
  \bibinfo{pages}{13244} (\bibinfo{year}{1992}).

\bibitem[{\citenamefont{Ma et~al.}(2012)\citenamefont{Ma, Pilati, Troyer, and
  Dai}}]{MaPilTroDai-NAT-12}
\bibinfo{author}{\bibfnamefont{P.~N.} \bibnamefont{Ma}},
  \bibinfo{author}{\bibfnamefont{S.}~\bibnamefont{Pilati}},
  \bibinfo{author}{\bibfnamefont{M.}~\bibnamefont{Troyer}}, \bibnamefont{and}
  \bibinfo{author}{\bibfnamefont{X.}~\bibnamefont{Dai}},
  \bibinfo{journal}{Nature Physics} \textbf{\bibinfo{volume}{8}},
  \bibinfo{pages}{601} (\bibinfo{year}{2012}).

\bibitem[{\citenamefont{Zecca et~al.}(2004)\citenamefont{Zecca, Gori-Giorgi,
  Moroni, and Bachelet}}]{ZecGorMorBac-PRB-04}
\bibinfo{author}{\bibfnamefont{L.}~\bibnamefont{Zecca}},
  \bibinfo{author}{\bibfnamefont{P.}~\bibnamefont{Gori-Giorgi}},
  \bibinfo{author}{\bibfnamefont{S.}~\bibnamefont{Moroni}}, \bibnamefont{and}
  \bibinfo{author}{\bibfnamefont{G.~B.} \bibnamefont{Bachelet}},
  \bibinfo{journal}{Phys. Rev. B} \textbf{\bibinfo{volume}{70}},
  \bibinfo{pages}{205127} (\bibinfo{year}{2004}).

\bibitem[{\citenamefont{Toulouse et~al.}(2004)\citenamefont{Toulouse, Savin,
  and Flad}}]{TouSavFla-IJQC-04}
\bibinfo{author}{\bibfnamefont{J.}~\bibnamefont{Toulouse}},
  \bibinfo{author}{\bibfnamefont{A.}~\bibnamefont{Savin}}, \bibnamefont{and}
  \bibinfo{author}{\bibfnamefont{H.-J.} \bibnamefont{Flad}},
  \bibinfo{journal}{Int. J. Quantum. Chem.} \textbf{\bibinfo{volume}{100}},
  \bibinfo{pages}{1047} (\bibinfo{year}{2004}).

\bibitem[{\citenamefont{Paziani et~al.}(2006)\citenamefont{Paziani, Moroni,
  Gori-Giorgi, and Bachelet}}]{PazMorGorBac-PRB-06}
\bibinfo{author}{\bibfnamefont{S.}~\bibnamefont{Paziani}},
  \bibinfo{author}{\bibfnamefont{S.}~\bibnamefont{Moroni}},
  \bibinfo{author}{\bibfnamefont{P.}~\bibnamefont{Gori-Giorgi}},
  \bibnamefont{and} \bibinfo{author}{\bibfnamefont{G.~B.}
  \bibnamefont{Bachelet}}, \bibinfo{journal}{Phys. Rev. B}
  \textbf{\bibinfo{volume}{{73}}}, \bibinfo{pages}{155111}
  (\bibinfo{year}{2006}).

\bibitem[{\citenamefont{Attaccalite et~al.}(2002)\citenamefont{Attaccalite,
  Moroni, Gori-Giorgi, and Bachelet}}]{AttMorGorBac-PRL-02}
\bibinfo{author}{\bibfnamefont{C.}~\bibnamefont{Attaccalite}},
  \bibinfo{author}{\bibfnamefont{S.}~\bibnamefont{Moroni}},
  \bibinfo{author}{\bibfnamefont{P.}~\bibnamefont{Gori-Giorgi}},
  \bibnamefont{and} \bibinfo{author}{\bibfnamefont{G.~B.}
  \bibnamefont{Bachelet}}, \bibinfo{journal}{Phys. Rev. Lett.}
  \textbf{\bibinfo{volume}{88}}, \bibinfo{pages}{256601}
  (\bibinfo{year}{2002}).

\bibitem[{\citenamefont{De~Palo et~al.}(2004)\citenamefont{De~Palo, Conti, and
  Moroni}}]{DepConMor-PRB-04}
\bibinfo{author}{\bibfnamefont{S.}~\bibnamefont{De~Palo}},
  \bibinfo{author}{\bibfnamefont{S.}~\bibnamefont{Conti}}, \bibnamefont{and}
  \bibinfo{author}{\bibfnamefont{S.}~\bibnamefont{Moroni}},
  \bibinfo{journal}{Physical Review B} \textbf{\bibinfo{volume}{69}},
  \bibinfo{pages}{035109} (\bibinfo{year}{2004}).

\bibitem[{\citenamefont{K{\"a}rkk{\"a}inen
  et~al.}(2003)\citenamefont{K{\"a}rkk{\"a}inen, Koskinen, Reimann, and
  Manninen}}]{KarKosReiMan-PRB-03}
\bibinfo{author}{\bibfnamefont{K.}~\bibnamefont{K{\"a}rkk{\"a}inen}},
  \bibinfo{author}{\bibfnamefont{M.}~\bibnamefont{Koskinen}},
  \bibinfo{author}{\bibfnamefont{S.}~\bibnamefont{Reimann}}, \bibnamefont{and}
  \bibinfo{author}{\bibfnamefont{M.}~\bibnamefont{Manninen}},
  \bibinfo{journal}{Physical Review B} \textbf{\bibinfo{volume}{68}},
  \bibinfo{pages}{205322} (\bibinfo{year}{2003}).

\bibitem[{\citenamefont{Helbig et~al.}(2011)\citenamefont{Helbig, Fuks, Casula,
  Verstraete, Marques, Tokatly, and Rubio}}]{HelFukCasVerMarTokRub-PRA-11}
\bibinfo{author}{\bibfnamefont{N.}~\bibnamefont{Helbig}},
  \bibinfo{author}{\bibfnamefont{J.~I.} \bibnamefont{Fuks}},
  \bibinfo{author}{\bibfnamefont{M.}~\bibnamefont{Casula}},
  \bibinfo{author}{\bibfnamefont{M.~J.} \bibnamefont{Verstraete}},
  \bibinfo{author}{\bibfnamefont{M.~A.} \bibnamefont{Marques}},
  \bibinfo{author}{\bibfnamefont{I.}~\bibnamefont{Tokatly}}, \bibnamefont{and}
  \bibinfo{author}{\bibfnamefont{A.}~\bibnamefont{Rubio}},
  \bibinfo{journal}{Physical Review A} \textbf{\bibinfo{volume}{83}},
  \bibinfo{pages}{032503} (\bibinfo{year}{2011}).

\bibitem[{\citenamefont{Grossi et~al.}(2017)\citenamefont{Grossi, Kooi,
  Giesbertz, Seidl, Cohen, Mori-S{\'a}nchez, and
  Gori-Giorgi}}]{GroKooGieSeiCohMorGor-JCTC-17}
\bibinfo{author}{\bibfnamefont{J.}~\bibnamefont{Grossi}},
  \bibinfo{author}{\bibfnamefont{D.~P.} \bibnamefont{Kooi}},
  \bibinfo{author}{\bibfnamefont{K.~J.~H.} \bibnamefont{Giesbertz}},
  \bibinfo{author}{\bibfnamefont{M.}~\bibnamefont{Seidl}},
  \bibinfo{author}{\bibfnamefont{A.~J.} \bibnamefont{Cohen}},
  \bibinfo{author}{\bibfnamefont{P.}~\bibnamefont{Mori-S{\'a}nchez}},
  \bibnamefont{and}
  \bibinfo{author}{\bibfnamefont{P.}~\bibnamefont{Gori-Giorgi}},
  \bibinfo{journal}{J. Chem. Theory Comput.} \textbf{\bibinfo{volume}{13}},
  \bibinfo{pages}{6089} (\bibinfo{year}{2017}).

\bibitem[{\citenamefont{Langreth and Perdew}(1975)}]{LanPer-SSC-75}
\bibinfo{author}{\bibfnamefont{D.~C.} \bibnamefont{Langreth}} \bibnamefont{and}
  \bibinfo{author}{\bibfnamefont{J.~P.} \bibnamefont{Perdew}},
  \bibinfo{journal}{Solid. State Commun.} \textbf{\bibinfo{volume}{17}},
  \bibinfo{pages}{1425} (\bibinfo{year}{1975}).

\bibitem[{\citenamefont{Seidl et~al.}(1999)\citenamefont{Seidl, Perdew, and
  Levy}}]{SeiPerLev-PRA-99}
\bibinfo{author}{\bibfnamefont{M.}~\bibnamefont{Seidl}},
  \bibinfo{author}{\bibfnamefont{J.~P.} \bibnamefont{Perdew}},
  \bibnamefont{and} \bibinfo{author}{\bibfnamefont{M.}~\bibnamefont{Levy}},
  \bibinfo{journal}{Physical Review A} \textbf{\bibinfo{volume}{59}},
  \bibinfo{pages}{51} (\bibinfo{year}{1999}).

\bibitem[{\citenamefont{Seidl et~al.}(2000)\citenamefont{Seidl, Perdew, and
  Kurth}}]{SeiPerKur-PRL-00}
\bibinfo{author}{\bibfnamefont{M.}~\bibnamefont{Seidl}},
  \bibinfo{author}{\bibfnamefont{J.~P.} \bibnamefont{Perdew}},
  \bibnamefont{and} \bibinfo{author}{\bibfnamefont{S.}~\bibnamefont{Kurth}},
  \bibinfo{journal}{Phys. Rev. Lett.} \textbf{\bibinfo{volume}{84}},
  \bibinfo{pages}{5070} (\bibinfo{year}{2000}).

\bibitem[{\citenamefont{Liu and Burke}(2009)}]{LiuBur-JCP-09}
\bibinfo{author}{\bibfnamefont{Z.~F.} \bibnamefont{Liu}} \bibnamefont{and}
  \bibinfo{author}{\bibfnamefont{K.}~\bibnamefont{Burke}}, \bibinfo{journal}{J.
  Chem. Phys.} \textbf{\bibinfo{volume}{{131}}}, \bibinfo{pages}{124124}
  (\bibinfo{year}{2009}).

\bibitem[{\citenamefont{Fabiano et~al.}(2016)\citenamefont{Fabiano,
  Gori-Giorgi, Seidl, and Della~Sala}}]{FabGorSeiDel-JCTC-16}
\bibinfo{author}{\bibfnamefont{E.}~\bibnamefont{Fabiano}},
  \bibinfo{author}{\bibfnamefont{P.}~\bibnamefont{Gori-Giorgi}},
  \bibinfo{author}{\bibfnamefont{M.}~\bibnamefont{Seidl}}, \bibnamefont{and}
  \bibinfo{author}{\bibfnamefont{F.}~\bibnamefont{Della~Sala}},
  \bibinfo{journal}{J. Chem. Theory. Comput.} \textbf{\bibinfo{volume}{12}},
  \bibinfo{pages}{4885} (\bibinfo{year}{2016}).

\bibitem[{\citenamefont{Giarrusso et~al.}(2018)\citenamefont{Giarrusso,
  Gori-Giorgi, Della~Sala, and Fabiano}}]{GiaGorDelFab-JCP-18}
\bibinfo{author}{\bibfnamefont{S.}~\bibnamefont{Giarrusso}},
  \bibinfo{author}{\bibfnamefont{P.}~\bibnamefont{Gori-Giorgi}},
  \bibinfo{author}{\bibfnamefont{F.}~\bibnamefont{Della~Sala}},
  \bibnamefont{and} \bibinfo{author}{\bibfnamefont{E.}~\bibnamefont{Fabiano}},
  \bibinfo{journal}{J. Chem. Phys.} \textbf{\bibinfo{volume}{148}},
  \bibinfo{pages}{134106} (\bibinfo{year}{2018}).

\bibitem[{\citenamefont{Vuckovic et~al.}(2018)\citenamefont{Vuckovic,
  Gori-Giorgi, Della~Sala, and Fabiano}}]{VucGorDelFab-JPCL-18}
\bibinfo{author}{\bibfnamefont{S.}~\bibnamefont{Vuckovic}},
  \bibinfo{author}{\bibfnamefont{P.}~\bibnamefont{Gori-Giorgi}},
  \bibinfo{author}{\bibfnamefont{F.}~\bibnamefont{Della~Sala}},
  \bibnamefont{and} \bibinfo{author}{\bibfnamefont{E.}~\bibnamefont{Fabiano}},
  \bibinfo{journal}{J. Phys. Chem. Lett.} \textbf{\bibinfo{volume}{9}},
  \bibinfo{pages}{3137} (\bibinfo{year}{2018}).

\bibitem[{\citenamefont{Fabiano et~al.}(2019)\citenamefont{Fabiano, Smiga,
  Giarrusso, Daas, Della~Sala, Grabowski, and
  Gori-Giorgi}}]{FabSmiGiaDaaDelGraGor-JCTC-19}
\bibinfo{author}{\bibfnamefont{E.}~\bibnamefont{Fabiano}},
  \bibinfo{author}{\bibfnamefont{S.}~\bibnamefont{Smiga}},
  \bibinfo{author}{\bibfnamefont{S.}~\bibnamefont{Giarrusso}},
  \bibinfo{author}{\bibfnamefont{T.~J.} \bibnamefont{Daas}},
  \bibinfo{author}{\bibfnamefont{F.}~\bibnamefont{Della~Sala}},
  \bibinfo{author}{\bibfnamefont{I.}~\bibnamefont{Grabowski}},
  \bibnamefont{and}
  \bibinfo{author}{\bibfnamefont{P.}~\bibnamefont{Gori-Giorgi}},
  \bibinfo{journal}{Journal of chemical theory and computation}
  \textbf{\bibinfo{volume}{15}}, \bibinfo{pages}{1006} (\bibinfo{year}{2019}).

\bibitem[{\citenamefont{Constantin}(2019)}]{Con-PRB-19}
\bibinfo{author}{\bibfnamefont{L.~A.} \bibnamefont{Constantin}},
  \bibinfo{journal}{Phys. Rev. B} \textbf{\bibinfo{volume}{99}},
  \bibinfo{pages}{085117} (\bibinfo{year}{2019}).

\bibitem[{\citenamefont{Vuckovic et~al.}(2020)\citenamefont{Vuckovic, Fabiano,
  Gori-Giorgi, and Burke}}]{VucFabGorBur-arxiv-20}
\bibinfo{author}{\bibfnamefont{S.}~\bibnamefont{Vuckovic}},
  \bibinfo{author}{\bibfnamefont{E.}~\bibnamefont{Fabiano}},
  \bibinfo{author}{\bibfnamefont{P.}~\bibnamefont{Gori-Giorgi}},
  \bibnamefont{and} \bibinfo{author}{\bibfnamefont{K.}~\bibnamefont{Burke}}
  (\bibinfo{year}{2020}), \eprint{arXiv:2001.06364}.

\bibitem[{\citenamefont{Zarenia et~al.}(2017)\citenamefont{Zarenia, Neilson,
  Partoens, and Peeters}}]{ZarNeiParPee-PRB-17}
\bibinfo{author}{\bibfnamefont{M.}~\bibnamefont{Zarenia}},
  \bibinfo{author}{\bibfnamefont{D.}~\bibnamefont{Neilson}},
  \bibinfo{author}{\bibfnamefont{B.}~\bibnamefont{Partoens}}, \bibnamefont{and}
  \bibinfo{author}{\bibfnamefont{F.~M.} \bibnamefont{Peeters}},
  \bibinfo{journal}{Phys. Rev. B} \textbf{\bibinfo{volume}{95}},
  \bibinfo{pages}{115438} (\bibinfo{year}{2017}),
  \urlprefix\url{https://link.aps.org/doi/10.1103/PhysRevB.95.115438}.

\bibitem[{\citenamefont{Malet et~al.}(2014)\citenamefont{Malet, Mirtschink,
  Giesbertz, Wagner, and Gori-Giorgi}}]{MalMirGieWagGor-PCCP-14}
\bibinfo{author}{\bibfnamefont{F.}~\bibnamefont{Malet}},
  \bibinfo{author}{\bibfnamefont{A.}~\bibnamefont{Mirtschink}},
  \bibinfo{author}{\bibfnamefont{K.~J.~H.} \bibnamefont{Giesbertz}},
  \bibinfo{author}{\bibfnamefont{L.~O.} \bibnamefont{Wagner}},
  \bibnamefont{and}
  \bibinfo{author}{\bibfnamefont{P.}~\bibnamefont{Gori-Giorgi}},
  \bibinfo{journal}{Phys. Chem. Chem. Phys.} \textbf{\bibinfo{volume}{16}},
  \bibinfo{pages}{14551} (\bibinfo{year}{2014}).

\bibitem[{\citenamefont{Weideman and Reddy}(2000)}]{WeiRed-ACM-00}
\bibinfo{author}{\bibfnamefont{J.~A.} \bibnamefont{Weideman}} \bibnamefont{and}
  \bibinfo{author}{\bibfnamefont{S.~C.} \bibnamefont{Reddy}},
  \bibinfo{journal}{ACM Transactions on Mathematical Software (TOMS)}
  \textbf{\bibinfo{volume}{26}}, \bibinfo{pages}{465} (\bibinfo{year}{2000}).

\bibitem[{\citenamefont{Gori-Giorgi and Seidl}(2010)}]{GorSei-PCCP-10}
\bibinfo{author}{\bibfnamefont{P.}~\bibnamefont{Gori-Giorgi}} \bibnamefont{and}
  \bibinfo{author}{\bibfnamefont{M.}~\bibnamefont{Seidl}},
  \bibinfo{journal}{Phys. Chem. Chem. Phys} \textbf{\bibinfo{volume}{12}},
  \bibinfo{pages}{14405} (\bibinfo{year}{2010}).

\end{thebibliography}
\end{document}